\documentclass[prd,twocolumn,showpacs,nofootinbib,amsmath,amssymb,floatfix,superscriptaddress,showkeys]{revtex4}
\usepackage{multirow}
\usepackage{graphicx}% Include figure files
\usepackage{epstopdf}
\usepackage{dcolumn}% Align table columns on decimal point
\usepackage{bm}% bold math
\usepackage{threeparttable}
\usepackage{subfigure}
\usepackage{color,txfonts}
\usepackage{ulem}
\definecolor{blue0}{rgb}{0,0,0.6}
\usepackage[colorlinks,linkcolor=blue0,anchorcolor=blue0,citecolor=blue0,urlcolor=blue0]{hyperref}

\newcommand{\jcap}{J. Cosmol. Astropart. Phys.}

\newcommand{\mnras}{Mon. Not. R. Astron. Soc.}

\newcommand{\aap}{Astron. Astrophys}

\newcommand{\apjl}{Astrophys. J.}

\newcommand{\beq}{\begin{equation}}
\newcommand{\eeq}{\end{equation}}
\newcommand{\beqa}{\begin{eqnarray}}
\newcommand{\eeqa}{\end{eqnarray}}

\begin{document}

\title{Constraints on ultracompact minihalos from the extragalactic gamma-ray background observation}

\author{Xing-Fu Zhang}
\affiliation{Laboratory for Relativistic Astrophysics, Department of Physics, Guangxi University, Nanning 530004, China}
\author{Ji-Gui Cheng}
\affiliation{Laboratory for Relativistic Astrophysics, Department of Physics, Guangxi University, Nanning 530004, China}
\author{Ben-Yang Zhu}
\affiliation{Laboratory for Relativistic Astrophysics, Department of Physics, Guangxi University, Nanning 530004, China}
\author{Tian-Ci Liu}
\affiliation{Laboratory for Relativistic Astrophysics, Department of Physics, Guangxi University, Nanning 530004, China}
\author{Yun-Feng Liang}
\email[]{liangyf@gxu.edu.cn}
\affiliation{Laboratory for Relativistic Astrophysics, Department of Physics, Guangxi University, Nanning 530004, China}
%\author{xxx}
%\affiliation{xxx}
\author{En-Wei Liang}
\email[]{lew@gxu.edu.cn}
\affiliation{Laboratory for Relativistic Astrophysics, Department of Physics, Guangxi University, Nanning 530004, China}

\date{\today}

\begin{abstract}
Ultracompact minihalo (UCMH) is a special type of dark matter halo with a very steep density profile which may form in the early universe seeded by an overdense region or a primordial black hole. Constraints on its abundance give valuable information on the power spectrum of primordial perturbation. In this work, we update the constraints on the UCMH abundance in the universe using the extragalactic gamma-ray background (EGB) observation.
Comparing to previous works, we adopt the updated Fermi-LAT EGB measurement and derive constraints based on a full consideration of the astrophysical contributions. With these improvements, we place constraints on UCMH abundance 1-2 orders of magnitude better than previous results.
With the background components considered, we can also attempt to search for possible  additional components beyond the known astrophysical contributions.

\end{abstract}

\maketitle
\section{Introduction}

The extragalactic gamma-ray background (EGB) is the total contribution of gamma-ray integrated flux from all objects in the history of the extragalactic universe, and was first detected by the SAS-2 satellite \cite{1978ApJ...222..833F,1982A&A...109..352T} and subsequently measured by the Energetic Gamma Ray Experiment Telescope (EGRET) \cite{1994JPhG...20.1089O,1998ApJ...494..523S,2002astro.ph..1515W}.
Better measurements on EGB were achieved by the Large Area Telescope (LAT) \cite{lat09} instrument installed on the Fermi satellite \cite{fermi10_egbobs,fermi15igrb}.  The integrated flux of Fermi-LAT observation above 100 MeV is $1.29\pm0.07\times10^{-5}\,{\rm ph/cm^2/s/sr}$ (Model B of  \cite{fermi15igrb}), consistent with those of EGRET, $1.14\pm0.05\,{\rm ph/cm^2/s/sr}$ \cite{awstrong04egret}. The latest Fermi-LAT observation shows that a power law function with an exponential cutoff ($dN/dE=I_{100}({E}/{\rm 100\,MeV})^{-\gamma}\exp(-{E}/{E_{\rm cut}})$) can well describe the EGB spectrum \cite{fermi15igrb}, with spectral index of $\gamma=2.28\pm0.01$ and cutoff energy of $E_{\rm cut}=267\pm37\,{\rm GeV}$ (model B of \cite{fermi15igrb}).

Fermi-LAT also provides more accurate observations of extragalactic sources \cite{fermi3fgl,fermi4fgl}, allowing for a better understanding of the compositions of the EGB.
It has been shown that the extragalactic gamma-ray background is mainly contributed by Blazars, radio galaxies (RG) and star-forming galaxies (SFG) \cite{inoue11RG,fermi12sfg,zeng13_egbFSRQ,Ajello14_bllac,DiMauro14_magn, ajello15egbdm,qu19_bllac,zhd21blazar,Roth21sfgNat}.
Most of extragalactic sources detected in the Fermi sky are Blazars \cite{fermi4lac}, which can be further classified into two subclasses, BL Lac and flat-spectrum radio quasars (FSRQs) \cite{1997MmSAI..68...47P}.
Blazars could emit gamma rays through inverse Compton scattering (ICS) and / or hadronic processes and their contribution to EGB has been widely discussed \cite{zeng13_egbFSRQ,Ajello14_bllac,ajello15egbdm,qu19_bllac,zhd21blazar}.
Radio galaxies, although with lower gamma-ray luminosity for individual sources, are more numerous in the whole sky. The contribution of RGs to the EGB can be studied via the correlation between radio and gamma-ray luminosities \cite{inoue11RG,DiMauro14_magn}. The $\gamma$-ray radiation of SFG arises from the decay of neutral mesons produced in the inelastic interaction of cosmic rays with the interstellar medium and interstellar radiation field \cite{fermi12sfg,ajello2020sfg}.
Above 100 MeV, RG and SFG each contributes about 10-30\% of the observed photon flux of EGB, while blazars contribute about $\sim50\%$ \cite{ajello15egbdm}.
In addition, the contributions to EGB from other sources or processes include gamma-ray bursts (GRB) \cite{2007ApJ...656..306C}, pulsars at high galactic latitudes \cite{2010JCAP...01..005F}, inter-galactic shocks \cite{2000Natur.405..156L,2000ApJ...545..572T}, cascade processes of high energy cosmic rays \cite{1995PhRvL..75.3052D}, and so on.

Except for the aforementioned components, another source that may contribute to EGB is the dark matter (DM) \cite{Abazajian10_igrbDM,ajello15egbdm,Ando15_egbDM,DiMauro15_igrbDM,fermi15_igbdm, liuwei17cpc,hooper19igrb,arbey20_igrbPBH}. The existence of DM has been confirmed by many astronomical and cosmological observations, and it is likely to account for $\sim$26\% of the total energy density of the Universe \cite{planck2015}. DM has the potential to emit gamma-ray signals through annihilation or decay. The flux depends on the interaction cross-section of DM particles and the DM abundance \cite{1964ocr..book.....G,2009ApJ...707..979R}. Therefore, the DM properties (cross section or abundance) can be constrained by requiring the expected flux not higher than the actual measurements of the EGB spectrum.

In this work, we will focus on a particular DM halo model, i.e. ultracompact minihalos (UCMH) \cite{ricotti09_ucmh,scott09_UCMHprl,yang2011,yang11_ucmhCMB,yang2020}, and constrain their abundance in the Universe with the EGB observation. The UCMH is characterized by a very steep density profile ($\rho\propto r^{-9/4}$). If DM consists of Weakly Interacting Massive Particles (WIMPs), the UCMHs will be gamma-ray emitters due to the DM annihilation within them and the profile makes them have high expected gamma-ray flux compared to the normal DM halo (e.g. NFW \cite{navarro97nfw}, Einasto \cite{einasto}).  Constraints on the abundance of UCMHs or primordial black hole (PBHs) may provide valuable information on the power spectrum of the primordial perturbulation at small scale \cite{Josan10ucmhgamma,Bringmann12ucmhPS,aslanyan16_InflationPRL,Nakama18_igrbPS}.

In idealized cases, the UCMH can form in the early universe when the primordial density perturbations are between 10$^{-3}$ and 0.3 (a PBH will be produced if the amplitude of the perturbation is $\delta>0.3$ \cite{carr2010pbh}).
However it has been shown that the postulated steep inner profile can not appear in realistic simulations since the required initial conditions (self-similarity, radial infall, isolation, etc) for forming UCMHs can only be satisfied in idealized cases \cite{Delos18_noucmh,Delos18_cheng,adamek19_pbhucmh}.
Alternatively, PBHs formed in the early Universe can accret DM particles due to gravity and form UCMHs (a mixed WIMP-PBH dark matter model) \cite{adamek19_pbhucmh,yang2020}.
In this work, we give constraints from an observational aspect, regardless of the exact mechanism of UCMH formation. Our constraints on the UCMH abundance can be directly converted into constrants on the PBHs in the mixed model \cite{yang2020}. Furthermore, the derived constraints are also valid for the mini-spike around a astrophysical black hole \cite{Belikov14spikeIGRB,Lacroix18bh,cheng20,xzq2021}.

Comparing to previous works \cite{yang2011,yang2020}, our studies contain the following improvements. We use the updated Fermi-LAT EGB observation to perform the analysis.
%which is still missing at present.
In addition, in the previous works of limiting the abundance of UCMH with the EGB observations \cite{yang2011,yang2020,Nakama18_igrbPS}, they usually used the inclusive energy spectrum to provide relatively conservative constraints without considering the astrophysical components. We will alternatively derive restrictions based on a full consideration of the astrophysical contributions to obtain more realistic (though not that conservative) results.
With the background components considered, we can also attempt to search for possible signals / additional components beyond the background. Another motivation for our study of UCMH is that this type of objects was recently suggested to be able to better (compared to the traditional density profiles, e.g. NFW, Einasto) interpret the tentative 1.4 TeV $e^+e^-$ excess of DAMPE (Dark Matter Particle Explorer) \cite{dampe17_nature,huang18_dampe,zhaoyi2019,cheng20}. We therefore examine whether such a probability can accommodate the abundance upper limits derived from the EGB observation.

Through out this papper, we use the cosmological parameters from Planck2015 \cite{planck2015}, i.e. $\Omega_m=0.31$, $\Omega_\Lambda=0.69$ and $H_0=67.74\,{\rm km\,Mpc^{-1}s^{-1}}$.

\section{Methed}
\subsection{The model expected gamma-ray signal from a single UCMH}
UCMHs are growing spherical DM halos which are seeded by an overdense region in the early universe with initial density perturbations greater than 0.01\% (or alternatively seeded by a PBH).
The mass of UCMHs $M_u$ depends on their formation time and can be described as \cite{ricotti09_ucmh,scott09_UCMHprl}:
\beq
M_u(z)=\delta_m\left(\frac{1+z_{\rm eq}}{1+z}\right)
\eeq
where $\delta_{m}$ is the mass of the perturbation at the redshift of matter-radiation equality ($1+z_{\rm eq}\approx3260$). Since the accretion will be prevented after $z=10$, we assume the UCMHs stoped growing at $z=10$, i.e. $M(z<10)=M(z=10)$ \cite{scott09_UCMHprl}.
Compared to the amplitude of perturbations seen in CMB observation ($\sim10^{-5}$), the requred value for forming UCMH ($>10^{-3}$) is large.
The non-Gaussian perturbations at phase transitions can enhance the amplitudes at small scale, therefore the UCMHs are more likely borned at the epoches of phase transitions. The UCMHs produced at three phase transitions are usually considered in literature \cite{scott09_UCMHprl,Josan10ucmhgamma,yang2011,yang11_ucmhCMB}: electroweak symmetry breaking, the QCD confinement and $e^+e^-$ annihilation.
The $\delta_{m}$ for (QCD, EW, $e^{+}e^{-}$) epoches are $\delta_{m,{\rm \{EW, QCD, e+e-\}}}$=\{$5.6\times10^{-19}$, $1.1\times10^{-9}$, 0.33\}$\,{\rm M_\odot}$ \cite{scott09_UCMHprl} and the current masses of UCMHs are $M_{u}(0)$= \{$1.6\times10^{-7}$, $0.2$, $1.2\times10^5$\}$\,{\rm M_\odot}$, respectively. In fact, the chossen of the $\delta_m$ does not affect the predicted EGB spectrum of UCMH \cite{yang2011}.

\begin{figure*}[htp]
\begin{center}
\includegraphics[width=0.8\textwidth]{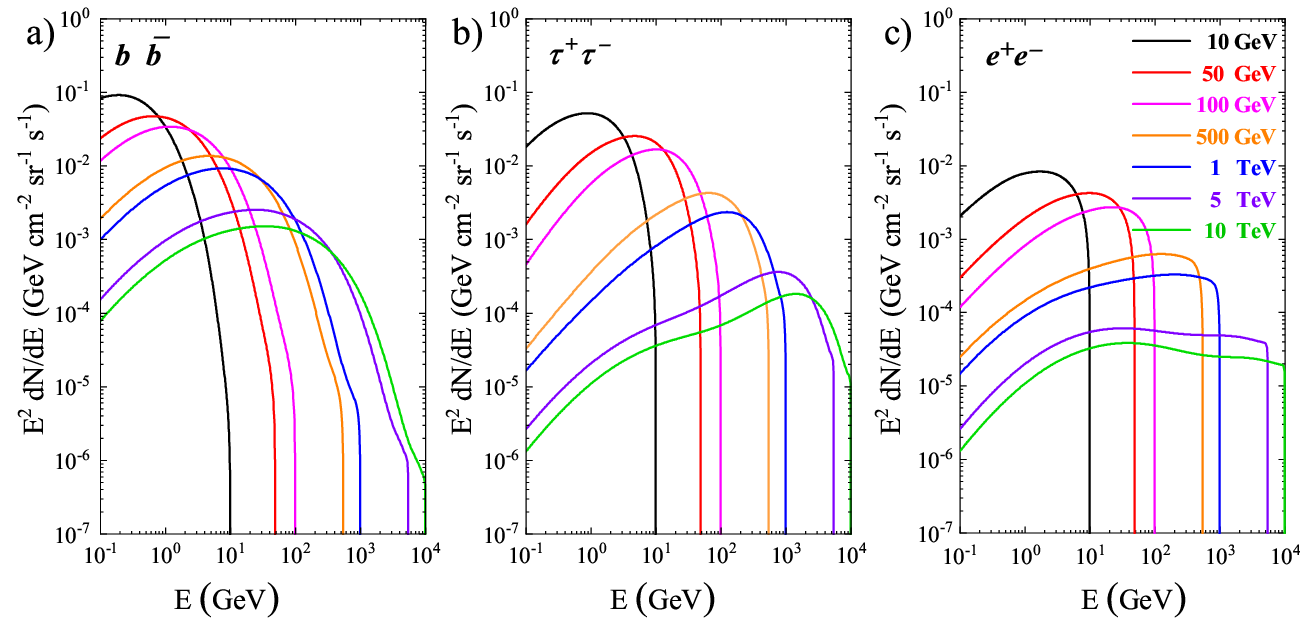}
\end{center}
\caption{The model-predicted EGB spectra from UCMHs ($f_u=1$) for different channels and DM masses.
The sharp cutoff at high energies close to $M_{\chi}$ could be a characteristic signature for DM search.}
\label{fig:ucmhsp}
\end{figure*}

UCMHs are predicted to form by the secondary infall of DM onto PBHs or initial DM overdensity produced by the primordial density perturbation.
The DM particles within the overdense region initially have an extremely small velocity dispersion.
UCMHs thus form via a spherically symmetric gravitational collapse (pure radial infall).
According to the secondary infall theory \cite{1984ApJ...281....1F,1985ApJS...58...39B}, the UCMHs will develop a self-similar power-law density profile $\rho\propto r^{-9/4}$. Such a steep profile is supported by both analytical solution \cite{1984ApJ...281....1F,1985ApJS...58...39B} and (idealized) N-body simulations \cite{vogelsberger09ucmhsim,ludlow10ucmhsim,Delos18_noucmh}.
Normalizing the $\rho(r)$ to make it have a halo mass of $M_u(z)$ within the truncated radius $R_u(z)$ gives the density profile of \cite{ricotti09_ucmh,scott09_UCMHprl}
\beq
\rho_u(r, z)=\frac{3 f_{\chi} M_u(z)}{16 \pi R_u(z)^{{3}/{4}} r^{{9}/{4}}}
\eeq
where $f_\chi=\Omega_\chi/(\Omega_b+\Omega_\chi)\approx0.83$ \cite{planck2015}. The profile truncated at a halo radius \cite{scott09_UCMHprl}
\beq
{R_u(z)}=0.019\left(\frac{1000}{z+1}\right)\left(\frac{M_u(z)}{\mathrm{M}_{\odot}}\right)^{{1}/{3}} {\rm pc}.
\eeq
Due to the DM annihilation, for the most inner region of the halo ($r<r_{\rm cut}$) the density is set to \cite{Ullio02dm}
\beq
\rho_{\max}(z)=\frac{m_{\chi}}{\langle\sigma v\rangle\left(t(z)-t_{i}\right)}
\eeq
where $m_\chi$ is the mass of DM particle, $\langle\sigma v\rangle$ the annihilation cross section, and $t(z)$ is the age of the Universe at redshift $z$. The $r_{\rm cut}$ is determined by requiring the $\rho_{\rm max}=\rho(r_{\rm cut})$.

If DM consists of WIMPs, it can produce gamma rays through annihilation or decay.
In this work, we mainly concern on the annihilation DM.
The expected gamma-ray flux emitted from a single UCMH can be expressed as
\beq
F(E)=\frac{1}{4\pi}\frac{\langle\sigma{v}\rangle}{2m_{\chi}^2}\frac{dN_\gamma}{dE}\times \iint_{\rm los}\rho_u^2(r)dsd\Omega
\eeq
where ${dN_\gamma}/{dE}$ is the photon yield per annihilation, which is calculated using PPPC4DMID \cite{pppc4}.

\subsection{Extragalactic $\gamma$-ray background from UCMHs}
For UCMHs with monochromatic mass function, the differential EGB energy spectrum contributed by UCMHs is expressed as \cite{bergstrom01_od,Ullio02dm,yang2020}:
\beq
\frac{d \phi_{\gamma}}{d E}=\frac{f_u\,\rho_{c,0}}{M_u(0)}\frac{c}{8 \pi}\frac{\langle\sigma v\rangle}{m_{\chi}^2} \int_{0}^{z_{u p}} d z \frac{e^{-\tau(E, z)}}{H(z)} \frac{d N_{\gamma}}{d E}(E',z) \int \rho_{u}^{2}(r,z) dV
\label{eq:egbsp}
\eeq
where $f_u$ is the present abundance of UCMHs (in terms of the fraction of the critical density $\rho_{c,0}$), $E'=E(1+z)$ is the photon energy at redshift $z$, $E$ is the observed photon energy, the $z_{\mathrm{up}}= m_{\chi} / E-1$ is the maximal redshift that a UCMH can contribute photons of energy $E$.
For the DM annihilation cross section, we adopt the thermal relic value ${\langle\sigma v\rangle}=3 \times 10^{-26} \mathrm{cm}^{3} / \mathrm{s}$ \cite{Steigman12thermalcs}; and for the Hubble parameter $H(z)=H_{0} \sqrt{\Omega_{M}(1+z)^{3}+\Omega_{\Lambda}}$, we use the cosmological parameters from Planck2015 \cite{planck2015}.
The $\tau$($E$,$z$) in Eq.(\ref{eq:egbsp}) is the optical depth, for which we consider the EBL absorption only and can be approximated by $\tau(E,z)\sim z/3.3\,({E}/{10\,\mathrm{GeV}})^{0.8}$
\cite{bergstrom01_od}. We use the approximation expression (rather than the models of e.g. \cite{DominguezA2011,Inoue13_ebl}) for better obtaining $\tau$ at high redshift (e.g. $z>10$).

In addition to the prompt gamma-ray emission, DM annihilation can produce energetic electrons/positrons, which generate gamma rays through inverse Compton scattering (ICS) off background radiation field.
In this work, we only consider the prompt gamma rays from DM annihilation but neglecting the secondary IC component. Tighter constraints are expected with the IC contribution included.
The contribution to the EGB from normal halos is also ignored, since it has been shown that the inclusion of them hardly affect the results \cite{yang2011} due to the much lower annihilation rate therein.
For the three typical channels $b \bar{b}, \tau^{+} \tau^{-}$ and $\mathrm{e}^{+} \mathrm{e}^{-}$, we show the DM-induced EGB spectra with $f_u=1$ in Fig.\ref{fig:ucmhsp}.

\begin{figure}[htp]
\begin{center}
\includegraphics[width=0.49\textwidth]{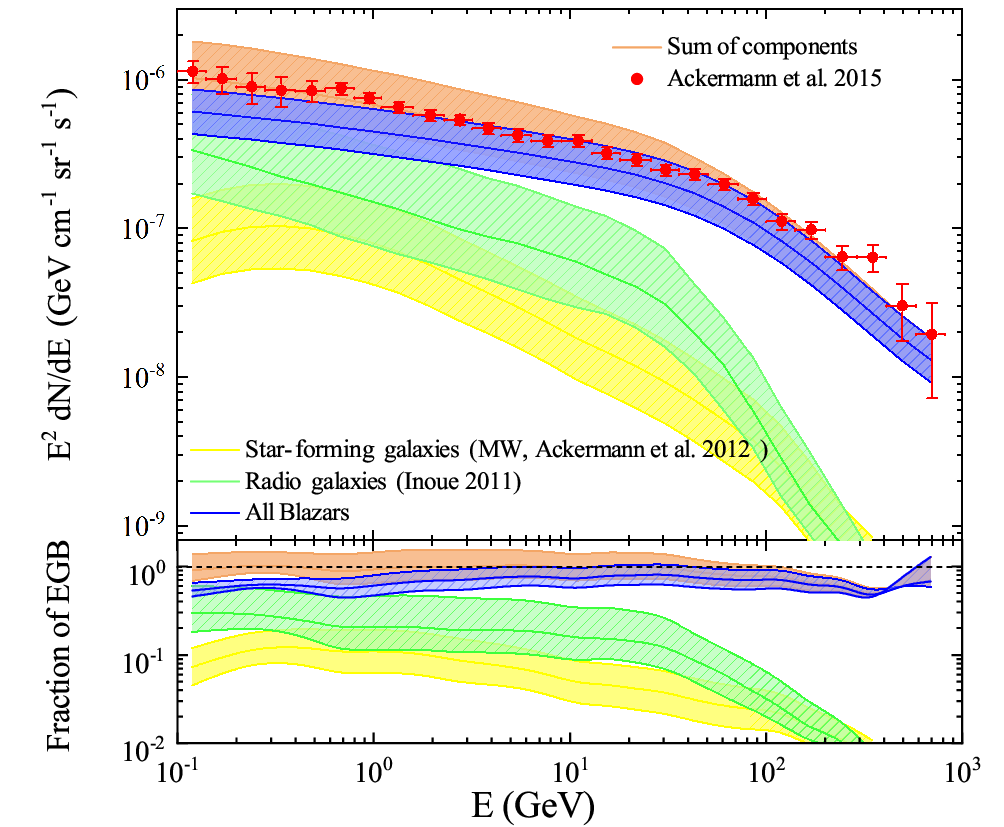}
\end{center}
\caption{The EGB spectrum observed by Fermi-LAT (red points) \cite{fermi15igrb} together with the predicted astrophysical contributions from blazars (blue band) \cite{ajello15egbdm}, radio galaxies (green band) \cite{inoue11RG} and star-forming galaxies (yellow band) \cite{fermi12sfg}. All the components have not been renomalized through the $\chi^2$ fitting, yet. The red band is the sum of all three components. The lower panel demonstrates the fraction of each astrophysical component contributing to the total EGB spectrum.}
\label{fig:egbsp}
\end{figure}

\subsection{Astrophysical components of the extragalactic gamma-ray background}
\label{sec:egbcal}
Compared with former researches on limiting UCMHs with EGB, one of the improvements is we consider the contributions to EGB from background astrophysical components. The previous works have shown that most of the EGB can be accounted for by the joint contributions of blazars (including both BL Lac objects and FSRQs), radio galaxies (RG) and star-forming galaxies (SFG). Above 100 MeV RG and SFG each contributes about 10-30\% of the observed photon flux of EGB, while blazars contributes about $\sim50\%$ \cite{ajello15egbdm}. The luminosity functions of these source populations can be derived from the resolved gamma-ray sources (for blazar) or from the relations between radio/infrared and gamma-ray luminosities (for SFG and RG). The contribution from the unresolved extragalactic sources can then be estimated by extrapolating the luminosity function (LF). In this paper, we consider these three types of sources as well. For SFG and RG, we directly use the EGB spectrum (and corresponding uncertainties) presented in \cite{fermi12sfg} (MW model) and \cite{inoue11RG}.
A newer result for the SFG contribution to the EGB has been reported in \cite{ajello2020sfg}. We also use the SFG spectrum in \cite{ajello2020sfg} (the one based on the IR luminosity function of \citet{Gruppioni:2013jna}) to test the main results of this paper and find that it only slightly affect the results since the SFG component accounts for merely $5\%$ of the total EGB.
For blazars we employ the formalism and parameters in \cite{Ajello14_bllac,ajello15egbdm}, which will be briefly re-introduced below.

The differential intensity (in unit of ${\rm ph\,cm^{-2} sr^{-1} s^{-1} GeV^{-1}}$) of the EGB contributed by the blazars with photon index ($1.0<\Gamma_{\gamma}<3.5$), redshift ($10^{-3}<z<6$) and gamma-ray luminosity ($10^{43}<L_{\gamma}<10^{52}$) can be computed by:
\begin{equation}
\begin{aligned}
F_{\rm EGB}(E)&=\int_{\Gamma_{\min }=1.0}^{\Gamma_{\max}=3.5} d \Gamma \int_{z_{\min }=10^{-3}}^{z_{\max }=6} d z \int_{L_{\gamma, \min }=10^{43}}^{L_{\gamma, \max }=10^{52}} d L_{\gamma}\\
&\times  \Phi\left(L_{\gamma}, z, \Gamma\right)\cdot f(E)\cdot \frac{d V}{d z d \Omega}
\label{eq:luminosity}
\end{aligned}
\end{equation}
where $dV/(dzd\Omega)$ is the differential comoving volume at redshift $z$, and the EBL modulated spectrum of blazars is
$$f(E;\Gamma,z,L_\gamma)=K\left[\left(\frac{E}{E_b}\right)^{1.7}+\left(\frac{E}{E_b}\right)^{2.6}\right]^{-1}\cdot e^{-\tau(E,z)}$$
with $\log E_b({\rm GeV})\approx9.25-4.11\Gamma$ and $K=L_\gamma/[4\pi d_L^2k\int Ef(E,K=1)dE]$, where $k$ is the $K$-correction term. For the optical depth term here we use the EBL model of \cite{DominguezA2011}.

The $\Phi$ in Eq. (\ref{eq:luminosity}) is the blazar LF, namely the number density of blazars at luminosity $L_\gamma$, redshift $z$ and spectral index $\Gamma$. We use the simplest pure density evolution (PDE) model of the LF, which reads
\beq
\Phi\left(L_{\gamma}, z, \Gamma\right)=\Phi\left(L_{\gamma}, z=0, \Gamma\right) \times e\left(z, L_{\gamma}\right)
\eeq
where the luminosity function at redshift $z=0$ is
\beq
\begin{aligned}
&\Phi\left(L_{\gamma}, z=0, \Gamma\right)=\frac{d N}{d L_{\gamma} d V d \Gamma} \\
&\quad=\frac{A}{\ln (10) L_{\gamma}} \left[\left(\frac{L_{\gamma}}{L_{*}}\right)^{\gamma_{1}}+\left(\frac{L_{\gamma}}{L_{*}}\right)^{\gamma_{2}}\right]^{-1} \cdot e^{-0.5\left[\Gamma-\mu\left(L_{\gamma}\right)\right]^{2} / \sigma^{2}}.
\end{aligned}
\eeq
The expressions of $e(z, L_{\gamma})$ and $\mu(L_{\gamma})$ can be found in \cite{ajello15egbdm}.

We plot the model expected EGB spectra for blazar, RG and SFG together with the Fermi-LAT EGB measurements in Fig. \ref{fig:egbsp}. Also shown is the proportion of each component in the total observed EGB.

\subsection{Limiting the abundance of UCMHs with the Fermi-LAT EGB observation}
If the UCMHs exist in the Universe, they are another type of extragalactice  gamma-ray emitters due to the DM annihialtaion \cite{scott09_UCMHprl}. The annihilation photons may contribute to the extragalactic gamma-ray background, it is practicable to limit the abundance of UCMHs with EGB observation. The latest EGB measurements at GeV energies are from the Fermi-LAT observation \cite{fermi15igrb}. The Fermi-LAT Collaboration adopted three different Galactic foreground models to obtain the EGB spectrum.
For our purpose they do not differ with each other significantly, and in this paper we use the foreground model B of \cite{fermi15igrb}.
%, which has the highest intensities among the three and will give the most conservative constraints.

To compare the models with the observation, the $\chi^{2}$ fitting method is used. We first obtain the best-fit astrophysical components without the DM model included by minimizing
\begin{equation}
\begin{aligned}
\chi^{2}=&\sum_{i=1}^{N} \frac{\left(F_{i,{\rm obs}}-\boldsymbol{\alpha}_{1} F_{i, 1}-\boldsymbol{\alpha}_{2}F_{i,2}-\boldsymbol{\alpha}_{3}F_{i,3}\right)^{2}}{\sigma_{i, {\rm obs}}^{2}}\\&+\sum_{j=1}^{3} \frac{\left(1-\boldsymbol{\alpha}_{i}\right)^{2}}{\delta_{j}^{2}}
\label{eq:withoutucmh}
\end{aligned}
\end{equation}
where $F_{i,{\rm obs}}$ and the $\sigma_{i, {\rm obs}}$ are the EGB spectrum measured by the Fermi-LAT (see Table 3 of \cite{fermi15igrb}).  The error bars $\sigma_{i, {\rm obs}}$ include the statistical uncertainty and systematic uncertainties from the effective area parameterization, as well as the CR background subtraction \cite{fermi15igrb}. The systematic uncertainty related to the modeling of the Galactic foreground is not further included, which may vary the intensity by $+15\%/-30\%$. However, we adopt the EGB spectrum having the highest intensities among the three benchmark foreground models in \cite{fermi15igrb} (i.e. the FG model B), which would give  relatively conservative constraints.
The $F_{i,1}$ ,$F_{i,2}$ ,$F_{i,3}$ in Eq. (\ref{eq:withoutucmh}) are the model-expected fluxes of the $i{-\text {th}}$ energy bin from blazar, RG and SFG, respectively, and the $\boldsymbol{\alpha}_{i}$ is a renormalization constant of each spectrum which is free to vary in the fit. The last term is introduced to ensure that the best-fit gamma-ray intensities do not deviate from their original values in the literature too much.
The $\delta_{j}$ is determined by the uncertainty band of each component as demonstrated in Fig. \ref{fig:egbsp} and we choose a mean value over all the energies.

Based on the best-fit astrophysical model, we add an additional UCMH component into the $\chi^2$ fit to constrain the UCMH abundance or search for possible signals. At this stage, the $\chi^2$ is defined as
\begin{equation}
\chi^{2}=\sum_{i=1}^{N} \frac{\left[F_{i,{\rm obs}}-\mathcal{A}F_{i,\rm astro}-f_uF_{i,\rm ucmh}\right]^{2}}{\sigma_{i,\rm obs}^{2}}
\label{eq:withucmh}
\end{equation}
where $F_{i,\rm astro}$ is the sum of the best-fit astrophysical contributions in the above step, and $F_{i,\rm ucmh}$ is the flux from UCMHs as calculated by Eq. (\ref{eq:egbsp}).

The best-fit chi-square value $\chi^2_{f_u}$
%under a certain $f_u$
will change along with the given normalization parameter of the UCMH component. The chi-square difference is $\Delta\chi^{2}=\chi^{2}_{f_u}-\chi^{2}_{f_u=0}$  where $\chi^{2}_{f_u=0}$ is the minimum $\chi^{2}$ under the background-only model. Because for a fixed DM mass $M_\chi$, the UCMH model have 1 more additional parameter than the background model, the chi-square difference follows $\Delta \chi^{2}\sim \chi^{2}(1)$ \cite{Chernoff1954}. The variance of the $\chi^2$ by 2.71 corresponds to an upper limit of the abundece at 95\% confidence level.

\begin{figure*}[htp]
\begin{center}
\includegraphics[width=0.47\textwidth]{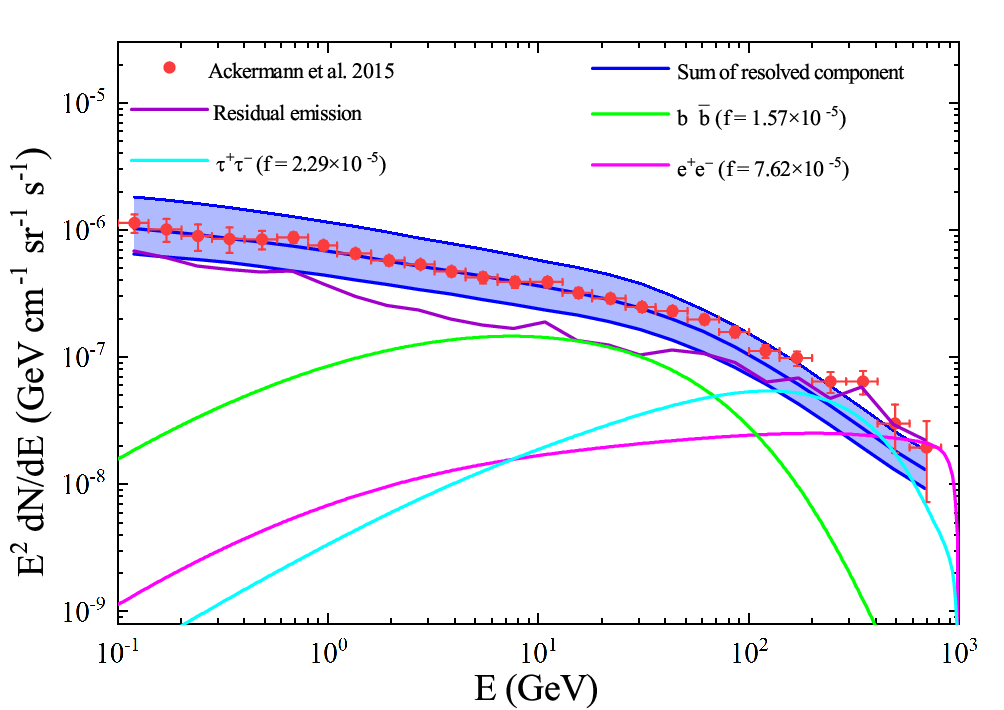}
\includegraphics[width=0.47\textwidth]{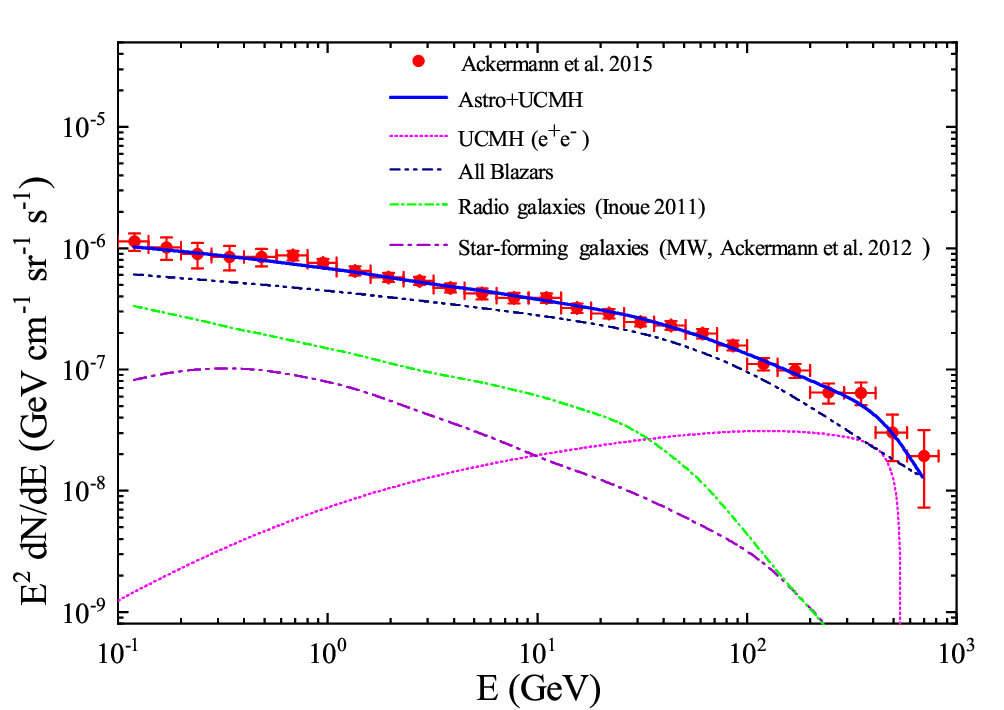}
\end{center}
\caption{Left panel: The best-fit EGB spectrum without a DM component included (blue line and relevant uncertainty band). The purple line is the conservative residuals subtracting the best-fit astrophysical contributions (see the main text for details). Also shown are the DM spectra for 1 TeV DM for $b \bar{b}$ (green line), $\tau^{+} \tau^{-}$ (cyan line) and $\mathrm{e}^{+} \mathrm{e}^{-}$ (magenta line) channels, respectively. In this plot, the amplitude of the three components are determined by requiring not to exceed the residual emission (i.e. a demonstration of our conservative methods).
Right panel: The best-fit total EGB spectrum containing the DM component (blue line) and the individual contributions.
%{\bf Here we use the SFG spectrum from Ref. \cite{}. The same results based on an updated SFG model can be found in \ref{}.}
}
\label{fig:conservative}
\end{figure*}

%*************************************************************************

\section{RESULT}
The fitted renormalization parameters $\alpha_i$ for the three background components and the 1$\sigma$ uncertainties are summarized in Table \ref{tab:tab1} (benchmark row).
For Blazar and SFG they are close to 1, while for RG a smaller renomalization parameter is required to fit the data. In Fig.\ref{fig:conservative} we exhibit the best-fit background-only EGB spectrum as well as the corresponding conservative residuals (see below). Also shown are the spectra of UCMH with $M_{\chi}$=1 TeV in different annihilation channels, which are required not to exceed the residuals in the plot.  We can see that below 50 GeV, the model match the data points well, while at energies of  $>50$ GeV it slightly underestimates the observation.

\begin{figure}[bth]
\begin{center}
\includegraphics[width=0.45\textwidth]{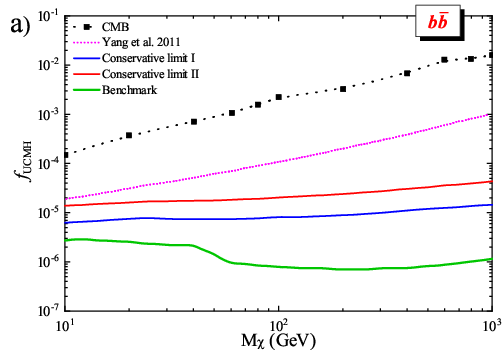}\label{ucmh-RG1}
\includegraphics[width=0.45\textwidth]{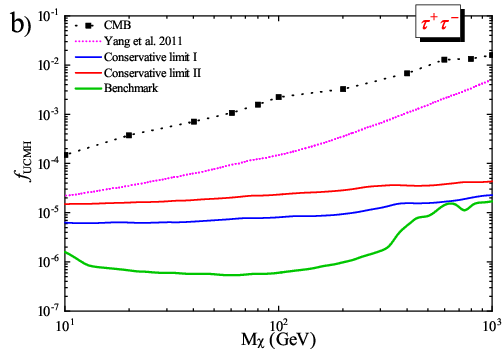}\label{ucmh-RG2}
\includegraphics[width=0.45\textwidth]{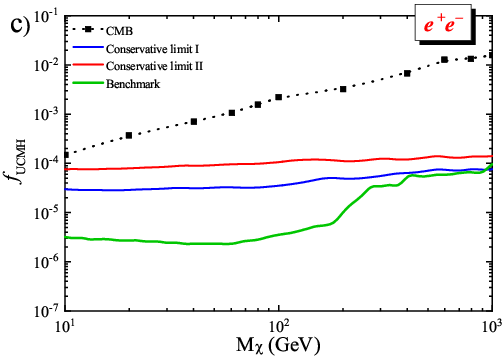}\label{ucmh-RG3}
\end{center}
\caption{The constraints on the UCMH abundance in the Universe obtained through the EGB analysis in this work (solid lines). As a comparison, we also plot the previous constraints based on EGB (dashed) and CMB (dotted) observations \cite{yang2011}. The three panels are for different annihilation channels as labeled in the plots.}\label{fig:constraints}
\end{figure}

According to the $\chi^2$ analysis (Eq. (\ref{eq:withucmh})), the upper limits on the UCMH abundance $f_u$ as a function of DM mass $M_\chi$ after containing astrophysical components in the fit are shown in Fig.\ref{fig:constraints} for $b\bar{b}$, $\tau\tau$, $e^+e^-$ channels.
For all three channels, we can place constraints on the abundance down to $\sim3\times10^{-6}$ in the range $M_\chi<100\,{\rm GeV}$, namely only $\lesssim3\times10^{-6}$ of the universe energy density could be in the form of UCMH, otherwise their predicted EGB emission will exceed the actual observation.
For the $\tau^+\tau^-$ and $e^+e^-$ channels, the constraints become weaker as the $M_\chi$ is increased to $>$300 GeV. This is due to the existence of residuals at this high energy range (see Fig. \ref{eq:withucmh}) which may be accounted for by including a DM component (see Section \ref{sec:disc1}).

\begin{table}[h]
\caption{Best-fit results for the models with only astrophysical components.}
\begin{tabular}{cccccc}
\hline
\hline
\noalign{\smallskip}\noalign{\smallskip}
Model\footnotemark[1] & Balzar &  RG  & SFG &$\chi^{2}$  \\\noalign{\smallskip}\noalign{\smallskip}
\hline
\noalign{\smallskip}\noalign{\smallskip}
benchmark & $1.146^{+0.031}_{-0.031}$ & $0.621^{+0.125}_{-0.125}$ &$1.202^{+0.251}_{-0.251}$  & 19.359   \\\noalign{\smallskip}\noalign{\smallskip}
MAGN & $1.001^{+0.029}_{-0.029}$& $0.723^{+0.091}_{-0.091}$& $1.489^{+0.251}_{-0.251}$ & 19.785  \\\noalign{\smallskip}\noalign{\smallskip}
RG($\Gamma=2.11$) & $1.165^{+0.037}_{-0.037}$& $0.521^{+0.101}_{-0.101}$& $1.453^{+0.251}_{-0.251}$ & 24.288  \\\noalign{\smallskip}\noalign{\smallskip}
SFG(PL) & $1.165^{+0.044}_{-0.044}$& $1.027^{+0.121}_{-0.121}$& $1.237^{+0.180}_{-0.180}$ & 19.558  \\\noalign{\smallskip}\noalign{\smallskip}
SFG2020 & $0.983^{+0.004}_{-0.004}$& $0.824^{+0.152}_{-0.152}$& $1.022^{+0.143}_{-0.143}$ & 21.685
\\\noalign{\smallskip}\noalign{\smallskip}
$\tau=1.49$ & $0.964^{+0.004}_{-0.004}$& $1.840^{+0.153}_{-0.153}$& $1.417^{+0.251}_{-0.251}$ & 11.443  \\\noalign{\smallskip}\noalign{\smallskip}
$\gamma_2=1.35$ & $0.710^{+0.018}_{-0.018}$& $1.180^{+0.127}_{-0.127}$& $1.309^{+0.251}_{-0.251}$ & 12.772  \\\noalign{\smallskip}\noalign{\smallskip}
\hline
\footnotetext[1]{See Sec. \ref{sec:disc1} for the description of the tested models.}
\end{tabular}
\label{tab:tab1}
\end{table}

\begin{figure}
\begin{center}
\includegraphics[width=0.45\textwidth]{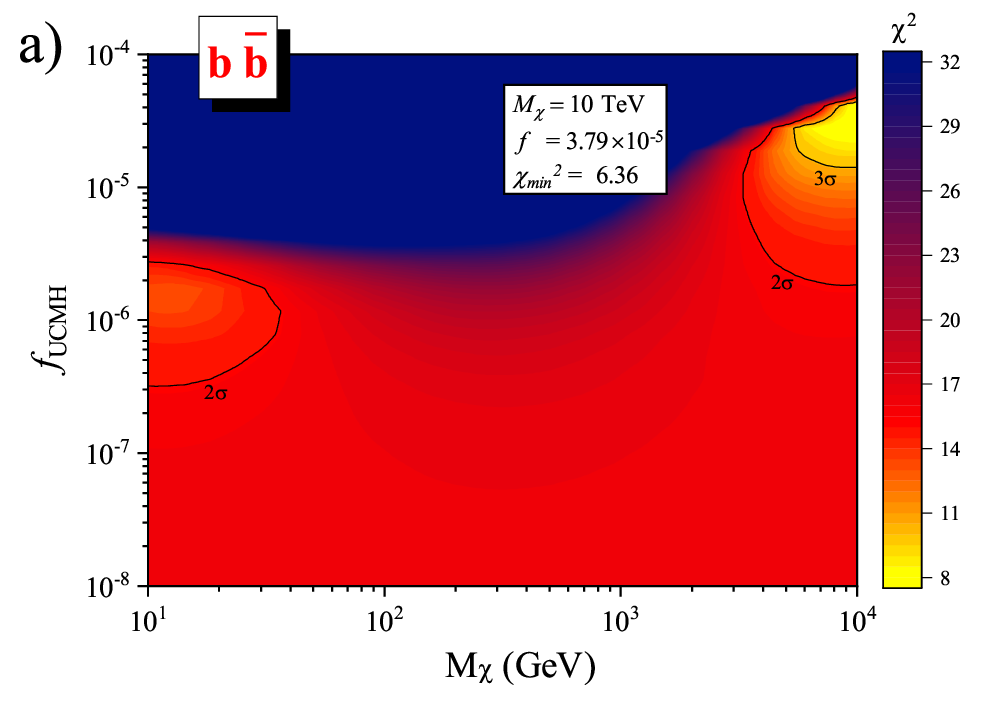}\label{ucmh-map1}
\includegraphics[width=0.45\textwidth]{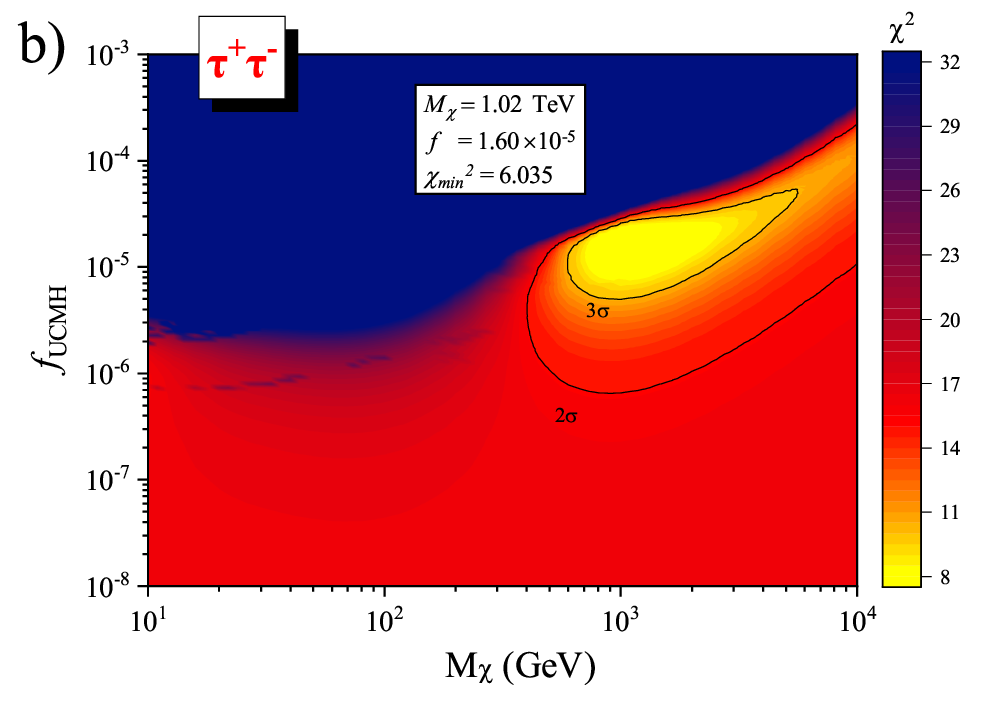}\label{ucmh-map2}
\includegraphics[width=0.45\textwidth]{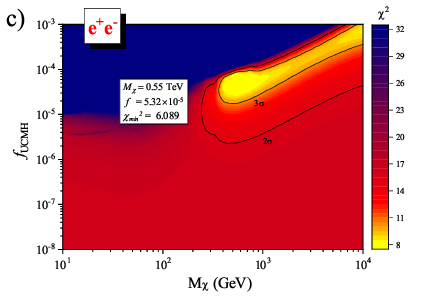}\label{ucmh-map3}
\end{center}
\caption{The $\chi^2$ maps as a function DM mass $M_\chi$ and UCMH abundance $f_u$ for three different annihilation channels.
The brightest point with minimum $\chi^2$ corresponds to the best fit to the observation and the fitted parameters have been shown in the plots.
%The black lines are the 1 to $3\,\sigma$ contours of the parameter uncertainty.
}\label{fig:ucmh-map}
\end{figure}

Compared with the previous results which are based on the 1-year Fermi-LAT EGB observation \cite{yang2011} (dashed line in Fig. \ref{fig:constraints}), we can see that our constraints are about $1-2$ orders of magnitude better. The improvement is owing to the use of the updated EGB observation and subtracting the astrophysical contributions.
The UCMH abundance $f_u$ in the universe can also be constrained by the CMB observation since in the early universe the particles emitted from the DM annihilation within UCMHs will influence the ionization and recombination before the structure formation \cite{yang11_ucmhCMB,yang2011}. As a comparison, the CMB constraints with WMAP-7 data \cite{yang2011} are shown in the Fig. \ref{fig:constraints} (dotted line), which is however not as stringent as the EGB limits.

In addition, we use two other approaches to set more conservative limits. The most conservative one is obtained by using an inclusive EGB spectrum with no any background subtracted (I). Less conservative limits (II) are given by the following prescription. We define the upper bound of the error bars of the EGB measurements as $F_{i,\rm up}$, while for the model we use the lower bound of the uncertainty bands $F_{i,\rm low}$, and $F_{i,\rm res}=F_{i,\rm up}-F_{i,\rm low}$ is considered a conservative residual after subtracting the background. Namely, for the observation we adopt the maximal values under the $1\sigma$ range, and for the model-expected one we use the minimum. Requiring that the EGB from UCMHs does not exceed the $F_{i,\rm res}$ gives the limits on the UCMH abundance. As is shown, even with the most conservative approach, the results have improved greatly than that of \cite{yang2011}, mainly due to the adoption of the new EGB observation.

\section{DISCUSSIONS}
\subsection{Searh for possible additional DM component}
\label{sec:disc1}

In contrast with previous analyses, we are able to search for possible UCMH signals in addition to the background components because the astrophysical contributions are considered in this work.
The search  is also based on the chi-square analysis of Eq. (\ref{eq:withucmh}).
A background model corresponds to $f_u=0$, while for the signal model $f_u$ is free to vary. Then the significance of the existence of a UCMH component is given by the chi-square difference $\Delta\chi^2$.
According to Wilks' theorem \cite{Chernoff1954}, the $\Delta\chi^2>9$ indicates the observed data rejecting the null model at a confidence level of $>3\sigma$, i.e. there may exist a possible signal. We scan for a series of DM masses with the EGB observation, and the related results are shown in Fig. \ref{fig:ucmh-map}.

In our analysis, we notice that the inclusion of a UCMH component improves the fit significantly.
The test-statistic ($TS$) of the additional DM composition can be estimated by the difference of the minimum chi-square values between the following two cases: the fitting of only considering the astrophysical components, and that with the addition of a UCMH composition, namely $TS=\Delta\chi^{2}=\tilde{\chi}^{2}(f_u=0)-\tilde{\chi}^{2}$. The $\chi^2$ with tilde denotes the minimum value in the fit.
The brightest points in Fig. \ref{fig:ucmh-map} give the best-fit $M_\chi$ and $f_u$ parameters.
We obtain the optimum DM masses of 10 TeV\footnote{The 10 TeV is the  upper boundary of the scanned $M_\chi$.}, 1.09 TeV and 0.55 TeV with {TS of 13.0, 13.3 and 13.3} for $b\bar{b}$, $\tau\tau$ and $e^+e^-$, respectively. The TS$>9$ suggests existing a tentative signal.

Although such a tentative excess is interesting, it is difficult to reliably claim that it comes from UCMHs, given the large uncertainties in the modeling of astrophysical components.
{We here demonstrate that the uncertainties in the astrophysical models have a great impact on the obtained significance.}
We note that the fitting is improved mainly because the addition of a UCMH component compensates for the residuals in the high energy range (see Fig. \ref{fig:conservative} right for the demonstration). In light of this, we focus on some alternative models that can increase the high energy flux of the EGB spectrum. We do the following checks.
%A. Replace the RG model of \cite{} by that of \cite{}; B. Consider a PL model instead of the MW model for the SFG component; C. For RG we use the EGB spectrum with another photon index of $\Gamma=2.11$. D. The uncertainties of some parameters are considered in the calculation of the contribution from Blazars.

\begin{figure}
\begin{center}
\includegraphics[width=0.49\textwidth]{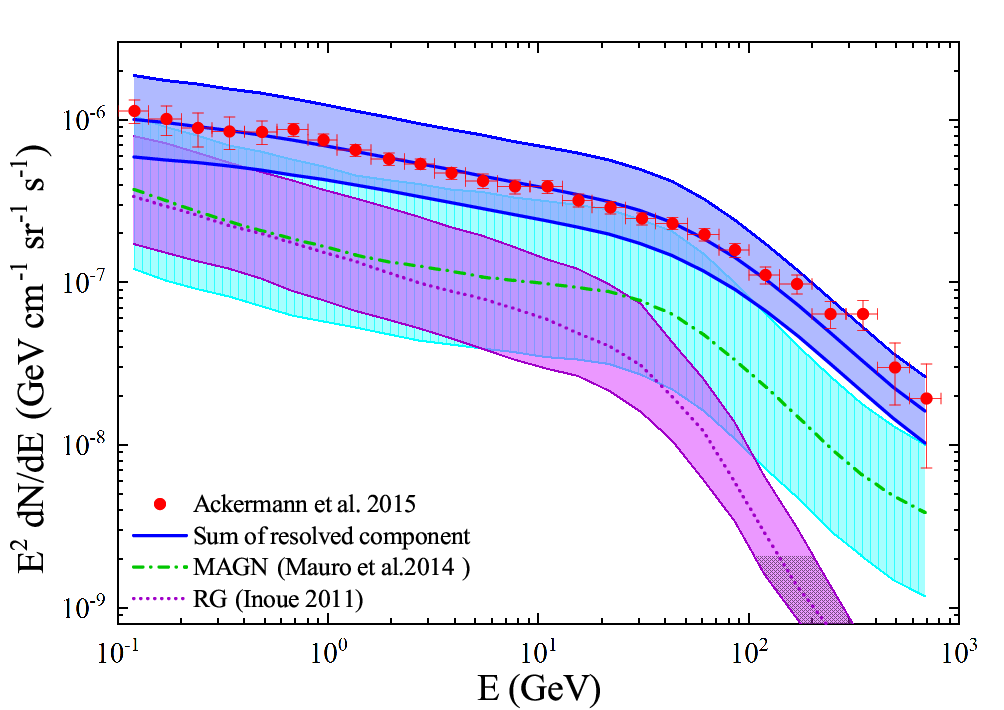}
\end{center}
\caption{We compare the RG models of \cite{DiMauro14_magn} and \cite{inoue11RG}. The RG model from \cite{inoue11RG} has higher EGB flux at higher energies. The blue line is the best-fit adopting the model of \cite{DiMauro14_magn} (DM component is not included).}
\label{fig:rgcompare}
\end{figure}

%*******************table.1****************************
\begin{table}[t]
\caption{Best-fit parameters for the models with DM included.}
\begin{tabular}{cccccc}
\hline
\hline\noalign{\smallskip}\noalign{\smallskip}
 Model &  Channel & $f_{u}$  &  $M_\chi\footnotemark[1]$ & TS\footnotemark[2] & $\chi^2$ \\\noalign{\smallskip}\noalign{\smallskip}
\hline\noalign{\smallskip}\noalign{\smallskip}
Benchmark    & $b\bar{b}$      & $3.79\times10^{-5}$ & $10$ & $12.99$ & $6.36$ \\\noalign{\smallskip}\noalign{\smallskip}
Benchmark    & $\tau^+\tau^-$  & $1.60\times10^{-5}$ & $1.09$ & $13.32$ & $6.03$ \\\noalign{\smallskip}\noalign{\smallskip}
Benchmark    & $e^+e^-$        & $5.23\times10^{-5}$ & $0.55$ & $13.27$ & $6.08$ \\\noalign{\smallskip}\noalign{\smallskip}
MAGN         & $\tau^+\tau^-$  & $1.50\times10^{-5}$ & $1.48$ & $12.69$ & $7.09$ \\\noalign{\smallskip}\noalign{\smallskip}
RG/$\Gamma=2.11$ & $\tau^+\tau^-$  & $1.45\times10^{-5}$ & $1.48$ & $14.78$ & $9.50$ \\\noalign{\smallskip}\noalign{\smallskip}
SFG/PL      & $\tau^+\tau^-$  & $1.85\times10^{-5}$ & $1.35$ & $11.44$ & $7.91$ \\\noalign{\smallskip}\noalign{\smallskip}
SFG2020 & $\tau^+\tau^-$  & $1.87\times10^{-5}$ & $1.23$ & $14.56$ & $7.12$
\\\noalign{\smallskip}\noalign{\smallskip}
$\tau=1.49$  & $\tau^+\tau^-$  & $1.17\times10^{-5}$ & $1.66$ & $5.27$ & $6.17$ \\\noalign{\smallskip}\noalign{\smallskip}
$\gamma_2=1.35$  & $\tau^+\tau^-$  & $3.94\times10^{-6}$ & $2.23$ & $5.05$ & $7.72$ \\\noalign{\smallskip}\noalign{\smallskip}
\hline\noalign{\smallskip}\noalign{\smallskip}
\footnotetext[1]{DM mass in unit of TeV.}
\footnotetext[2]{TS value of the UCMH component.}
\end{tabular}
\label{tab:tab2}
\end{table}

The reference \cite{DiMauro15_igrbDM} has also searched for a probable DM component that could be hidden beneath the EGB. We note that they did not report the presence of a tentative additional component in the $>$100 GeV energy range\footnote{Note that they focused on the DM halo of the Milky Way rather than the extragalactic UCMHs here.  However if the additional UCMH component does exists, it will be partly revealed in their results since both (UCMH and MW halo) spectra have more or less similar bump-like shape.}.  One difference between their work and ours is that for the RG component, they employ the energy spectrum of \cite{DiMauro14_magn} instead of \cite{inoue11RG} in our study. By comparison (Fig. \ref{fig:rgcompare}), it can be found that the RG spectrum predicted by \cite{DiMauro14_magn} has a higher energy flux than \cite{inoue11RG} at energies greater than 100 GeV. We use the RG model of \cite{DiMauro14_magn} to check our results with the models of the other components unchanged. The results reveal that even when the RG model is replaced, the fitting still gives a relatively high TS of the tentative DM component (see MAGN model in Table \ref{tab:tab1} \&  \ref{tab:tab2}).

In addition, when modeling the SFG component, different assumptions of the average spectrum of the source population will lead to different EGB spectra of SFGs \cite{fermi12sfg}. Our benchmark results adopt the MW model (i.e., assuming all SFGs are Milky Way-like), but at 100 GeV the PL model (all SFGs share the same spectrum as those detected by Fermi-LAT) is higher than the MW model by a factor of $\sim$10. We therefore examine the outcome of taking this SFG/PL model.
Further, we notice that an updated result for the SFG contribution to the EGB has been reported in \cite{ajello2020sfg}. They derive the SFG spectrum based on the detection of 11 SFGs and the emission from unresolved SFGs with the 10-year Fermi-LAT data. We test the analysis with this SFG model and find results consistent with our benchmark ones. The related results are shown in the SFG2020 row of the two tables.

The large uncertainties in the LF parameters will induce a significant uncertainty of the predicted EGB spectrum. For RG we examine the spectral uncertainty introduced by the photon index parameter. We consider a harder photon index ($\Gamma=2.11$) of the source population in the luminosity function (see Fig. 4 of \cite{inoue11RG}).
The corresponding results are shown in Table \ref{tab:tab1} \&  \ref{tab:tab2} (labeled as RG/$\Gamma=2.11$).
For the Blazar component, with the formalism described in Section \ref{sec:egbcal}, we test the uncertainties associated with all 10 parameters of the PDE LF.
The $\tau$ and $\gamma_2$ parameters are found to have the greatest influence on the obtained TS value when the parameter values are changed within their uncertainty range.
The TS reduces to $\sim$5.3 and $\sim$5.1 for the $\tau=1.49$ and $\gamma_2=1.35$ models, respectively.

According to these test, we conclude that the results from the EGB analysis are currently subject to considerable uncertainty and we can not claim the presence of additional components despite obtaining a relatively high TS value.

\subsection{The UCMH contribution to the $e^+e^-$ energy density near the Earth}

At last we discuss the implication of our constraints to the DAMPE 1.4 TeV excess.
One of the most intriguing structrue displayed in the DAMPE $e^+e^-$ spectrum is the peak-like excess at $\sim$1.4 TeV with a significance of $\sim3.7\sigma$ which may be caused by the monochromatic injection of electrons due to the DM annihilation within nearby DM halos  \cite{dampe17_nature,yuan2017,caojj18_dampe,huang18_dampe,panxu18,zhaoyi2019}. In the DM scenario, the DM annihilation is accompanied with production of gamma-ray photons.
While the normal DM halo models (like NFW \cite{navarro97nfw}, Einasto \cite{einasto}) are challenged by the gamma-ray observations \cite{Ghosh18,Belotsky19pdu}, the DM annihilation within nearby UCMHs can provide a better interpretation to the excess \cite{huang18_dampe,zhaoyi2019,cheng20}.
Assuming that the local fraction of the DM in the form of UCMHs is identical to that in the whole universe, the above constraints can be used to exame the UCMH interpretation of the 1.4 TeV excess.

Here we especially consider the channel of $\chi\chi\rightarrow e^+e^-$. The number of electrons/positrons emitted per unit time and energy from a UCMH is
\beq
\frac{dN_e}{dEdt}=2R\cdot\delta(E-E')
\eeq
with $R$ the annihilation rate of the DM particles within the UCMH
\beq
R=\frac{\langle\sigma v\rangle}{2 m_{\chi}^{2}} \int \rho^{2} \mathrm{~d} V.
\eeq
We then have the injection rate of
\beq
Q(E,r) = \frac{{f_u} \rho_{\chi,\rm local}}{M_u(0)} \frac{dN_e}{dEdt}
\eeq
where $\rho_{\chi,\rm local}= 0.4\,{\rm GeV/cm^{3}}$ \cite{localdmdens} is the DM density near the location of the Earth.
By solving the propagation equation of electrons one can obtain the energy density $\omega_e$ contributed by UCMHs for a given $f_u$. The narrow peak of the tentative exess requires the distance of the source is within $d$=0.3kpc for avoiding the widden of the peak due to cooling effect \cite{yuan2017,huang18_dampe}. Assuming the DM density of the field halo does not vary a lot within the region of propagation length (i.e. assuming the UCMHs distributed evenly near the Earth), we can reasonably neglect the diffusion term, and the number density of the electrons provided by UCMHs can be approximated by
\beq
\frac{d n_{e}}{d E}=\frac{1}{b(E)} \int_{E}^{\infty} Q\left(E^{\prime}, r\right) d E^{\prime},
\label{eq:edens}
\eeq
where the $b(E)\approx b_2(E/{\rm GeV})^2$ is the electron cooling rate, for which we consider only the synchrotron and ICS losses $b_2=1.0\times10^{-16}\,{\rm GeV/s}$ \cite{Atoyan95,yuan2017}.

The measured energy density of the 1.4 TeV peak is estimated to be about $1.2\times10^{-18}\,{\rm erg\,cm^{-3}}$ \cite{yuan2017}. However, using Eq. (\ref{eq:edens}) and the upper limits of the UCMH abundance in Fig. \ref{fig:constraints}, we obtain an upper limits of the energy density of $\omega_{\rm up}\sim6.25\times10^{-19}\,{\rm erg/cm^3}$ for the ${e}^{+}{e}^{-}$ channel.
This indicates that the UCMHs formed in the transition epoches with a density profile of $\gamma=-9/4$ is hard to interpret the DAMPE 1.4 TeV excess if the UCMH abundance near the Earth is the same as that in the whole universe. It has also been shown that such type ($\gamma=-9/4$) of UCMH is not supported by the realistic simulations \cite{Delos18_noucmh,Delos18_cheng,adamek19_pbhucmh}. To still use UCMH to account for the 1.4 TeV excess, possible solution is that the UCMHs are in the form of $\gamma=-3/2$ as expected by the simulations. The shallower density profile reduces the annihilation rate in the UCMHs, making the upper limits of the abundance deduced from the EGB observation much weaker. Note that the $\gamma=-3/2$ UCMHs could also behave as point-like sources in the Fermi-LAT gamma-ray sky \cite{cheng20}, and would not be constrained by the gamma-ray observation.
Another posibility is the UCMH abundance near the Earth is higher than the average value in the whole universe.

\section{summary}

In this work, we revisit the analysis of constraining the UCMH abundance with EGB observations, using the latest measurements at 0.1-820 GeV energies by Fermi-LAT. Except for the use of updated data, another improvement of this work is that we take into account the astrophysical contributions in the EGB and subtract them before setting the constraints in order to obtain more strict limits on the abundance. With these improvements, we find that our results are 1-2 orders of magnitude better than previous. Even adopting a conservative method of using the inclusive EGB spectrum as \cite{yang2011}, our results are substantially stronger due to the use of the new EGB observation \cite{fermi15igrb}. Thus, the constraints presented in the work are currently the most serious ones for the UCMHs with monochromatic mass function. Though some $N$-body simulations do not support the existence of UCMHs, our results can also apply to the dressed PBH \cite{adamek19_pbhucmh,yang2020}.

In addition to deriving constraints, we also search for possible DM components after subtracting the astrophysical contributions. {We find that in our benchmark model (see Table \ref{tab:tab2}), the $\chi^2$ analysis shows that the significance of existing a UCMH component reaches $3.6\sigma$ (i.e., ${\rm TS}=13.3$) for the $\tau^+\tau^-$ channel. For $b\bar{b}$ and $e^+e^-$, the TS values are 13.0 and 13.3, respectively.} However, we point out that the uncertainty of the astrophysical models is large and it is hard to claim the existence of an additional component at present. {The TS value can be reduced to as low as $\sim5.3$ if we change the astrophysical models.} Observing more resolved extragalactic sources in the future with next generation gamma-ray telescopes (especially for the SFG and RG components) will be helpful to better determine the gamma-ray luminosity function of these source classes and is crucial for the better determination of whether existing additional components in the EGB.

%################################################################################
%################################
\begin{acknowledgments}
We thank the kindly suggetion from the anonymous referee.
We thank Yupeng Yang and Houdun Zeng for their helpful discussions.
This work is supported by the National Natural Science Foundation of China (Nos. 11851304, U1738136, 11533003, U1938106, 11703094) and the Guangxi Science Foundation (2017AD22006,2019AC20334,
    2018GXNSFDA281033) and Bagui Young Scholars Program (LHJ).
\end{acknowledgments}
\bibliographystyle{apsrev4-1-lyf}
\bibliography{ucmh}

\begin{thebibliography}{79}%
\makeatletter
\providecommand \@ifxundefined [1]{%
 \@ifx{#1\undefined}
}%
\providecommand \@ifnum [1]{%
 \ifnum #1\expandafter \@firstoftwo
 \else \expandafter \@secondoftwo
 \fi
}%
\providecommand \@ifx [1]{%
 \ifx #1\expandafter \@firstoftwo
 \else \expandafter \@secondoftwo
 \fi
}%
\providecommand \natexlab [1]{#1}%
\providecommand \enquote  [1]{``#1''}%
\providecommand \bibnamefont  [1]{#1}%
\providecommand \bibfnamefont [1]{#1}%
\providecommand \citenamefont [1]{#1}%
\providecommand \href@noop [0]{\@secondoftwo}%
\providecommand \href [0]{\begingroup \@sanitize@url \@href}%
\providecommand \@href[1]{\@@startlink{#1}\@@href}%
\providecommand \@@href[1]{\endgroup#1\@@endlink}%
\providecommand \@sanitize@url [0]{\catcode `\\12\catcode `\$12\catcode
  `\&12\catcode `\#12\catcode `\^12\catcode `\_12\catcode `\%12\relax}%
\providecommand \@@startlink[1]{}%
\providecommand \@@endlink[0]{}%
\providecommand \url  [0]{\begingroup\@sanitize@url \@url }%
\providecommand \@url [1]{\endgroup\@href {#1}{\urlprefix }}%
\providecommand \urlprefix  [0]{URL }%
\providecommand \Eprint [0]{\href }%
\providecommand \doibase [0]{http://dx.doi.org/}%
\providecommand \selectlanguage [0]{\@gobble}%
\providecommand \bibinfo  [0]{\@secondoftwo}%
\providecommand \bibfield  [0]{\@secondoftwo}%
\providecommand \translation [1]{[#1]}%
\providecommand \BibitemOpen [0]{}%
\providecommand \bibitemStop [0]{}%
\providecommand \bibitemNoStop [0]{.\EOS\space}%
\providecommand \EOS [0]{\spacefactor3000\relax}%
\providecommand \BibitemShut  [1]{\csname bibitem#1\endcsname}%
\let\auto@bib@innerbib\@empty
%</preamble>
\bibitem [{\citenamefont {{Fichtel}}\ \emph {{\it et~al.}}(1978)\citenamefont
  {{Fichtel}}, \citenamefont {{Simpson}},\ and\ \citenamefont
  {{Thompson}}}]{1978ApJ...222..833F}%
  \BibitemOpen
  \bibfield  {author} {\bibinfo {author} {\bibfnamefont {C.~E.}\ \bibnamefont
  {{Fichtel}}}, \bibinfo {author} {\bibfnamefont {G.~A.}\ \bibnamefont
  {{Simpson}}}, and\ \bibinfo {author} {\bibfnamefont {D.~J.}\ \bibnamefont
  {{Thompson}}},\ }\bibfield  {title} {\enquote {\bibinfo {title} {{Diffuse
  gamma radiation.}}}\ }\href {\doibase 10.1086/156202} {\bibfield  {journal}
  {\bibinfo  {journal} {The Astrophysical Journal}\ }\textbf {\bibinfo {volume}
  {222}},\ \bibinfo {pages} {833} (\bibinfo {year} {1978})}\BibitemShut
  {NoStop}%
\bibitem [{\citenamefont {{Thompson}}\ and\ \citenamefont
  {{Fichtel}}(1982)}]{1982A&A...109..352T}%
  \BibitemOpen
  \bibfield  {author} {\bibinfo {author} {\bibfnamefont {D.~J.}\ \bibnamefont
  {{Thompson}}} and\ \bibinfo {author} {\bibfnamefont {C.~E.}\ \bibnamefont
  {{Fichtel}}},\ }\bibfield  {title} {\enquote {\bibinfo {title}
  {{Extragalactic gamma radiation - Use of galaxy counts as a galactic
  tracer}},}\ }\href@noop {} {\bibfield  {journal} {\bibinfo  {journal}
  {Astronomy and Astrophysics}\ }\textbf {\bibinfo {volume} {109}},\ \bibinfo
  {pages} {352} (\bibinfo {year} {1982})}\BibitemShut {NoStop}%
\bibitem [{\citenamefont {{Osborne}}\ \emph {{\it et~al.}}(1994)\citenamefont
  {{Osborne}}, \citenamefont {{Wolfendale}},\ and\ \citenamefont
  {{Zhang}}}]{1994JPhG...20.1089O}%
  \BibitemOpen
  \bibfield  {author} {\bibinfo {author} {\bibfnamefont {J.~L.}\ \bibnamefont
  {{Osborne}}}, \bibinfo {author} {\bibfnamefont {A.~W.}\ \bibnamefont
  {{Wolfendale}}}, and\ \bibinfo {author} {\bibfnamefont {L.}~\bibnamefont
  {{Zhang}}},\ }\bibfield  {title} {\enquote {\bibinfo {title} {{The diffuse
  flux of energetic extragalactic gamma rays}},}\ }\href {\doibase
  10.1088/0954-3899/20/7/010} {\bibfield  {journal} {\bibinfo  {journal}
  {Journal of Physics G Nuclear Physics}\ }\textbf {\bibinfo {volume} {20}},\
  \bibinfo {pages} {1089} (\bibinfo {year} {1994})}\BibitemShut {NoStop}%
\bibitem [{\citenamefont {{Sreekumar}}\ \emph {{\it et~al.}}(1998)\citenamefont
  {{Sreekumar}}, \citenamefont {{Bertsch}}, \citenamefont {{Dingus}},
  \citenamefont {{Esposito}}, \citenamefont {{Fichtel}}, \citenamefont
  {{Hartman}}, \citenamefont {{Hunter}}, \citenamefont {{Kanbach}},
  \citenamefont {{Kniffen}}, \citenamefont {{Lin}}, \citenamefont
  {{Mayer-Hasselwander}}, \citenamefont {{Michelson}}, \citenamefont {{von
  Montigny}}, \citenamefont {{M{\"u}cke}}, \citenamefont {{Mukherjee}},
  \citenamefont {{Nolan}}, \citenamefont {{Pohl}}, \citenamefont {{Reimer}},
  \citenamefont {{Schneid}}, \citenamefont {{Stacy}}, \citenamefont
  {{Stecker}}, \citenamefont {{Thompson}},\ and\ \citenamefont
  {{Willis}}}]{1998ApJ...494..523S}%
  \BibitemOpen
  \bibfield  {author} {\bibinfo {author} {\bibfnamefont {P.}~\bibnamefont
  {{Sreekumar}}} {\it et~al.},\ }\bibfield  {title} {\enquote {\bibinfo {title}
  {{EGRET Observations of the Extragalactic Gamma-Ray Emission}},}\ }\href
  {\doibase 10.1086/305222} {\bibfield  {journal} {\bibinfo  {journal} {The
  Astrophysical Journal}\ }\textbf {\bibinfo {volume} {494}},\ \bibinfo {pages}
  {523} (\bibinfo {year} {1998})},\ \Eprint
  {http://arxiv.org/abs/astro-ph/9709257}{arXiv:astro-ph/9709257}\BibitemShut
  {NoStop}%
\bibitem [{\citenamefont {{Willis}}(2002)}]{2002astro.ph..1515W}%
  \BibitemOpen
  \bibfield  {author} {\bibinfo {author} {\bibfnamefont {T.~D.}\ \bibnamefont
  {{Willis}}},\ }\bibfield  {title} {\enquote {\bibinfo {title} {{Observations
  of the Isotropic Diffuse Gamma-ray Background with the EGRET Telescope}},}\
  }\href@noop {} {\bibfield  {journal} {\bibinfo  {journal} {Ph.D. Thesis}\ ,\
  \bibinfo {eid} {astro-ph/0201515}} (\bibinfo {year} {2002})},\ \Eprint
  {http://arxiv.org/abs/astro-ph/0201515}{arXiv:astro-ph/0201515}\BibitemShut
  {NoStop}%
\bibitem [{\citenamefont {{Atwood}}\ \emph {{\it et~al.}}(2009)\citenamefont
  {{Atwood}}, \citenamefont {{Abdo}}, \citenamefont {{Ackermann}},
  \citenamefont {{Althouse}}, \citenamefont {{Anderson}}, \citenamefont
  {{Axelsson}}, \citenamefont {{Baldini}}, \citenamefont {{Ballet}},
  \citenamefont {{Band}}, \citenamefont {{Barbiellini}},\ and\ \citenamefont
  {et~al.}}]{lat09}%
  \BibitemOpen
  \bibfield  {author} {\bibinfo {author} {\bibfnamefont {W.~B.}\ \bibnamefont
  {{Atwood}}} {\it et~al.},\ }\bibfield  {title} {\enquote {\bibinfo {title}
  {{The Large Area Telescope on the Fermi Gamma-Ray Space Telescope
  Mission}},}\ }\href {\doibase 10.1088/0004-637X/697/2/1071} {\bibfield
  {journal} {\bibinfo  {journal} {The Astrophysical Journal}\ }\textbf
  {\bibinfo {volume} {697}},\ \bibinfo {pages} {1071} (\bibinfo {year}
  {2009})},\ \Eprint
  {http://arxiv.org/abs/0902.1089}{arXiv:0902.1089}\BibitemShut {NoStop}%
\bibitem [{\citenamefont {{Abdo}}\ \emph {{\it et~al.}}(2010)\citenamefont
  {{Abdo}}, \citenamefont {{Ackermann}}, \citenamefont {{Ajello}},
  \citenamefont {{Atwood}}, \citenamefont {{Baldini}}, \citenamefont
  {{Ballet}}, \citenamefont {{Barbiellini}}, \citenamefont {{Bastieri}},
  \citenamefont {{Baughman}}, \citenamefont {{Bechtol}}, \citenamefont
  {{Bellazzini}}, \citenamefont {{Berenji}}, \citenamefont {{Blandford}},
  \citenamefont {{Bloom}}, \citenamefont {{Bonamente}}, \citenamefont
  {{Borgland}}, \citenamefont {{Bregeon}}, \citenamefont {{Brez}},
  \citenamefont {{Brigida}}, \citenamefont {{Bruel}}, \citenamefont
  {{Burnett}}, \citenamefont {{Buson}}, \citenamefont {{Caliandro}},
  \citenamefont {{Cameron}}, \citenamefont {{Caraveo}}, \citenamefont
  {{Casandjian}}, \citenamefont {{Cavazzuti}}, \citenamefont {{Cecchi}},
  \citenamefont {{{\c{C}}elik}}, \citenamefont {{Charles}}, \citenamefont
  {{Chekhtman}}, \citenamefont {{Cheung}}, \citenamefont {{Chiang}},
  \citenamefont {{Ciprini}}, \citenamefont {{Claus}}, \citenamefont
  {{Cohen-Tanugi}}, \citenamefont {{Cominsky}}, \citenamefont {{Conrad}},
  \citenamefont {{Cutini}}, \citenamefont {{Dermer}}, \citenamefont {{de
  Angelis}}, \citenamefont {{de Palma}}, \citenamefont {{Digel}}, \citenamefont
  {{di Bernardo}}, \citenamefont {{do Couto e Silva}}, \citenamefont {{Drell}},
  \citenamefont {{Drlica-Wagner}}, \citenamefont {{Dubois}}, \citenamefont
  {{Dumora}}, \citenamefont {{Farnier}}, \citenamefont {{Favuzzi}},
  \citenamefont {{Fegan}}, \citenamefont {{Focke}}, \citenamefont {{Fortin}},
  \citenamefont {{Frailis}}, \citenamefont {{Fukazawa}}, \citenamefont
  {{Funk}}, \citenamefont {{Fusco}}, \citenamefont {{Gaggero}}, \citenamefont
  {{Gargano}}, \citenamefont {{Gasparrini}}, \citenamefont {{Gehrels}},
  \citenamefont {{Germani}}, \citenamefont {{Giebels}}, \citenamefont
  {{Giglietto}}, \citenamefont {{Giommi}}, \citenamefont {{Giordano}},
  \citenamefont {{Glanzman}}, \citenamefont {{Godfrey}}, \citenamefont
  {{Grenier}}, \citenamefont {{Grondin}}, \citenamefont {{Grove}},
  \citenamefont {{Guillemot}}, \citenamefont {{Guiriec}}, \citenamefont
  {{Gustafsson}}, \citenamefont {{Hanabata}}, \citenamefont {{Harding}},
  \citenamefont {{Hayashida}}, \citenamefont {{Hughes}}, \citenamefont
  {{Itoh}}, \citenamefont {{Jackson}}, \citenamefont {{J{\'o}hannesson}},
  \citenamefont {{Johnson}}, \citenamefont {{Johnson}}, \citenamefont
  {{Johnson}}, \citenamefont {{Johnson}}, \citenamefont {{Kamae}},
  \citenamefont {{Katagiri}}, \citenamefont {{Kataoka}}, \citenamefont
  {{Kawai}}, \citenamefont {{Kerr}}, \citenamefont {{Kn{\"o}dlseder}},
  \citenamefont {{Kocian}}, \citenamefont {{Kuehn}}, \citenamefont {{Kuss}},
  \citenamefont {{Lande}}, \citenamefont {{Latronico}}, \citenamefont
  {{Lemoine-Goumard}}, \citenamefont {{Longo}}, \citenamefont {{Loparco}},
  \citenamefont {{Lott}}, \citenamefont {{Lovellette}}, \citenamefont
  {{Lubrano}}, \citenamefont {{Madejski}}, \citenamefont {{Makeev}},
  \citenamefont {{Mazziotta}}, \citenamefont {{McConville}}, \citenamefont
  {{McEnery}}, \citenamefont {{Meurer}}, \citenamefont {{Michelson}},
  \citenamefont {{Mitthumsiri}}, \citenamefont {{Mizuno}}, \citenamefont
  {{Moiseev}}, \citenamefont {{Monte}}, \citenamefont {{Monzani}},
  \citenamefont {{Morselli}}, \citenamefont {{Moskalenko}}, \citenamefont
  {{Murgia}}, \citenamefont {{Nolan}}, \citenamefont {{Norris}}, \citenamefont
  {{Nuss}}, \citenamefont {{Ohsugi}}, \citenamefont {{Omodei}}, \citenamefont
  {{Orlando}}, \citenamefont {{Ormes}}, \citenamefont {{Paneque}},
  \citenamefont {{Panetta}}, \citenamefont {{Parent}}, \citenamefont
  {{Pelassa}}, \citenamefont {{Pepe}}, \citenamefont {{Pesce-Rollins}},
  \citenamefont {{Piron}}, \citenamefont {{Porter}}, \citenamefont
  {{Rain{\`o}}}, \citenamefont {{Rando}}, \citenamefont {{Razzano}},
  \citenamefont {{Reimer}}, \citenamefont {{Reimer}}, \citenamefont
  {{Reposeur}}, \citenamefont {{Ritz}}, \citenamefont {{Rochester}},
  \citenamefont {{Rodriguez}}, \citenamefont {{Roth}}, \citenamefont {{Ryde}},
  \citenamefont {{Sadrozinski}}, \citenamefont {{Sanchez}}, \citenamefont
  {{Sander}}, \citenamefont {{Parkinson}}, \citenamefont {{Scargle}},
  \citenamefont {{Sellerholm}}, \citenamefont {{Sgr{\`o}}}, \citenamefont
  {{Shaw}}, \citenamefont {{Siskind}}, \citenamefont {{Smith}}, \citenamefont
  {{Smith}}, \citenamefont {{Spandre}}, \citenamefont {{Spinelli}},
  \citenamefont {{Starck}}, \citenamefont {{Strickman}}, \citenamefont
  {{Strong}}, \citenamefont {{Suson}}, \citenamefont {{Tajima}}, \citenamefont
  {{Takahashi}}, \citenamefont {{Takahashi}}, \citenamefont {{Tanaka}},
  \citenamefont {{Thayer}}, \citenamefont {{Thayer}}, \citenamefont
  {{Thompson}}, \citenamefont {{Tibaldo}}, \citenamefont {{Torres}},
  \citenamefont {{Tosti}}, \citenamefont {{Tramacere}}, \citenamefont
  {{Uchiyama}}, \citenamefont {{Usher}}, \citenamefont {{Vasileiou}},
  \citenamefont {{Vilchez}}, \citenamefont {{Vitale}}, \citenamefont {{Waite}},
  \citenamefont {{Wang}}, \citenamefont {{Winer}}, \citenamefont {{Wood}},
  \citenamefont {{Ylinen}}, \citenamefont {{Ziegler}},\ and\ \citenamefont
  {{Fermi LAT Collaboration}}}]{fermi10_egbobs}%
  \BibitemOpen
  \bibfield  {author} {\bibinfo {author} {\bibfnamefont {A.~A.}\ \bibnamefont
  {{Abdo}}} {\it et~al.},\ }\bibfield  {title} {\enquote {\bibinfo {title}
  {{Spectrum of the Isotropic Diffuse Gamma-Ray Emission Derived from
  First-Year Fermi Large Area Telescope Data}},}\ }\href {\doibase
  10.1103/PhysRevLett.104.101101} {\bibfield  {journal} {\bibinfo  {journal}
  {Physical Review Letters}\ }\textbf {\bibinfo {volume} {104}},\ \bibinfo
  {eid} {101101} (\bibinfo {year} {2010})},\ \Eprint
  {http://arxiv.org/abs/1002.3603}{arXiv:1002.3603}\BibitemShut {NoStop}%
\bibitem [{\citenamefont {{Ackermann}}\ \emph {{\it et~al.}}(2015)\citenamefont
  {{Ackermann}}, \citenamefont {{Ajello}}, \citenamefont {{Albert}},
  \citenamefont {{Atwood}}, \citenamefont {{Baldini}}, \citenamefont
  {{Ballet}}, \citenamefont {{Barbiellini}}, \citenamefont {{Bastieri}},
  \citenamefont {{Bechtol}}, \citenamefont {{Bellazzini}}, \citenamefont
  {{Bissaldi}}, \citenamefont {{Blandford}}, \citenamefont {{Bloom}},
  \citenamefont {{Bottacini}}, \citenamefont {{Brandt}}, \citenamefont
  {{Bregeon}}, \citenamefont {{Bruel}}, \citenamefont {{Buehler}},
  \citenamefont {{Buson}}, \citenamefont {{Caliandro}}, \citenamefont
  {{Cameron}}, \citenamefont {{Caragiulo}}, \citenamefont {{Caraveo}},
  \citenamefont {{Cavazzuti}}, \citenamefont {{Cecchi}}, \citenamefont
  {{Charles}}, \citenamefont {{Chekhtman}}, \citenamefont {{Chiang}},
  \citenamefont {{Chiaro}}, \citenamefont {{Ciprini}}, \citenamefont {{Claus}},
  \citenamefont {{Cohen-Tanugi}}, \citenamefont {{Conrad}}, \citenamefont
  {{Cuoco}}, \citenamefont {{Cutini}}, \citenamefont {{D'Ammando}},
  \citenamefont {{de Angelis}}, \citenamefont {{de Palma}}, \citenamefont
  {{Dermer}}, \citenamefont {{Digel}}, \citenamefont {{Silva}}, \citenamefont
  {{Drell}}, \citenamefont {{Favuzzi}}, \citenamefont {{Ferrara}},
  \citenamefont {{Focke}}, \citenamefont {{Franckowiak}}, \citenamefont
  {{Fukazawa}}, \citenamefont {{Funk}}, \citenamefont {{Fusco}}, \citenamefont
  {{Gargano}}, \citenamefont {{Gasparrini}}, \citenamefont {{Germani}},
  \citenamefont {{Giglietto}}, \citenamefont {{Giommi}}, \citenamefont
  {{Giordano}}, \citenamefont {{Giroletti}}, \citenamefont {{Godfrey}},
  \citenamefont {{Gomez-Vargas}}, \citenamefont {{Grenier}}, \citenamefont
  {{Guiriec}}, \citenamefont {{Gustafsson}}, \citenamefont {{Hadasch}},
  \citenamefont {{Hayashi}}, \citenamefont {{Hays}}, \citenamefont {{Hewitt}},
  \citenamefont {{Ippoliti}}, \citenamefont {{Jogler}}, \citenamefont
  {{J{\'o}hannesson}}, \citenamefont {{Johnson}}, \citenamefont {{Johnson}},
  \citenamefont {{Kamae}}, \citenamefont {{Kataoka}}, \citenamefont
  {{Kn{\"o}dlseder}}, \citenamefont {{Kuss}}, \citenamefont {{Larsson}},
  \citenamefont {{Latronico}}, \citenamefont {{Li}}, \citenamefont {{Li}},
  \citenamefont {{Longo}}, \citenamefont {{Loparco}}, \citenamefont {{Lott}},
  \citenamefont {{Lovellette}}, \citenamefont {{Lubrano}}, \citenamefont
  {{Madejski}}, \citenamefont {{Manfreda}}, \citenamefont {{Massaro}},
  \citenamefont {{Mayer}}, \citenamefont {{Mazziotta}}, \citenamefont
  {{McEnery}}, \citenamefont {{Michelson}}, \citenamefont {{Mitthumsiri}},
  \citenamefont {{Mizuno}}, \citenamefont {{Moiseev}}, \citenamefont
  {{Monzani}}, \citenamefont {{Morselli}}, \citenamefont {{Moskalenko}},
  \citenamefont {{Murgia}}, \citenamefont {{Nemmen}}, \citenamefont {{Nuss}},
  \citenamefont {{Ohsugi}}, \citenamefont {{Omodei}}, \citenamefont
  {{Orlando}}, \citenamefont {{Ormes}}, \citenamefont {{Paneque}},
  \citenamefont {{Panetta}}, \citenamefont {{Perkins}}, \citenamefont
  {{Pesce-Rollins}}, \citenamefont {{Piron}}, \citenamefont {{Pivato}},
  \citenamefont {{Porter}}, \citenamefont {{Rain{\`o}}}, \citenamefont
  {{Rando}}, \citenamefont {{Razzano}}, \citenamefont {{Razzaque}},
  \citenamefont {{Reimer}}, \citenamefont {{Reimer}}, \citenamefont
  {{Reposeur}}, \citenamefont {{Ritz}}, \citenamefont {{Romani}}, \citenamefont
  {{S{\'a}nchez-Conde}}, \citenamefont {{Schaal}}, \citenamefont {{Schulz}},
  \citenamefont {{Sgr{\`o}}}, \citenamefont {{Siskind}}, \citenamefont
  {{Spandre}}, \citenamefont {{Spinelli}}, \citenamefont {{Strong}},
  \citenamefont {{Suson}}, \citenamefont {{Takahashi}}, \citenamefont
  {{Thayer}}, \citenamefont {{Thayer}}, \citenamefont {{Tibaldo}},
  \citenamefont {{Tinivella}}, \citenamefont {{Torres}}, \citenamefont
  {{Tosti}}, \citenamefont {{Troja}}, \citenamefont {{Uchiyama}}, \citenamefont
  {{Vianello}}, \citenamefont {{Werner}}, \citenamefont {{Winer}},
  \citenamefont {{Wood}}, \citenamefont {{Wood}}, \citenamefont {{Zaharijas}},\
  and\ \citenamefont {{Zimmer}}}]{fermi15igrb}%
  \BibitemOpen
  \bibfield  {author} {\bibinfo {author} {\bibfnamefont {M.}~\bibnamefont
  {{Ackermann}}} {\it et~al.},\ }\bibfield  {title} {\enquote {\bibinfo {title}
  {{The Spectrum of Isotropic Diffuse Gamma-Ray Emission between 100 MeV and
  820 GeV}},}\ }\href {\doibase 10.1088/0004-637X/799/1/86} {\bibfield
  {journal} {\bibinfo  {journal} {The Astrophysical Journal}\ }\textbf
  {\bibinfo {volume} {799}},\ \bibinfo {eid} {86} (\bibinfo {year} {2015})},\
  \Eprint {http://arxiv.org/abs/1410.3696}{arXiv:1410.3696}\BibitemShut
  {NoStop}%
\bibitem [{\citenamefont {{Strong}}\ \emph {{\it et~al.}}(2004)\citenamefont
  {{Strong}}, \citenamefont {{Moskalenko}},\ and\ \citenamefont
  {{Reimer}}}]{awstrong04egret}%
  \BibitemOpen
  \bibfield  {author} {\bibinfo {author} {\bibfnamefont {A.~W.}\ \bibnamefont
  {{Strong}}}, \bibinfo {author} {\bibfnamefont {I.~V.}\ \bibnamefont
  {{Moskalenko}}}, and\ \bibinfo {author} {\bibfnamefont {O.}~\bibnamefont
  {{Reimer}}},\ }\bibfield  {title} {\enquote {\bibinfo {title} {{A New
  Determination of the Extragalactic Diffuse Gamma-Ray Background from EGRET
  Data}},}\ }\href {\doibase 10.1086/423196} {\bibfield  {journal} {\bibinfo
  {journal} {The Astrophysical Journal}\ }\textbf {\bibinfo {volume} {613}},\
  \bibinfo {pages} {956} (\bibinfo {year} {2004})},\ \Eprint
  {http://arxiv.org/abs/astro-ph/0405441}{arXiv:astro-ph/0405441}\BibitemShut
  {NoStop}%
\bibitem [{\citenamefont {{Acero}}\ \emph {{\it et~al.}}(2015)\citenamefont
  {{Acero}}, \citenamefont {{Ackermann}}, \citenamefont {{Ajello}},
  \citenamefont {{Albert}}, \citenamefont {{Atwood}}, \citenamefont
  {{Axelsson}}, \citenamefont {{Baldini}}, \citenamefont {{Ballet}},
  \citenamefont {{Barbiellini}}, \citenamefont {{Bastieri}}, \citenamefont
  {{Belfiore}}, \citenamefont {{Bellazzini}}, \citenamefont {{Bissaldi}},
  \citenamefont {{Blandford}}, \citenamefont {{Bloom}}, \citenamefont
  {{Bogart}}, \citenamefont {{Bonino}}, \citenamefont {{Bottacini}},
  \citenamefont {{Bregeon}}, \citenamefont {{Britto}}, \citenamefont {{Bruel}},
  \citenamefont {{Buehler}}, \citenamefont {{Burnett}}, \citenamefont
  {{Buson}}, \citenamefont {{Caliandro}}, \citenamefont {{Cameron}},
  \citenamefont {{Caputo}}, \citenamefont {{Caragiulo}}, \citenamefont
  {{Caraveo}}, \citenamefont {{Casandjian}}, \citenamefont {{Cavazzuti}},
  \citenamefont {{Charles}}, \citenamefont {{Chaves}}, \citenamefont
  {{Chekhtman}}, \citenamefont {{Cheung}}, \citenamefont {{Chiang}},
  \citenamefont {{Chiaro}}, \citenamefont {{Ciprini}}, \citenamefont {{Claus}},
  \citenamefont {{Cohen-Tanugi}}, \citenamefont {{Cominsky}}, \citenamefont
  {{Conrad}}, \citenamefont {{Cutini}}, \citenamefont {{D'Ammando}},
  \citenamefont {{de Angelis}}, \citenamefont {{DeKlotz}}, \citenamefont {{de
  Palma}}, \citenamefont {{Desiante}}, \citenamefont {{Digel}}, \citenamefont
  {{Di Venere}}, \citenamefont {{Drell}}, \citenamefont {{Dubois}},
  \citenamefont {{Dumora}}, \citenamefont {{Favuzzi}}, \citenamefont {{Fegan}},
  \citenamefont {{Ferrara}}, \citenamefont {{Finke}}, \citenamefont
  {{Franckowiak}}, \citenamefont {{Fukazawa}}, \citenamefont {{Funk}},
  \citenamefont {{Fusco}}, \citenamefont {{Gargano}}, \citenamefont
  {{Gasparrini}}, \citenamefont {{Giebels}}, \citenamefont {{Giglietto}},
  \citenamefont {{Giommi}}, \citenamefont {{Giordano}}, \citenamefont
  {{Giroletti}}, \citenamefont {{Glanzman}}, \citenamefont {{Godfrey}},
  \citenamefont {{Grenier}}, \citenamefont {{Grondin}}, \citenamefont
  {{Grove}}, \citenamefont {{Guillemot}}, \citenamefont {{Guiriec}},
  \citenamefont {{Hadasch}}, \citenamefont {{Harding}}, \citenamefont {{Hays}},
  \citenamefont {{Hewitt}}, \citenamefont {{Hill}}, \citenamefont {{Horan}},
  \citenamefont {{Iafrate}}, \citenamefont {{Jogler}}, \citenamefont
  {{J{\'o}hannesson}}, \citenamefont {{Johnson}}, \citenamefont {{Johnson}},
  \citenamefont {{Johnson}}, \citenamefont {{Johnson}}, \citenamefont
  {{Kamae}}, \citenamefont {{Kataoka}}, \citenamefont {{Katsuta}},
  \citenamefont {{Kuss}}, \citenamefont {{La Mura}}, \citenamefont {{Landriu}},
  \citenamefont {{Larsson}}, \citenamefont {{Latronico}}, \citenamefont
  {{Lemoine-Goumard}}, \citenamefont {{Li}}, \citenamefont {{Li}},
  \citenamefont {{Longo}}, \citenamefont {{Loparco}}, \citenamefont {{Lott}},
  \citenamefont {{Lovellette}}, \citenamefont {{Lubrano}}, \citenamefont
  {{Madejski}}, \citenamefont {{Massaro}}, \citenamefont {{Mayer}},
  \citenamefont {{Mazziotta}}, \citenamefont {{McEnery}}, \citenamefont
  {{Michelson}}, \citenamefont {{Mirabal}}, \citenamefont {{Mizuno}},
  \citenamefont {{Moiseev}}, \citenamefont {{Mongelli}}, \citenamefont
  {{Monzani}}, \citenamefont {{Morselli}}, \citenamefont {{Moskalenko}},
  \citenamefont {{Murgia}}, \citenamefont {{Nuss}}, \citenamefont {{Ohno}},
  \citenamefont {{Ohsugi}}, \citenamefont {{Omodei}}, \citenamefont
  {{Orienti}}, \citenamefont {{Orlando}}, \citenamefont {{Ormes}},
  \citenamefont {{Paneque}}, \citenamefont {{Panetta}}, \citenamefont
  {{Perkins}}, \citenamefont {{Pesce-Rollins}}, \citenamefont {{Piron}},
  \citenamefont {{Pivato}}, \citenamefont {{Porter}}, \citenamefont
  {{Racusin}}, \citenamefont {{Rando}}, \citenamefont {{Razzano}},
  \citenamefont {{Razzaque}}, \citenamefont {{Reimer}}, \citenamefont
  {{Reimer}}, \citenamefont {{Reposeur}}, \citenamefont {{Rochester}},
  \citenamefont {{Romani}}, \citenamefont {{Salvetti}}, \citenamefont
  {{S{\'a}nchez-Conde}}, \citenamefont {{Saz Parkinson}}, \citenamefont
  {{Schulz}}, \citenamefont {{Siskind}}, \citenamefont {{Smith}}, \citenamefont
  {{Spada}}, \citenamefont {{Spandre}}, \citenamefont {{Spinelli}},
  \citenamefont {{Stephens}}, \citenamefont {{Strong}}, \citenamefont
  {{Suson}}, \citenamefont {{Takahashi}}, \citenamefont {{Takahashi}},
  \citenamefont {{Tanaka}}, \citenamefont {{Thayer}}, \citenamefont {{Thayer}},
  \citenamefont {{Thompson}}, \citenamefont {{Tibaldo}}, \citenamefont
  {{Tibolla}}, \citenamefont {{Torres}}, \citenamefont {{Torresi}},
  \citenamefont {{Tosti}}, \citenamefont {{Troja}}, \citenamefont {{Van
  Klaveren}}, \citenamefont {{Vianello}}, \citenamefont {{Winer}},
  \citenamefont {{Wood}}, \citenamefont {{Wood}}, \citenamefont {{Zimmer}},\
  and\ \citenamefont {{Fermi-LAT Collaboration}}}]{fermi3fgl}%
  \BibitemOpen
  \bibfield  {author} {\bibinfo {author} {\bibfnamefont {F.}~\bibnamefont
  {{Acero}}} {\it et~al.},\ }\bibfield  {title} {\enquote {\bibinfo {title}
  {{Fermi Large Area Telescope Third Source Catalog}},}\ }\href {\doibase
  10.1088/0067-0049/218/2/23} {\bibfield  {journal} {\bibinfo  {journal} {The
  Astrophysical Journal Supplement Series}\ }\textbf {\bibinfo {volume}
  {218}},\ \bibinfo {eid} {23} (\bibinfo {year} {2015})},\ \Eprint
  {http://arxiv.org/abs/1501.02003}{arXiv:1501.02003}\BibitemShut {NoStop}%
\bibitem [{\citenamefont {{Abdollahi}}\ \emph {{\it et~al.}}(2020)\citenamefont
  {{Abdollahi}}, \citenamefont {{Acero}}, \citenamefont {{Ackermann}},
  \citenamefont {{Ajello}}, \citenamefont {{Atwood}}, \citenamefont
  {{Axelsson}}, \citenamefont {{Baldini}}, \citenamefont {{Ballet}},
  \citenamefont {{Barbiellini}}, \citenamefont {{Bastieri}}, \citenamefont
  {{Becerra Gonzalez}}, \citenamefont {{Bellazzini}}, \citenamefont
  {{Berretta}}, \citenamefont {{Bissaldi}}, \citenamefont {{Blandford}},
  \citenamefont {{Bloom}}, \citenamefont {{Bonino}}, \citenamefont
  {{Bottacini}}, \citenamefont {{Brandt}}, \citenamefont {{Bregeon}},
  \citenamefont {{Bruel}}, \citenamefont {{Buehler}}, \citenamefont
  {{Burnett}}, \citenamefont {{Buson}}, \citenamefont {{Cameron}},
  \citenamefont {{Caputo}}, \citenamefont {{Caraveo}}, \citenamefont
  {{Casandjian}}, \citenamefont {{Castro}}, \citenamefont {{Cavazzuti}},
  \citenamefont {{Charles}}, \citenamefont {{Chaty}}, \citenamefont {{Chen}},
  \citenamefont {{Cheung}}, \citenamefont {{Chiaro}}, \citenamefont
  {{Ciprini}}, \citenamefont {{Cohen-Tanugi}}, \citenamefont {{Cominsky}},
  \citenamefont {{Coronado-Bl{\'a}zquez}}, \citenamefont {{Costantin}},
  \citenamefont {{Cuoco}}, \citenamefont {{Cutini}}, \citenamefont
  {{D'Ammando}}, \citenamefont {{DeKlotz}}, \citenamefont {{de la Torre
  Luque}}, \citenamefont {{de Palma}}, \citenamefont {{Desai}}, \citenamefont
  {{Digel}}, \citenamefont {{Di Lalla}}, \citenamefont {{Di Mauro}},
  \citenamefont {{Di Venere}}, \citenamefont {{Dom{\'\i}nguez}}, \citenamefont
  {{Dumora}}, \citenamefont {{Fana Dirirsa}}, \citenamefont {{Fegan}},
  \citenamefont {{Ferrara}}, \citenamefont {{Franckowiak}}, \citenamefont
  {{Fukazawa}}, \citenamefont {{Funk}}, \citenamefont {{Fusco}}, \citenamefont
  {{Gargano}}, \citenamefont {{Gasparrini}}, \citenamefont {{Giglietto}},
  \citenamefont {{Giommi}}, \citenamefont {{Giordano}}, \citenamefont
  {{Giroletti}}, \citenamefont {{Glanzman}}, \citenamefont {{Green}},
  \citenamefont {{Grenier}}, \citenamefont {{Griffin}}, \citenamefont
  {{Grondin}}, \citenamefont {{Grove}}, \citenamefont {{Guiriec}},
  \citenamefont {{Harding}}, \citenamefont {{Hayashi}}, \citenamefont {{Hays}},
  \citenamefont {{Hewitt}}, \citenamefont {{Horan}}, \citenamefont
  {{J{\'o}hannesson}}, \citenamefont {{Johnson}}, \citenamefont {{Kamae}},
  \citenamefont {{Kerr}}, \citenamefont {{Kocevski}}, \citenamefont
  {{Kovac'evic'}}, \citenamefont {{Kuss}}, \citenamefont {{Landriu}},
  \citenamefont {{Larsson}}, \citenamefont {{Latronico}}, \citenamefont
  {{Lemoine-Goumard}}, \citenamefont {{Li}}, \citenamefont {{Liodakis}},
  \citenamefont {{Longo}}, \citenamefont {{Loparco}}, \citenamefont {{Lott}},
  \citenamefont {{Lovellette}}, \citenamefont {{Lubrano}}, \citenamefont
  {{Madejski}}, \citenamefont {{Maldera}}, \citenamefont {{Malyshev}},
  \citenamefont {{Manfreda}}, \citenamefont {{Marchesini}}, \citenamefont
  {{Marcotulli}}, \citenamefont {{Mart{\'\i}-Devesa}}, \citenamefont
  {{Martin}}, \citenamefont {{Massaro}}, \citenamefont {{Mazziotta}},
  \citenamefont {{McEnery}}, \citenamefont {{Mereu}}, \citenamefont {{Meyer}},
  \citenamefont {{Michelson}}, \citenamefont {{Mirabal}}, \citenamefont
  {{Mizuno}}, \citenamefont {{Monzani}}, \citenamefont {{Morselli}},
  \citenamefont {{Moskalenko}}, \citenamefont {{Negro}}, \citenamefont
  {{Nuss}}, \citenamefont {{Ojha}}, \citenamefont {{Omodei}}, \citenamefont
  {{Orienti}}, \citenamefont {{Orlando}}, \citenamefont {{Ormes}},
  \citenamefont {{Palatiello}}, \citenamefont {{Paliya}}, \citenamefont
  {{Paneque}}, \citenamefont {{Pei}}, \citenamefont {{Pe{\~n}a-Herazo}},
  \citenamefont {{Perkins}}, \citenamefont {{Persic}}, \citenamefont
  {{Pesce-Rollins}}, \citenamefont {{Petrosian}}, \citenamefont {{Petrov}},
  \citenamefont {{Piron}}, \citenamefont {{Poon}}, \citenamefont {{Porter}},
  \citenamefont {{Principe}}, \citenamefont {{Rain{\`o}}}, \citenamefont
  {{Rando}}, \citenamefont {{Razzano}}, \citenamefont {{Razzaque}},
  \citenamefont {{Reimer}}, \citenamefont {{Reimer}}, \citenamefont {{Remy}},
  \citenamefont {{Reposeur}}, \citenamefont {{Romani}}, \citenamefont {{Saz
  Parkinson}}, \citenamefont {{Schinzel}}, \citenamefont {{Serini}},
  \citenamefont {{Sgr{\`o}}}, \citenamefont {{Siskind}}, \citenamefont
  {{Smith}}, \citenamefont {{Spandre}}, \citenamefont {{Spinelli}},
  \citenamefont {{Strong}}, \citenamefont {{Suson}}, \citenamefont {{Tajima}},
  \citenamefont {{Takahashi}}, \citenamefont {{Tak}}, \citenamefont {{Thayer}},
  \citenamefont {{Thompson}}, \citenamefont {{Tibaldo}}, \citenamefont
  {{Torres}}, \citenamefont {{Torresi}}, \citenamefont {{Valverde}},
  \citenamefont {{Van Klaveren}}, \citenamefont {{van Zyl}}, \citenamefont
  {{Wood}}, \citenamefont {{Yassine}},\ and\ \citenamefont
  {{Zaharijas}}}]{fermi4fgl}%
  \BibitemOpen
  \bibfield  {author} {\bibinfo {author} {\bibfnamefont {S.}~\bibnamefont
  {{Abdollahi}}} {\it et~al.},\ }\bibfield  {title} {\enquote {\bibinfo {title}
  {{Fermi Large Area Telescope Fourth Source Catalog}},}\ }\href {\doibase
  10.3847/1538-4365/ab6bcb} {\bibfield  {journal} {\bibinfo  {journal}
  {Astrophys. J. Suppl.}\ }\textbf {\bibinfo {volume} {247}},\ \bibinfo {eid}
  {33} (\bibinfo {year} {2020})},\ \Eprint
  {http://arxiv.org/abs/1902.10045}{arXiv:1902.10045}\BibitemShut {NoStop}%
\bibitem [{\citenamefont {{Inoue}}(2011)}]{inoue11RG}%
  \BibitemOpen
  \bibfield  {author} {\bibinfo {author} {\bibfnamefont {Y.}~\bibnamefont
  {{Inoue}}},\ }\bibfield  {title} {\enquote {\bibinfo {title} {{Contribution
  of Gamma-Ray-loud Radio Galaxies' Core Emissions to the Cosmic MeV and GeV
  Gamma-Ray Background Radiation}},}\ }\href {\doibase
  10.1088/0004-637X/733/1/66} {\bibfield  {journal} {\bibinfo  {journal} {The
  Astrophysical Journal}\ }\textbf {\bibinfo {volume} {733}},\ \bibinfo {eid}
  {66} (\bibinfo {year} {2011})},\ \Eprint
  {http://arxiv.org/abs/1103.3946}{arXiv:1103.3946}\BibitemShut {NoStop}%
\bibitem [{\citenamefont {{Ackermann}}\ \emph {{\it et~al.}}(2012)\citenamefont
  {{Ackermann}}, \citenamefont {{Ajello}}, \citenamefont {{Allafort}},
  \citenamefont {{Baldini}}, \citenamefont {{Ballet}}, \citenamefont
  {{Bastieri}}, \citenamefont {{Bechtol}}, \citenamefont {{Bellazzini}},
  \citenamefont {{Berenji}}, \citenamefont {{Bloom}}, \citenamefont
  {{Bonamente}}, \citenamefont {{Borgland}}, \citenamefont {{Bouvier}},
  \citenamefont {{Bregeon}}, \citenamefont {{Brigida}}, \citenamefont
  {{Bruel}}, \citenamefont {{Buehler}}, \citenamefont {{Buson}}, \citenamefont
  {{Caliandro}}, \citenamefont {{Cameron}}, \citenamefont {{Caraveo}},
  \citenamefont {{Casandjian}}, \citenamefont {{Cecchi}}, \citenamefont
  {{Charles}}, \citenamefont {{Chekhtman}}, \citenamefont {{Cheung}},
  \citenamefont {{Chiang}}, \citenamefont {{Cillis}}, \citenamefont
  {{Ciprini}}, \citenamefont {{Claus}}, \citenamefont {{Cohen-Tanugi}},
  \citenamefont {{Conrad}}, \citenamefont {{Cutini}}, \citenamefont {{de
  Palma}}, \citenamefont {{Dermer}}, \citenamefont {{Digel}}, \citenamefont
  {{Silva}}, \citenamefont {{Drell}}, \citenamefont {{Drlica-Wagner}},
  \citenamefont {{Favuzzi}}, \citenamefont {{Fegan}}, \citenamefont {{Fortin}},
  \citenamefont {{Fukazawa}}, \citenamefont {{Funk}}, \citenamefont {{Fusco}},
  \citenamefont {{Gargano}}, \citenamefont {{Gasparrini}}, \citenamefont
  {{Germani}}, \citenamefont {{Giglietto}}, \citenamefont {{Giordano}},
  \citenamefont {{Glanzman}}, \citenamefont {{Godfrey}}, \citenamefont
  {{Grenier}}, \citenamefont {{Guiriec}}, \citenamefont {{Gustafsson}},
  \citenamefont {{Hadasch}}, \citenamefont {{Hayashida}}, \citenamefont
  {{Hays}}, \citenamefont {{Hughes}}, \citenamefont {{J{\'o}hannesson}},
  \citenamefont {{Johnson}}, \citenamefont {{Kamae}}, \citenamefont
  {{Katagiri}}, \citenamefont {{Kataoka}}, \citenamefont {{Kn{\"o}dlseder}},
  \citenamefont {{Kuss}}, \citenamefont {{Lande}}, \citenamefont {{Longo}},
  \citenamefont {{Loparco}}, \citenamefont {{Lott}}, \citenamefont
  {{Lovellette}}, \citenamefont {{Lubrano}}, \citenamefont {{Madejski}},
  \citenamefont {{Martin}}, \citenamefont {{Mazziotta}}, \citenamefont
  {{McEnery}}, \citenamefont {{Michelson}}, \citenamefont {{Mizuno}},
  \citenamefont {{Monte}}, \citenamefont {{Monzani}}, \citenamefont
  {{Morselli}}, \citenamefont {{Moskalenko}}, \citenamefont {{Murgia}},
  \citenamefont {{Nishino}}, \citenamefont {{Norris}}, \citenamefont {{Nuss}},
  \citenamefont {{Ohno}}, \citenamefont {{Ohsugi}}, \citenamefont {{Okumura}},
  \citenamefont {{Omodei}}, \citenamefont {{Orlando}}, \citenamefont {{Ozaki}},
  \citenamefont {{Parent}}, \citenamefont {{Persic}}, \citenamefont
  {{Pesce-Rollins}}, \citenamefont {{Petrosian}}, \citenamefont
  {{Pierbattista}}, \citenamefont {{Piron}}, \citenamefont {{Pivato}},
  \citenamefont {{Porter}}, \citenamefont {{Rain{\`o}}}, \citenamefont
  {{Rando}}, \citenamefont {{Razzano}}, \citenamefont {{Reimer}}, \citenamefont
  {{Reimer}}, \citenamefont {{Ritz}}, \citenamefont {{Roth}}, \citenamefont
  {{Sbarra}}, \citenamefont {{Sgr{\`o}}}, \citenamefont {{Siskind}},
  \citenamefont {{Spandre}}, \citenamefont {{Spinelli}}, \citenamefont
  {{Stawarz}}, \citenamefont {{Strong}}, \citenamefont {{Takahashi}},
  \citenamefont {{Tanaka}}, \citenamefont {{Thayer}}, \citenamefont
  {{Tibaldo}}, \citenamefont {{Tinivella}}, \citenamefont {{Torres}},
  \citenamefont {{Tosti}}, \citenamefont {{Troja}}, \citenamefont {{Uchiyama}},
  \citenamefont {{Vandenbroucke}}, \citenamefont {{Vianello}}, \citenamefont
  {{Vitale}}, \citenamefont {{Waite}}, \citenamefont {{Wood}},\ and\
  \citenamefont {{Yang}}}]{fermi12sfg}%
  \BibitemOpen
  \bibfield  {author} {\bibinfo {author} {\bibfnamefont {M.}~\bibnamefont
  {{Ackermann}}} {\it et~al.},\ }\bibfield  {title} {\enquote {\bibinfo {title}
  {{GeV Observations of Star-forming Galaxies with the Fermi Large Area
  Telescope}},}\ }\href {\doibase 10.1088/0004-637X/755/2/164} {\bibfield
  {journal} {\bibinfo  {journal} {The Astrophysical Journal}\ }\textbf
  {\bibinfo {volume} {755}},\ \bibinfo {eid} {164} (\bibinfo {year} {2012})},\
  \Eprint {http://arxiv.org/abs/1206.1346}{arXiv:1206.1346}\BibitemShut
  {NoStop}%
\bibitem [{\citenamefont {{Zeng}}\ \emph {{\it et~al.}}(2013)\citenamefont
  {{Zeng}}, \citenamefont {{Yan}},\ and\ \citenamefont
  {{Zhang}}}]{zeng13_egbFSRQ}%
  \BibitemOpen
  \bibfield  {author} {\bibinfo {author} {\bibfnamefont {H.}~\bibnamefont
  {{Zeng}}}, \bibinfo {author} {\bibfnamefont {D.}~\bibnamefont {{Yan}}}, and\
  \bibinfo {author} {\bibfnamefont {L.}~\bibnamefont {{Zhang}}},\ }\bibfield
  {title} {\enquote {\bibinfo {title} {{A revisit of gamma-ray luminosity
  function and contribution to the extragalactic diffuse gamma-ray background
  for Fermi FSRQs}},}\ }\href {\doibase 10.1093/mnras/stt223} {\bibfield
  {journal} {\bibinfo  {journal} {Monthly Notices of the Royal Astronomical
  Society}\ }\textbf {\bibinfo {volume} {431}},\ \bibinfo {pages} {997}
  (\bibinfo {year} {2013})}\BibitemShut {NoStop}%
\bibitem [{\citenamefont {{Ajello}}\ \emph {{\it et~al.}}(2014)\citenamefont
  {{Ajello}}, \citenamefont {{Romani}}, \citenamefont {{Gasparrini}},
  \citenamefont {{Shaw}}, \citenamefont {{Bolmer}}, \citenamefont {{Cotter}},
  \citenamefont {{Finke}}, \citenamefont {{Greiner}}, \citenamefont {{Healey}},
  \citenamefont {{King}}, \citenamefont {{Max-Moerbeck}}, \citenamefont
  {{Michelson}}, \citenamefont {{Potter}}, \citenamefont {{Rau}}, \citenamefont
  {{Readhead}}, \citenamefont {{Richards}},\ and\ \citenamefont
  {{Schady}}}]{Ajello14_bllac}%
  \BibitemOpen
  \bibfield  {author} {\bibinfo {author} {\bibfnamefont {M.}~\bibnamefont
  {{Ajello}}} {\it et~al.},\ }\bibfield  {title} {\enquote {\bibinfo {title}
  {{The Cosmic Evolution of Fermi BL Lacertae Objects}},}\ }\href {\doibase
  10.1088/0004-637X/780/1/73} {\bibfield  {journal} {\bibinfo  {journal} {The
  Astrophysical Journal}\ }\textbf {\bibinfo {volume} {780}},\ \bibinfo {eid}
  {73} (\bibinfo {year} {2014})},\ \Eprint
  {http://arxiv.org/abs/1310.0006}{arXiv:1310.0006}\BibitemShut {NoStop}%
\bibitem [{\citenamefont {{Di Mauro}}\ \emph {{\it et~al.}}(2014)\citenamefont
  {{Di Mauro}}, \citenamefont {{Calore}}, \citenamefont {{Donato}},
  \citenamefont {{Ajello}},\ and\ \citenamefont
  {{Latronico}}}]{DiMauro14_magn}%
  \BibitemOpen
  \bibfield  {author} {\bibinfo {author} {\bibfnamefont {M.}~\bibnamefont {{Di
  Mauro}}}, \bibinfo {author} {\bibfnamefont {F.}~\bibnamefont {{Calore}}},
  \bibinfo {author} {\bibfnamefont {F.}~\bibnamefont {{Donato}}}, \bibinfo
  {author} {\bibfnamefont {M.}~\bibnamefont {{Ajello}}}, and\ \bibinfo {author}
  {\bibfnamefont {L.}~\bibnamefont {{Latronico}}},\ }\bibfield  {title}
  {\enquote {\bibinfo {title} {{Diffuse {\ensuremath{\gamma}}-Ray Emission from
  Misaligned Active Galactic Nuclei}},}\ }\href {\doibase
  10.1088/0004-637X/780/2/161} {\bibfield  {journal} {\bibinfo  {journal}
  {Astrophys. J.}\ }\textbf {\bibinfo {volume} {780}},\ \bibinfo {eid} {161}
  (\bibinfo {year} {2014})},\ \Eprint
  {http://arxiv.org/abs/1304.0908}{arXiv:1304.0908}\BibitemShut {NoStop}%
\bibitem [{\citenamefont {{Ajello}}\ \emph {{\it et~al.}}(2015)\citenamefont
  {{Ajello}}, \citenamefont {{Gasparrini}}, \citenamefont
  {{S{\'a}nchez-Conde}}, \citenamefont {{Zaharijas}}, \citenamefont
  {{Gustafsson}}, \citenamefont {{Cohen-Tanugi}}, \citenamefont {{Dermer}},
  \citenamefont {{Inoue}}, \citenamefont {{Hartmann}}, \citenamefont
  {{Ackermann}}, \citenamefont {{Bechtol}}, \citenamefont {{Franckowiak}},
  \citenamefont {{Reimer}}, \citenamefont {{Romani}},\ and\ \citenamefont
  {{Strong}}}]{ajello15egbdm}%
  \BibitemOpen
  \bibfield  {author} {\bibinfo {author} {\bibfnamefont {M.}~\bibnamefont
  {{Ajello}}} {\it et~al.},\ }\bibfield  {title} {\enquote {\bibinfo {title}
  {{The Origin of the Extragalactic Gamma-Ray Background and Implications for
  Dark Matter Annihilation}},}\ }\href {\doibase 10.1088/2041-8205/800/2/L27}
  {\bibfield  {journal} {\bibinfo  {journal} {The Astrophysical Journal}\
  }\textbf {\bibinfo {volume} {800}},\ \bibinfo {eid} {L27} (\bibinfo {year}
  {2015})},\ \Eprint
  {http://arxiv.org/abs/1501.05301}{arXiv:1501.05301}\BibitemShut {NoStop}%
\bibitem [{\citenamefont {{Qu}}\ \emph {{\it et~al.}}(2019)\citenamefont
  {{Qu}}, \citenamefont {{Zeng}},\ and\ \citenamefont {{Yan}}}]{qu19_bllac}%
  \BibitemOpen
  \bibfield  {author} {\bibinfo {author} {\bibfnamefont {Y.}~\bibnamefont
  {{Qu}}}, \bibinfo {author} {\bibfnamefont {H.}~\bibnamefont {{Zeng}}}, and\
  \bibinfo {author} {\bibfnamefont {D.}~\bibnamefont {{Yan}}},\ }\bibfield
  {title} {\enquote {\bibinfo {title} {{Gamma-ray luminosity function of BL Lac
  objects and contribution to the extragalactic gamma-ray background}},}\
  }\href {\doibase 10.1093/mnras/stz2651} {\bibfield  {journal} {\bibinfo
  {journal} {\mnras}\ }\textbf {\bibinfo {volume} {490}},\ \bibinfo {pages}
  {758} (\bibinfo {year} {2019})},\ \Eprint
  {http://arxiv.org/abs/1909.07542}{arXiv:1909.07542}\BibitemShut {NoStop}%
\bibitem [{\citenamefont {{Zeng}}\ \emph {{\it et~al.}}(2021)\citenamefont
  {{Zeng}}, \citenamefont {{Petrosian}},\ and\ \citenamefont
  {{Yi}}}]{zhd21blazar}%
  \BibitemOpen
  \bibfield  {author} {\bibinfo {author} {\bibfnamefont {H.}~\bibnamefont
  {{Zeng}}}, \bibinfo {author} {\bibfnamefont {V.}~\bibnamefont {{Petrosian}}},
  and\ \bibinfo {author} {\bibfnamefont {T.}~\bibnamefont {{Yi}}},\ }\bibfield
  {title} {\enquote {\bibinfo {title} {{Cosmological Evolution of Fermi Large
  Area Telescope Gamma-Ray Blazars Using Novel Nonparametric Methods}},}\
  }\href {\doibase 10.3847/1538-4357/abf65e} {\bibfield  {journal} {\bibinfo
  {journal} {Astrophys. J.}\ }\textbf {\bibinfo {volume} {913}},\ \bibinfo
  {eid} {120} (\bibinfo {year} {2021})},\ \Eprint
  {http://arxiv.org/abs/2104.04686}{arXiv:2104.04686}\BibitemShut {NoStop}%
\bibitem [{\citenamefont {Roth}\ \emph {{\it et~al.}}(2021)\citenamefont
  {Roth}, \citenamefont {Krumholz}, \citenamefont {Crocker},\ and\
  \citenamefont {Celli}}]{Roth21sfgNat}%
  \BibitemOpen
  \bibfield  {author} {\bibinfo {author} {\bibfnamefont {M.~A.}\ \bibnamefont
  {Roth}}, \bibinfo {author} {\bibfnamefont {M.~R.}\ \bibnamefont {Krumholz}},
  \bibinfo {author} {\bibfnamefont {R.~M.}\ \bibnamefont {Crocker}}, and\
  \bibinfo {author} {\bibfnamefont {S.}~\bibnamefont {Celli}},\ }\bibfield
  {title} {\enquote {\bibinfo {title} {{The diffuse $\gamma$-ray background is
  dominated by star-forming galaxies}},}\ }\href {\doibase
  10.1038/s41586-021-03802-x} {\bibfield  {journal} {\bibinfo  {journal}
  {Nature}\ }\textbf {\bibinfo {volume} {597}},\ \bibinfo {pages} {341}
  (\bibinfo {year} {2021})},\ \Eprint
  {http://arxiv.org/abs/2109.07598}{arXiv:2109.07598}\BibitemShut {NoStop}%
\bibitem [{\citenamefont {{Ajello}}\ \emph {{\it
  et~al.}}(2020{\natexlab{a}})\citenamefont {{Ajello}}, \citenamefont
  {{Angioni}}, \citenamefont {{Axelsson}}, \citenamefont {{Ballet}},
  \citenamefont {{Barbiellini}}, \citenamefont {{Bastieri}}, \citenamefont
  {{Becerra Gonzalez}}, \citenamefont {{Bellazzini}}, \citenamefont
  {{Bissaldi}}, \citenamefont {{Bloom}}, \citenamefont {{Bonino}},
  \citenamefont {{Bottacini}}, \citenamefont {{Bruel}}, \citenamefont
  {{Buson}}, \citenamefont {{Cafardo}}, \citenamefont {{Cameron}},
  \citenamefont {{Cavazzuti}}, \citenamefont {{Chen}}, \citenamefont
  {{Cheung}}, \citenamefont {{Ciprini}}, \citenamefont {{Costantin}},
  \citenamefont {{Cutini}}, \citenamefont {{D'Ammando}}, \citenamefont {{de la
  Torre Luque}}, \citenamefont {{de Menezes}}, \citenamefont {{de Palma}},
  \citenamefont {{Desai}}, \citenamefont {{Di Lalla}}, \citenamefont {{Di
  Venere}}, \citenamefont {{Dom{\'\i}nguez}}, \citenamefont {{Dirirsa}},
  \citenamefont {{Ferrara}}, \citenamefont {{Finke}}, \citenamefont
  {{Franckowiak}}, \citenamefont {{Fukazawa}}, \citenamefont {{Funk}},
  \citenamefont {{Fusco}}, \citenamefont {{Gargano}}, \citenamefont
  {{Garrappa}}, \citenamefont {{Gasparrini}}, \citenamefont {{Giglietto}},
  \citenamefont {{Giordano}}, \citenamefont {{Giroletti}}, \citenamefont
  {{Green}}, \citenamefont {{Grenier}}, \citenamefont {{Guiriec}},
  \citenamefont {{Harita}}, \citenamefont {{Hays}}, \citenamefont {{Horan}},
  \citenamefont {{Itoh}}, \citenamefont {{J{\'o}hannesson}}, \citenamefont
  {{Kovac'evic'}}, \citenamefont {{Krauss}}, \citenamefont {{Kreter}},
  \citenamefont {{Kuss}}, \citenamefont {{Larsson}}, \citenamefont {{Leto}},
  \citenamefont {{Li}}, \citenamefont {{Liodakis}}, \citenamefont {{Longo}},
  \citenamefont {{Loparco}}, \citenamefont {{Lott}}, \citenamefont
  {{Lovellette}}, \citenamefont {{Lubrano}}, \citenamefont {{Madejski}},
  \citenamefont {{Maldera}}, \citenamefont {{Manfreda}}, \citenamefont
  {{Mart{\'\i}-Devesa}}, \citenamefont {{Massaro}}, \citenamefont
  {{Mazziotta}}, \citenamefont {{Mereu}}, \citenamefont {{Meyer}},
  \citenamefont {{Migliori}}, \citenamefont {{Mirabal}}, \citenamefont
  {{Mizuno}}, \citenamefont {{Monzani}}, \citenamefont {{Morselli}},
  \citenamefont {{Moskalenko}}, \citenamefont {{Negro}}, \citenamefont
  {{Nemmen}}, \citenamefont {{Nuss}}, \citenamefont {{Ojha}}, \citenamefont
  {{Ojha}}, \citenamefont {{Omodei}}, \citenamefont {{Orienti}}, \citenamefont
  {{Orlando}}, \citenamefont {{Ormes}}, \citenamefont {{Paliya}}, \citenamefont
  {{Pei}}, \citenamefont {{Pe{\~n}a-Herazo}}, \citenamefont {{Persic}},
  \citenamefont {{Pesce-Rollins}}, \citenamefont {{Petrov}}, \citenamefont
  {{Piron}}, \citenamefont {{Poon}}, \citenamefont {{Principe}}, \citenamefont
  {{Rain{\`o}}}, \citenamefont {{Rando}}, \citenamefont {{Rani}}, \citenamefont
  {{Razzano}}, \citenamefont {{Razzaque}}, \citenamefont {{Reimer}},
  \citenamefont {{Reimer}}, \citenamefont {{Schinzel}}, \citenamefont
  {{Serini}}, \citenamefont {{Sgr{\`o}}}, \citenamefont {{Siskind}},
  \citenamefont {{Spandre}}, \citenamefont {{Spinelli}}, \citenamefont
  {{Suson}}, \citenamefont {{Tachibana}}, \citenamefont {{Thompson}},
  \citenamefont {{Torres}}, \citenamefont {{Torresi}}, \citenamefont {{Troja}},
  \citenamefont {{Valverde}}, \citenamefont {{van Zyl}},\ and\ \citenamefont
  {{Yassine}}}]{fermi4lac}%
  \BibitemOpen
  \bibfield  {author} {\bibinfo {author} {\bibfnamefont {M.}~\bibnamefont
  {{Ajello}}} {\it et~al.},\ }\bibfield  {title} {\enquote {\bibinfo {title}
  {{The Fourth Catalog of Active Galactic Nuclei Detected by the Fermi Large
  Area Telescope}},}\ }\href {\doibase 10.3847/1538-4357/ab791e} {\bibfield
  {journal} {\bibinfo  {journal} {Astrophys. J.}\ }\textbf {\bibinfo {volume}
  {892}},\ \bibinfo {eid} {105} (\bibinfo {year} {2020}{\natexlab{a}})},\
  \Eprint {http://arxiv.org/abs/1905.10771}{arXiv:1905.10771}\BibitemShut
  {NoStop}%
\bibitem [{\citenamefont {{Padovani}}(1997)}]{1997MmSAI..68...47P}%
  \BibitemOpen
  \bibfield  {author} {\bibinfo {author} {\bibfnamefont {P.}~\bibnamefont
  {{Padovani}}},\ }\bibfield  {title} {\enquote {\bibinfo {title} {{Unified
  schemes for radio-loud AGN: recent results.}}}\ }\href@noop {} {\bibfield
  {journal} {\bibinfo  {journal} {Memorie della Societa Astronomica Italiana}\
  }\textbf {\bibinfo {volume} {68}},\ \bibinfo {pages} {47} (\bibinfo {year}
  {1997})},\ \Eprint
  {http://arxiv.org/abs/astro-ph/9701074}{arXiv:astro-ph/9701074}\BibitemShut
  {NoStop}%
\bibitem [{\citenamefont {{Ajello}}\ \emph {{\it
  et~al.}}(2020{\natexlab{b}})\citenamefont {{Ajello}}, \citenamefont {{Di
  Mauro}}, \citenamefont {{Paliya}},\ and\ \citenamefont
  {{Garrappa}}}]{ajello2020sfg}%
  \BibitemOpen
  \bibfield  {author} {\bibinfo {author} {\bibfnamefont {M.}~\bibnamefont
  {{Ajello}}}, \bibinfo {author} {\bibfnamefont {M.}~\bibnamefont {{Di
  Mauro}}}, \bibinfo {author} {\bibfnamefont {V.~S.}\ \bibnamefont {{Paliya}}},
  and\ \bibinfo {author} {\bibfnamefont {S.}~\bibnamefont {{Garrappa}}},\
  }\bibfield  {title} {\enquote {\bibinfo {title} {{The
  {\ensuremath{\gamma}}-Ray Emission of Star-forming Galaxies}},}\ }\href
  {\doibase 10.3847/1538-4357/ab86a6} {\bibfield  {journal} {\bibinfo
  {journal} {Astrophys. J.}\ }\textbf {\bibinfo {volume} {894}},\ \bibinfo
  {eid} {88} (\bibinfo {year} {2020}{\natexlab{b}})},\ \Eprint
  {http://arxiv.org/abs/2003.05493}{arXiv:2003.05493}\BibitemShut {NoStop}%
\bibitem [{\citenamefont {{Casanova}}\ \emph {{\it et~al.}}(2007)\citenamefont
  {{Casanova}}, \citenamefont {{Dingus}},\ and\ \citenamefont
  {{Zhang}}}]{2007ApJ...656..306C}%
  \BibitemOpen
  \bibfield  {author} {\bibinfo {author} {\bibfnamefont {S.}~\bibnamefont
  {{Casanova}}}, \bibinfo {author} {\bibfnamefont {B.~L.}\ \bibnamefont
  {{Dingus}}}, and\ \bibinfo {author} {\bibfnamefont {B.}~\bibnamefont
  {{Zhang}}},\ }\bibfield  {title} {\enquote {\bibinfo {title} {{Contribution
  of GRB Emission to the GeV Extragalactic Diffuse Gamma-Ray Flux}},}\ }\href
  {\doibase 10.1086/510613} {\bibfield  {journal} {\bibinfo  {journal} {The
  Astrophysical Journal}\ }\textbf {\bibinfo {volume} {656}},\ \bibinfo {pages}
  {306} (\bibinfo {year} {2007})}\BibitemShut {NoStop}%
\bibitem [{\citenamefont {{Faucher-Gigu{\`e}re}}\ and\ \citenamefont
  {{Loeb}}(2010)}]{2010JCAP...01..005F}%
  \BibitemOpen
  \bibfield  {author} {\bibinfo {author} {\bibfnamefont {C.-A.}\ \bibnamefont
  {{Faucher-Gigu{\`e}re}}} and\ \bibinfo {author} {\bibfnamefont
  {A.}~\bibnamefont {{Loeb}}},\ }\bibfield  {title} {\enquote {\bibinfo {title}
  {{The pulsar contribution to the gamma-ray background}},}\ }\href {\doibase
  10.1088/1475-7516/2010/01/005} {\bibfield  {journal} {\bibinfo  {journal}
  {Journal of Cosmology and Astroparticle Physics}\ }\textbf {\bibinfo {volume}
  {2010}},\ \bibinfo {eid} {005} (\bibinfo {year} {2010})},\ \Eprint
  {http://arxiv.org/abs/0904.3102}{arXiv:0904.3102}\BibitemShut {NoStop}%
\bibitem [{\citenamefont {{Loeb}}\ and\ \citenamefont
  {{Waxman}}(2000)}]{2000Natur.405..156L}%
  \BibitemOpen
  \bibfield  {author} {\bibinfo {author} {\bibfnamefont {A.}~\bibnamefont
  {{Loeb}}} and\ \bibinfo {author} {\bibfnamefont {E.}~\bibnamefont
  {{Waxman}}},\ }\bibfield  {title} {\enquote {\bibinfo {title} {{Cosmic
  {\ensuremath{\gamma}}-ray background from structure formation in the
  intergalactic medium}},}\ }\href {\doibase 10.1038/35012018} {\bibfield
  {journal} {\bibinfo  {journal} {Nature}\ }\textbf {\bibinfo {volume} {405}},\
  \bibinfo {pages} {156} (\bibinfo {year} {2000})},\ \Eprint
  {http://arxiv.org/abs/astro-ph/0003447}{arXiv:astro-ph/0003447}\BibitemShut
  {NoStop}%
\bibitem [{\citenamefont {{Totani}}\ and\ \citenamefont
  {{Kitayama}}(2000)}]{2000ApJ...545..572T}%
  \BibitemOpen
  \bibfield  {author} {\bibinfo {author} {\bibfnamefont {T.}~\bibnamefont
  {{Totani}}} and\ \bibinfo {author} {\bibfnamefont {T.}~\bibnamefont
  {{Kitayama}}},\ }\bibfield  {title} {\enquote {\bibinfo {title} {{Forming
  Clusters of Galaxies as the Origin of Unidentified GEV Gamma-Ray Sources}},}\
  }\href {\doibase 10.1086/317872} {\bibfield  {journal} {\bibinfo  {journal}
  {The Astrophysical Journal}\ }\textbf {\bibinfo {volume} {545}},\ \bibinfo
  {pages} {572} (\bibinfo {year} {2000})},\ \Eprint
  {http://arxiv.org/abs/astro-ph/0006176}{arXiv:astro-ph/0006176}\BibitemShut
  {NoStop}%
\bibitem [{\citenamefont {{Dar}}\ and\ \citenamefont
  {{Shaviv}}(1995)}]{1995PhRvL..75.3052D}%
  \BibitemOpen
  \bibfield  {author} {\bibinfo {author} {\bibfnamefont {A.}~\bibnamefont
  {{Dar}}} and\ \bibinfo {author} {\bibfnamefont {N.~J.}\ \bibnamefont
  {{Shaviv}}},\ }\bibfield  {title} {\enquote {\bibinfo {title} {{Origin of the
  High Energy Extragalactic Diffuse Gamma Ray Background}},}\ }\href {\doibase
  10.1103/PhysRevLett.75.3052} {\bibfield  {journal} {\bibinfo  {journal}
  {Physical Review Letters}\ }\textbf {\bibinfo {volume} {75}},\ \bibinfo
  {pages} {3052} (\bibinfo {year} {1995})},\ \Eprint
  {http://arxiv.org/abs/astro-ph/9501079}{arXiv:astro-ph/9501079}\BibitemShut
  {NoStop}%
\bibitem [{\citenamefont {{Abazajian}}\ \emph {{\it et~al.}}(2010)\citenamefont
  {{Abazajian}}, \citenamefont {{Agrawal}}, \citenamefont {{Chacko}},\ and\
  \citenamefont {{Kilic}}}]{Abazajian10_igrbDM}%
  \BibitemOpen
  \bibfield  {author} {\bibinfo {author} {\bibfnamefont {K.~N.}\ \bibnamefont
  {{Abazajian}}}, \bibinfo {author} {\bibfnamefont {P.}~\bibnamefont
  {{Agrawal}}}, \bibinfo {author} {\bibfnamefont {Z.}~\bibnamefont {{Chacko}}},
  and\ \bibinfo {author} {\bibfnamefont {C.}~\bibnamefont {{Kilic}}},\
  }\bibfield  {title} {\enquote {\bibinfo {title} {{Conservative constraints on
  dark matter from the Fermi-LAT isotropic diffuse gamma-ray background
  spectrum}},}\ }\href {\doibase 10.1088/1475-7516/2010/11/041} {\bibfield
  {journal} {\bibinfo  {journal} {\jcap}\ }\textbf {\bibinfo {volume} {2010}},\
  \bibinfo {eid} {041} (\bibinfo {year} {2010})},\ \Eprint
  {http://arxiv.org/abs/1002.3820}{arXiv:1002.3820}\BibitemShut {NoStop}%
\bibitem [{\citenamefont {{Ando}}\ and\ \citenamefont
  {{Ishiwata}}(2015)}]{Ando15_egbDM}%
  \BibitemOpen
  \bibfield  {author} {\bibinfo {author} {\bibfnamefont {S.}~\bibnamefont
  {{Ando}}} and\ \bibinfo {author} {\bibfnamefont {K.}~\bibnamefont
  {{Ishiwata}}},\ }\bibfield  {title} {\enquote {\bibinfo {title} {{Constraints
  on decaying dark matter from the extragalactic gamma-ray background}},}\
  }\href {\doibase 10.1088/1475-7516/2015/05/024} {\bibfield  {journal}
  {\bibinfo  {journal} {\jcap}\ }\textbf {\bibinfo {volume} {2015}},\ \bibinfo
  {eid} {024} (\bibinfo {year} {2015})},\ \Eprint
  {http://arxiv.org/abs/1502.02007}{arXiv:1502.02007}\BibitemShut {NoStop}%
\bibitem [{\citenamefont {{Di Mauro}}\ and\ \citenamefont
  {{Donato}}(2015)}]{DiMauro15_igrbDM}%
  \BibitemOpen
  \bibfield  {author} {\bibinfo {author} {\bibfnamefont {M.}~\bibnamefont {{Di
  Mauro}}} and\ \bibinfo {author} {\bibfnamefont {F.}~\bibnamefont
  {{Donato}}},\ }\bibfield  {title} {\enquote {\bibinfo {title} {{Composition
  of the Fermi-LAT isotropic gamma-ray background intensity: Emission from
  extragalactic point sources and dark matter annihilations}},}\ }\href
  {\doibase 10.1103/PhysRevD.91.123001} {\bibfield  {journal} {\bibinfo
  {journal} {Physical Review D}\ }\textbf {\bibinfo {volume} {91}},\ \bibinfo
  {eid} {123001} (\bibinfo {year} {2015})},\ \Eprint
  {http://arxiv.org/abs/1501.05316}{arXiv:1501.05316}\BibitemShut {NoStop}%
\bibitem [{\citenamefont {{Fermi LAT Collaboration}}(2015)}]{fermi15_igbdm}%
  \BibitemOpen
  \bibfield  {author} {\bibinfo {author} {\bibnamefont {{Fermi LAT
  Collaboration}}},\ }\bibfield  {title} {\enquote {\bibinfo {title} {{Limits
  on dark matter annihilation signals from the Fermi LAT 4-year measurement of
  the isotropic gamma-ray background}},}\ }\href {\doibase
  10.1088/1475-7516/2015/09/008} {\bibfield  {journal} {\bibinfo  {journal}
  {\jcap}\ }\textbf {\bibinfo {volume} {2015}},\ \bibinfo {eid} {008} (\bibinfo
  {year} {2015})},\ \Eprint
  {http://arxiv.org/abs/1501.05464}{arXiv:1501.05464}\BibitemShut {NoStop}%
\bibitem [{\citenamefont {{Liu}}\ \emph {{\it et~al.}}(2017)\citenamefont
  {{Liu}}, \citenamefont {{Bi}}, \citenamefont {{Lin}},\ and\ \citenamefont
  {{Yin}}}]{liuwei17cpc}%
  \BibitemOpen
  \bibfield  {author} {\bibinfo {author} {\bibfnamefont {W.}~\bibnamefont
  {{Liu}}}, \bibinfo {author} {\bibfnamefont {X.-J.}\ \bibnamefont {{Bi}}},
  \bibinfo {author} {\bibfnamefont {S.-J.}\ \bibnamefont {{Lin}}}, and\
  \bibinfo {author} {\bibfnamefont {P.-F.}\ \bibnamefont {{Yin}}},\ }\bibfield
  {title} {\enquote {\bibinfo {title} {{Constraints on dark matter annihilation
  and decay from the isotropic gamma-ray background}},}\ }\href {\doibase
  10.1088/1674-1137/41/4/045104} {\bibfield  {journal} {\bibinfo  {journal}
  {Chinese Physics C}\ }\textbf {\bibinfo {volume} {41}},\ \bibinfo {eid}
  {045104} (\bibinfo {year} {2017})},\ \Eprint
  {http://arxiv.org/abs/1602.01012}{arXiv:1602.01012}\BibitemShut {NoStop}%
\bibitem [{\citenamefont {{Blanco}}\ and\ \citenamefont
  {{Hooper}}(2019)}]{hooper19igrb}%
  \BibitemOpen
  \bibfield  {author} {\bibinfo {author} {\bibfnamefont {C.}~\bibnamefont
  {{Blanco}}} and\ \bibinfo {author} {\bibfnamefont {D.}~\bibnamefont
  {{Hooper}}},\ }\bibfield  {title} {\enquote {\bibinfo {title} {{Constraints
  on decaying dark matter from the isotropic gamma-ray background}},}\ }\href
  {\doibase 10.1088/1475-7516/2019/03/019} {\bibfield  {journal} {\bibinfo
  {journal} {\jcap}\ }\textbf {\bibinfo {volume} {2019}},\ \bibinfo {eid} {019}
  (\bibinfo {year} {2019})},\ \Eprint
  {http://arxiv.org/abs/1811.05988}{arXiv:1811.05988}\BibitemShut {NoStop}%
\bibitem [{\citenamefont {{Arbey}}\ \emph {{\it et~al.}}(2020)\citenamefont
  {{Arbey}}, \citenamefont {{Auffinger}},\ and\ \citenamefont
  {{Silk}}}]{arbey20_igrbPBH}%
  \BibitemOpen
  \bibfield  {author} {\bibinfo {author} {\bibfnamefont {A.}~\bibnamefont
  {{Arbey}}}, \bibinfo {author} {\bibfnamefont {J.}~\bibnamefont
  {{Auffinger}}}, and\ \bibinfo {author} {\bibfnamefont {J.}~\bibnamefont
  {{Silk}}},\ }\bibfield  {title} {\enquote {\bibinfo {title} {{Constraining
  primordial black hole masses with the isotropic gamma ray background}},}\
  }\href {\doibase 10.1103/PhysRevD.101.023010} {\bibfield  {journal} {\bibinfo
   {journal} {Physical Review D}\ }\textbf {\bibinfo {volume} {101}},\ \bibinfo
  {eid} {023010} (\bibinfo {year} {2020})},\ \Eprint
  {http://arxiv.org/abs/1906.04750}{arXiv:1906.04750}\BibitemShut {NoStop}%
\bibitem [{\citenamefont {{Planck Collaboration}}\ \emph {{\it
  et~al.}}(2016)\citenamefont {{Planck Collaboration}}, \citenamefont {{Ade}},
  \citenamefont {{Aghanim}}, \citenamefont {{Arnaud}}, \citenamefont
  {{Ashdown}}, \citenamefont {{Aumont}}, \citenamefont {{Baccigalupi}},
  \citenamefont {{Banday}}, \citenamefont {{Barreiro}}, \citenamefont
  {{Bartlett}}, \citenamefont {{Bartolo}}, \citenamefont {{Battaner}},
  \citenamefont {{Battye}}, \citenamefont {{Benabed}}, \citenamefont
  {{Beno{\^\i}t}}, \citenamefont {{Benoit-L{\'e}vy}}, \citenamefont
  {{Bernard}}, \citenamefont {{Bersanelli}}, \citenamefont {{Bielewicz}},
  \citenamefont {{Bock}}, \citenamefont {{Bonaldi}}, \citenamefont
  {{Bonavera}}, \citenamefont {{Bond}}, \citenamefont {{Borrill}},
  \citenamefont {{Bouchet}}, \citenamefont {{Boulanger}}, \citenamefont
  {{Bucher}}, \citenamefont {{Burigana}}, \citenamefont {{Butler}},
  \citenamefont {{Calabrese}}, \citenamefont {{Cardoso}}, \citenamefont
  {{Catalano}}, \citenamefont {{Challinor}}, \citenamefont {{Chamballu}},
  \citenamefont {{Chary}}, \citenamefont {{Chiang}}, \citenamefont {{Chluba}},
  \citenamefont {{Christensen}}, \citenamefont {{Church}}, \citenamefont
  {{Clements}}, \citenamefont {{Colombi}}, \citenamefont {{Colombo}},
  \citenamefont {{Combet}}, \citenamefont {{Coulais}}, \citenamefont {{Crill}},
  \citenamefont {{Curto}}, \citenamefont {{Cuttaia}}, \citenamefont {{Danese}},
  \citenamefont {{Davies}}, \citenamefont {{Davis}}, \citenamefont {{de
  Bernardis}}, \citenamefont {{de Rosa}}, \citenamefont {{de Zotti}},
  \citenamefont {{Delabrouille}}, \citenamefont {{D{\'e}sert}}, \citenamefont
  {{Di Valentino}}, \citenamefont {{Dickinson}}, \citenamefont {{Diego}},
  \citenamefont {{Dolag}}, \citenamefont {{Dole}}, \citenamefont {{Donzelli}},
  \citenamefont {{Dor{\'e}}}, \citenamefont {{Douspis}}, \citenamefont
  {{Ducout}}, \citenamefont {{Dunkley}}, \citenamefont {{Dupac}}, \citenamefont
  {{Efstathiou}}, \citenamefont {{Elsner}}, \citenamefont {{En{\ss}lin}},
  \citenamefont {{Eriksen}}, \citenamefont {{Farhang}}, \citenamefont
  {{Fergusson}}, \citenamefont {{Finelli}}, \citenamefont {{Forni}},
  \citenamefont {{Frailis}}, \citenamefont {{Fraisse}}, \citenamefont
  {{Franceschi}}, \citenamefont {{Frejsel}}, \citenamefont {{Galeotta}},
  \citenamefont {{Galli}}, \citenamefont {{Ganga}}, \citenamefont {{Gauthier}},
  \citenamefont {{Gerbino}}, \citenamefont {{Ghosh}}, \citenamefont {{Giard}},
  \citenamefont {{Giraud-H{\'e}raud}}, \citenamefont {{Giusarma}},
  \citenamefont {{Gjerl{\o}w}}, \citenamefont {{Gonz{\'a}lez-Nuevo}},
  \citenamefont {{G{\'o}rski}}, \citenamefont {{Gratton}}, \citenamefont
  {{Gregorio}}, \citenamefont {{Gruppuso}}, \citenamefont {{Gudmundsson}},
  \citenamefont {{Hamann}}, \citenamefont {{Hansen}}, \citenamefont {{Hanson}},
  \citenamefont {{Harrison}}, \citenamefont {{Helou}}, \citenamefont
  {{Henrot-Versill{\'e}}}, \citenamefont {{Hern{\'a}ndez-Monteagudo}},
  \citenamefont {{Herranz}}, \citenamefont {{Hildebrandt}}, \citenamefont
  {{Hivon}}, \citenamefont {{Hobson}}, \citenamefont {{Holmes}}, \citenamefont
  {{Hornstrup}}, \citenamefont {{Hovest}}, \citenamefont {{Huang}},
  \citenamefont {{Huffenberger}}, \citenamefont {{Hurier}}, \citenamefont
  {{Jaffe}}, \citenamefont {{Jaffe}}, \citenamefont {{Jones}}, \citenamefont
  {{Juvela}}, \citenamefont {{Keih{\"a}nen}}, \citenamefont {{Keskitalo}},
  \citenamefont {{Kisner}}, \citenamefont {{Kneissl}}, \citenamefont
  {{Knoche}}, \citenamefont {{Knox}}, \citenamefont {{Kunz}}, \citenamefont
  {{Kurki-Suonio}}, \citenamefont {{Lagache}}, \citenamefont
  {{L{\"a}hteenm{\"a}ki}}, \citenamefont {{Lamarre}}, \citenamefont
  {{Lasenby}}, \citenamefont {{Lattanzi}}, \citenamefont {{Lawrence}},
  \citenamefont {{Leahy}}, \citenamefont {{Leonardi}}, \citenamefont
  {{Lesgourgues}}, \citenamefont {{Levrier}}, \citenamefont {{Lewis}},
  \citenamefont {{Liguori}}, \citenamefont {{Lilje}}, \citenamefont
  {{Linden-V{\o}rnle}}, \citenamefont {{L{\'o}pez-Caniego}}, \citenamefont
  {{Lubin}}, \citenamefont {{Mac{\'\i}as-P{\'e}rez}}, \citenamefont {{Maggio}},
  \citenamefont {{Maino}}, \citenamefont {{Mandolesi}}, \citenamefont
  {{Mangilli}}, \citenamefont {{Marchini}}, \citenamefont {{Maris}},
  \citenamefont {{Martin}}, \citenamefont {{Martinelli}}, \citenamefont
  {{Mart{\'\i}nez-Gonz{\'a}lez}}, \citenamefont {{Masi}}, \citenamefont
  {{Matarrese}}, \citenamefont {{McGehee}}, \citenamefont {{Meinhold}},
  \citenamefont {{Melchiorri}}, \citenamefont {{Melin}}, \citenamefont
  {{Mendes}}, \citenamefont {{Mennella}}, \citenamefont {{Migliaccio}},
  \citenamefont {{Millea}}, \citenamefont {{Mitra}}, \citenamefont
  {{Miville-Desch{\^e}nes}}, \citenamefont {{Moneti}}, \citenamefont
  {{Montier}}, \citenamefont {{Morgante}}, \citenamefont {{Mortlock}},
  \citenamefont {{Moss}}, \citenamefont {{Munshi}}, \citenamefont {{Murphy}},
  \citenamefont {{Naselsky}}, \citenamefont {{Nati}}, \citenamefont {{Natoli}},
  \citenamefont {{Netterfield}}, \citenamefont {{N{\o}rgaard-Nielsen}},
  \citenamefont {{Noviello}}, \citenamefont {{Novikov}}, \citenamefont
  {{Novikov}}, \citenamefont {{Oxborrow}}, \citenamefont {{Paci}},
  \citenamefont {{Pagano}}, \citenamefont {{Pajot}}, \citenamefont
  {{Paladini}}, \citenamefont {{Paoletti}}, \citenamefont {{Partridge}},
  \citenamefont {{Pasian}}, \citenamefont {{Patanchon}}, \citenamefont
  {{Pearson}}, \citenamefont {{Perdereau}}, \citenamefont {{Perotto}},
  \citenamefont {{Perrotta}}, \citenamefont {{Pettorino}}, \citenamefont
  {{Piacentini}}, \citenamefont {{Piat}}, \citenamefont {{Pierpaoli}},
  \citenamefont {{Pietrobon}}, \citenamefont {{Plaszczynski}}, \citenamefont
  {{Pointecouteau}}, \citenamefont {{Polenta}}, \citenamefont {{Popa}},
  \citenamefont {{Pratt}}, \citenamefont {{Pr{\'e}zeau}}, \citenamefont
  {{Prunet}}, \citenamefont {{Puget}}, \citenamefont {{Rachen}}, \citenamefont
  {{Reach}}, \citenamefont {{Rebolo}}, \citenamefont {{Reinecke}},
  \citenamefont {{Remazeilles}}, \citenamefont {{Renault}}, \citenamefont
  {{Renzi}}, \citenamefont {{Ristorcelli}}, \citenamefont {{Rocha}},
  \citenamefont {{Rosset}}, \citenamefont {{Rossetti}}, \citenamefont
  {{Roudier}}, \citenamefont {{Rouill{\'e} d'Orfeuil}}, \citenamefont
  {{Rowan-Robinson}}, \citenamefont {{Rubi{\~n}o-Mart{\'\i}n}}, \citenamefont
  {{Rusholme}}, \citenamefont {{Said}}, \citenamefont {{Salvatelli}},
  \citenamefont {{Salvati}}, \citenamefont {{Sandri}}, \citenamefont
  {{Santos}}, \citenamefont {{Savelainen}}, \citenamefont {{Savini}},
  \citenamefont {{Scott}}, \citenamefont {{Seiffert}}, \citenamefont {{Serra}},
  \citenamefont {{Shellard}}, \citenamefont {{Spencer}}, \citenamefont
  {{Spinelli}}, \citenamefont {{Stolyarov}}, \citenamefont {{Stompor}},
  \citenamefont {{Sudiwala}}, \citenamefont {{Sunyaev}}, \citenamefont
  {{Sutton}}, \citenamefont {{Suur-Uski}}, \citenamefont {{Sygnet}},
  \citenamefont {{Tauber}}, \citenamefont {{Terenzi}}, \citenamefont
  {{Toffolatti}}, \citenamefont {{Tomasi}}, \citenamefont {{Tristram}},
  \citenamefont {{Trombetti}}, \citenamefont {{Tucci}}, \citenamefont
  {{Tuovinen}}, \citenamefont {{T{\"u}rler}}, \citenamefont {{Umana}},
  \citenamefont {{Valenziano}}, \citenamefont {{Valiviita}}, \citenamefont
  {{Van Tent}}, \citenamefont {{Vielva}}, \citenamefont {{Villa}},
  \citenamefont {{Wade}}, \citenamefont {{Wandelt}}, \citenamefont {{Wehus}},
  \citenamefont {{White}}, \citenamefont {{White}}, \citenamefont
  {{Wilkinson}}, \citenamefont {{Yvon}}, \citenamefont {{Zacchei}},\ and\
  \citenamefont {{Zonca}}}]{planck2015}%
  \BibitemOpen
  \bibfield  {author} {\bibinfo {author} {\bibnamefont {{Planck
  Collaboration}}} {\it et~al.},\ }\bibfield  {title} {\enquote {\bibinfo
  {title} {{Planck 2015 results. XIII. Cosmological parameters}},}\ }\href
  {\doibase 10.1051/0004-6361/201525830} {\bibfield  {journal} {\bibinfo
  {journal} {\aap}\ }\textbf {\bibinfo {volume} {594}},\ \bibinfo {eid} {A13}
  (\bibinfo {year} {2016})},\ \Eprint
  {http://arxiv.org/abs/1502.01589}{arXiv:1502.01589}\BibitemShut {NoStop}%
\bibitem [{\citenamefont {{Ginzburg}}\ and\ \citenamefont
  {{Syrovatskii}}(1964)}]{1964ocr..book.....G}%
  \BibitemOpen
  \bibfield  {author} {\bibinfo {author} {\bibfnamefont {V.~L.}\ \bibnamefont
  {{Ginzburg}}} and\ \bibinfo {author} {\bibfnamefont {S.~I.}\ \bibnamefont
  {{Syrovatskii}}},\ }\href@noop {} {\emph {\bibinfo {title} {{The Origin of
  Cosmic Rays}}}}\ (\bibinfo {year} {1964})\BibitemShut {NoStop}%
\bibitem [{\citenamefont {{Ricotti}}\ and\ \citenamefont
  {{Gould}}(2009{\natexlab{a}})}]{2009ApJ...707..979R}%
  \BibitemOpen
  \bibfield  {author} {\bibinfo {author} {\bibfnamefont {M.}~\bibnamefont
  {{Ricotti}}} and\ \bibinfo {author} {\bibfnamefont {A.}~\bibnamefont
  {{Gould}}},\ }\bibfield  {title} {\enquote {\bibinfo {title} {{A New Probe of
  Dark Matter and High-Energy Universe Using Microlensing}},}\ }\href {\doibase
  10.1088/0004-637X/707/2/979} {\bibfield  {journal} {\bibinfo  {journal} {The
  Astrophysical Journal}\ }\textbf {\bibinfo {volume} {707}},\ \bibinfo {pages}
  {979} (\bibinfo {year} {2009}{\natexlab{a}})},\ \Eprint
  {http://arxiv.org/abs/0908.0735}{arXiv:0908.0735}\BibitemShut {NoStop}%
\bibitem [{\citenamefont {{Ricotti}}\ and\ \citenamefont
  {{Gould}}(2009{\natexlab{b}})}]{ricotti09_ucmh}%
  \BibitemOpen
  \bibfield  {author} {\bibinfo {author} {\bibfnamefont {M.}~\bibnamefont
  {{Ricotti}}} and\ \bibinfo {author} {\bibfnamefont {A.}~\bibnamefont
  {{Gould}}},\ }\bibfield  {title} {\enquote {\bibinfo {title} {{A New Probe of
  Dark Matter and High-Energy Universe Using Microlensing}},}\ }\href {\doibase
  10.1088/0004-637X/707/2/979} {\bibfield  {journal} {\bibinfo  {journal} {The
  Astrophysical Journal}\ }\textbf {\bibinfo {volume} {707}},\ \bibinfo {pages}
  {979} (\bibinfo {year} {2009}{\natexlab{b}})},\ \Eprint
  {http://arxiv.org/abs/0908.0735}{arXiv:0908.0735}\BibitemShut {NoStop}%
\bibitem [{\citenamefont {{Scott}}\ and\ \citenamefont
  {{Sivertsson}}(2009)}]{scott09_UCMHprl}%
  \BibitemOpen
  \bibfield  {author} {\bibinfo {author} {\bibfnamefont {P.}~\bibnamefont
  {{Scott}}} and\ \bibinfo {author} {\bibfnamefont {S.}~\bibnamefont
  {{Sivertsson}}},\ }\bibfield  {title} {\enquote {\bibinfo {title} {{Gamma
  Rays from Ultracompact Primordial Dark Matter Minihalos}},}\ }\href {\doibase
  10.1103/PhysRevLett.103.211301} {\bibfield  {journal} {\bibinfo  {journal}
  {Physical Review Letters}\ }\textbf {\bibinfo {volume} {103}},\ \bibinfo
  {eid} {211301} (\bibinfo {year} {2009})},\ \Eprint
  {http://arxiv.org/abs/0908.4082}{arXiv:0908.4082}\BibitemShut {NoStop}%
\bibitem [{\citenamefont {{Yang}}\ \emph {{\it
  et~al.}}(2011{\natexlab{a}})\citenamefont {{Yang}}, \citenamefont {{Feng}},
  \citenamefont {{Huang}}, \citenamefont {{Chen}}, \citenamefont {{Lu}},\ and\
  \citenamefont {{Zong}}}]{yang2011}%
  \BibitemOpen
  \bibfield  {author} {\bibinfo {author} {\bibfnamefont {Y.}~\bibnamefont
  {{Yang}}}, \bibinfo {author} {\bibfnamefont {L.}~\bibnamefont {{Feng}}},
  \bibinfo {author} {\bibfnamefont {X.}~\bibnamefont {{Huang}}}, \bibinfo
  {author} {\bibfnamefont {X.}~\bibnamefont {{Chen}}}, \bibinfo {author}
  {\bibfnamefont {T.}~\bibnamefont {{Lu}}}, and\ \bibinfo {author}
  {\bibfnamefont {H.}~\bibnamefont {{Zong}}},\ }\bibfield  {title} {\enquote
  {\bibinfo {title} {{Constraints on ultracompact minihalos from extragalactic
  {\ensuremath{\gamma}}-ray background}},}\ }\href {\doibase
  10.1088/1475-7516/2011/12/020} {\bibfield  {journal} {\bibinfo  {journal}
  {Journal of Cosmology and Astroparticle Physics}\ }\textbf {\bibinfo {volume}
  {2011}},\ \bibinfo {eid} {020} (\bibinfo {year} {2011}{\natexlab{a}})},\
  \Eprint {http://arxiv.org/abs/1112.6229}{arXiv:1112.6229}\BibitemShut
  {NoStop}%
\bibitem [{\citenamefont {{Yang}}\ \emph {{\it
  et~al.}}(2011{\natexlab{b}})\citenamefont {{Yang}}, \citenamefont {{Huang}},
  \citenamefont {{Chen}},\ and\ \citenamefont {{Zong}}}]{yang11_ucmhCMB}%
  \BibitemOpen
  \bibfield  {author} {\bibinfo {author} {\bibfnamefont {Y.}~\bibnamefont
  {{Yang}}}, \bibinfo {author} {\bibfnamefont {X.}~\bibnamefont {{Huang}}},
  \bibinfo {author} {\bibfnamefont {X.}~\bibnamefont {{Chen}}}, and\ \bibinfo
  {author} {\bibfnamefont {H.}~\bibnamefont {{Zong}}},\ }\bibfield  {title}
  {\enquote {\bibinfo {title} {{New constraints on primordial minihalo
  abundance using cosmic microwave background observations}},}\ }\href
  {\doibase 10.1103/PhysRevD.84.043506} {\bibfield  {journal} {\bibinfo
  {journal} {\prd}\ }\textbf {\bibinfo {volume} {84}},\ \bibinfo {eid} {043506}
  (\bibinfo {year} {2011}{\natexlab{b}})},\ \Eprint
  {http://arxiv.org/abs/1109.0156}{arXiv:1109.0156}\BibitemShut {NoStop}%
\bibitem [{\citenamefont {{Yang}}(2020)}]{yang2020}%
  \BibitemOpen
  \bibfield  {author} {\bibinfo {author} {\bibfnamefont {Y.}~\bibnamefont
  {{Yang}}},\ }\bibfield  {title} {\enquote {\bibinfo {title} {{The abundance
  of primordial black holes from the global 21cm signal and extragalactic
  gamma-ray background}},}\ }\href {\doibase 10.1140/epjp/s13360-020-00710-3}
  {\bibfield  {journal} {\bibinfo  {journal} {European Physical Journal Plus}\
  }\textbf {\bibinfo {volume} {135}},\ \bibinfo {eid} {690} (\bibinfo {year}
  {2020})},\ \Eprint
  {http://arxiv.org/abs/2008.11859}{arXiv:2008.11859}\BibitemShut {NoStop}%
\bibitem [{\citenamefont {{Navarro}}\ \emph {{\it et~al.}}(1997)\citenamefont
  {{Navarro}}, \citenamefont {{Frenk}},\ and\ \citenamefont
  {{White}}}]{navarro97nfw}%
  \BibitemOpen
  \bibfield  {author} {\bibinfo {author} {\bibfnamefont {J.~F.}\ \bibnamefont
  {{Navarro}}}, \bibinfo {author} {\bibfnamefont {C.~S.}\ \bibnamefont
  {{Frenk}}}, and\ \bibinfo {author} {\bibfnamefont {S.~D.~M.}\ \bibnamefont
  {{White}}},\ }\bibfield  {title} {\enquote {\bibinfo {title} {{A Universal
  Density Profile from Hierarchical Clustering}},}\ }\href {\doibase
  10.1086/304888} {\bibfield  {journal} {\bibinfo  {journal} {Astrophys. J.}\
  }\textbf {\bibinfo {volume} {490}},\ \bibinfo {pages} {493} (\bibinfo {year}
  {1997})},\ \Eprint
  {http://arxiv.org/abs/astro-ph/9611107}{astro-ph/9611107}\BibitemShut
  {NoStop}%
\bibitem [{\citenamefont {{Springel}}\ \emph {{\it et~al.}}(2008)\citenamefont
  {{Springel}}, \citenamefont {{Wang}}, \citenamefont {{Vogelsberger}},
  \citenamefont {{Ludlow}}, \citenamefont {{Jenkins}}, \citenamefont {{Helmi}},
  \citenamefont {{Navarro}}, \citenamefont {{Frenk}},\ and\ \citenamefont
  {{White}}}]{einasto}%
  \BibitemOpen
  \bibfield  {author} {\bibinfo {author} {\bibfnamefont {V.}~\bibnamefont
  {{Springel}}}, \bibinfo {author} {\bibfnamefont {J.}~\bibnamefont {{Wang}}},
  \bibinfo {author} {\bibfnamefont {M.}~\bibnamefont {{Vogelsberger}}},
  \bibinfo {author} {\bibfnamefont {A.}~\bibnamefont {{Ludlow}}}, \bibinfo
  {author} {\bibfnamefont {A.}~\bibnamefont {{Jenkins}}}, \bibinfo {author}
  {\bibfnamefont {A.}~\bibnamefont {{Helmi}}}, \bibinfo {author} {\bibfnamefont
  {J.~F.}\ \bibnamefont {{Navarro}}}, \bibinfo {author} {\bibfnamefont {C.~S.}\
  \bibnamefont {{Frenk}}}, and\ \bibinfo {author} {\bibfnamefont {S.~D.~M.}\
  \bibnamefont {{White}}},\ }\bibfield  {title} {\enquote {\bibinfo {title}
  {{The Aquarius Project: the subhaloes of galactic haloes}},}\ }\href
  {\doibase 10.1111/j.1365-2966.2008.14066.x} {\bibfield  {journal} {\bibinfo
  {journal} {\mnras}\ }\textbf {\bibinfo {volume} {391}},\ \bibinfo {pages}
  {1685} (\bibinfo {year} {2008})},\ \Eprint
  {http://arxiv.org/abs/0809.0898}{arXiv:0809.0898}\BibitemShut {NoStop}%
\bibitem [{\citenamefont {{Josan}}\ and\ \citenamefont
  {{Green}}(2010)}]{Josan10ucmhgamma}%
  \BibitemOpen
  \bibfield  {author} {\bibinfo {author} {\bibfnamefont {A.~S.}\ \bibnamefont
  {{Josan}}} and\ \bibinfo {author} {\bibfnamefont {A.~M.}\ \bibnamefont
  {{Green}}},\ }\bibfield  {title} {\enquote {\bibinfo {title} {{Gamma rays
  from ultracompact minihalos: Potential constraints on the primordial
  curvature perturbation}},}\ }\href {\doibase 10.1103/PhysRevD.82.083527}
  {\bibfield  {journal} {\bibinfo  {journal} {\prd}\ }\textbf {\bibinfo
  {volume} {82}},\ \bibinfo {eid} {083527} (\bibinfo {year} {2010})},\ \Eprint
  {http://arxiv.org/abs/1006.4970}{arXiv:1006.4970}\BibitemShut {NoStop}%
\bibitem [{\citenamefont {{Bringmann}}\ \emph {{\it et~al.}}(2012)\citenamefont
  {{Bringmann}}, \citenamefont {{Scott}},\ and\ \citenamefont
  {{Akrami}}}]{Bringmann12ucmhPS}%
  \BibitemOpen
  \bibfield  {author} {\bibinfo {author} {\bibfnamefont {T.}~\bibnamefont
  {{Bringmann}}}, \bibinfo {author} {\bibfnamefont {P.}~\bibnamefont
  {{Scott}}}, and\ \bibinfo {author} {\bibfnamefont {Y.}~\bibnamefont
  {{Akrami}}},\ }\bibfield  {title} {\enquote {\bibinfo {title} {{Improved
  constraints on the primordial power spectrum at small scales from
  ultracompact minihalos}},}\ }\href {\doibase 10.1103/PhysRevD.85.125027}
  {\bibfield  {journal} {\bibinfo  {journal} {\prd}\ }\textbf {\bibinfo
  {volume} {85}},\ \bibinfo {eid} {125027} (\bibinfo {year} {2012})},\ \Eprint
  {http://arxiv.org/abs/1110.2484}{arXiv:1110.2484}\BibitemShut {NoStop}%
\bibitem [{\citenamefont {{Aslanyan}}\ \emph {{\it et~al.}}(2016)\citenamefont
  {{Aslanyan}}, \citenamefont {{Price}}, \citenamefont {{Adams}}, \citenamefont
  {{Bringmann}}, \citenamefont {{Clark}}, \citenamefont {{Easther}},
  \citenamefont {{Lewis}},\ and\ \citenamefont
  {{Scott}}}]{aslanyan16_InflationPRL}%
  \BibitemOpen
  \bibfield  {author} {\bibinfo {author} {\bibfnamefont {G.}~\bibnamefont
  {{Aslanyan}}}, \bibinfo {author} {\bibfnamefont {L.~C.}\ \bibnamefont
  {{Price}}}, \bibinfo {author} {\bibfnamefont {J.}~\bibnamefont {{Adams}}},
  \bibinfo {author} {\bibfnamefont {T.}~\bibnamefont {{Bringmann}}}, \bibinfo
  {author} {\bibfnamefont {H.~A.}\ \bibnamefont {{Clark}}}, \bibinfo {author}
  {\bibfnamefont {R.}~\bibnamefont {{Easther}}}, \bibinfo {author}
  {\bibfnamefont {G.~F.}\ \bibnamefont {{Lewis}}}, and\ \bibinfo {author}
  {\bibfnamefont {P.}~\bibnamefont {{Scott}}},\ }\bibfield  {title} {\enquote
  {\bibinfo {title} {{Ultracompact Minihalos as Probes of Inflationary
  Cosmology}},}\ }\href {\doibase 10.1103/PhysRevLett.117.141102} {\bibfield
  {journal} {\bibinfo  {journal} {\prl}\ }\textbf {\bibinfo {volume} {117}},\
  \bibinfo {eid} {141102} (\bibinfo {year} {2016})},\ \Eprint
  {http://arxiv.org/abs/1512.04597}{arXiv:1512.04597}\BibitemShut {NoStop}%
\bibitem [{\citenamefont {{Nakama}}\ \emph {{\it et~al.}}(2018)\citenamefont
  {{Nakama}}, \citenamefont {{Suyama}}, \citenamefont {{Kohri}},\ and\
  \citenamefont {{Hiroshima}}}]{Nakama18_igrbPS}%
  \BibitemOpen
  \bibfield  {author} {\bibinfo {author} {\bibfnamefont {T.}~\bibnamefont
  {{Nakama}}}, \bibinfo {author} {\bibfnamefont {T.}~\bibnamefont {{Suyama}}},
  \bibinfo {author} {\bibfnamefont {K.}~\bibnamefont {{Kohri}}}, and\ \bibinfo
  {author} {\bibfnamefont {N.}~\bibnamefont {{Hiroshima}}},\ }\bibfield
  {title} {\enquote {\bibinfo {title} {{Constraints on small-scale primordial
  power by annihilation signals from extragalactic dark matter minihalos}},}\
  }\href {\doibase 10.1103/PhysRevD.97.023539} {\bibfield  {journal} {\bibinfo
  {journal} {\prd}\ }\textbf {\bibinfo {volume} {97}},\ \bibinfo {eid} {023539}
  (\bibinfo {year} {2018})},\ \Eprint
  {http://arxiv.org/abs/1712.08820}{arXiv:1712.08820}\BibitemShut {NoStop}%
\bibitem [{\citenamefont {{Carr}}\ \emph {{\it et~al.}}(2010)\citenamefont
  {{Carr}}, \citenamefont {{Kohri}}, \citenamefont {{Sendouda}},\ and\
  \citenamefont {{Yokoyama}}}]{carr2010pbh}%
  \BibitemOpen
  \bibfield  {author} {\bibinfo {author} {\bibfnamefont {B.~J.}\ \bibnamefont
  {{Carr}}}, \bibinfo {author} {\bibfnamefont {K.}~\bibnamefont {{Kohri}}},
  \bibinfo {author} {\bibfnamefont {Y.}~\bibnamefont {{Sendouda}}}, and\
  \bibinfo {author} {\bibfnamefont {J.}~\bibnamefont {{Yokoyama}}},\ }\bibfield
   {title} {\enquote {\bibinfo {title} {{New cosmological constraints on
  primordial black holes}},}\ }\href {\doibase 10.1103/PhysRevD.81.104019}
  {\bibfield  {journal} {\bibinfo  {journal} {Physical Review D}\ }\textbf
  {\bibinfo {volume} {81}},\ \bibinfo {eid} {104019} (\bibinfo {year}
  {2010})},\ \Eprint
  {http://arxiv.org/abs/0912.5297}{arXiv:0912.5297}\BibitemShut {NoStop}%
\bibitem [{\citenamefont {{Delos}}\ \emph {{\it
  et~al.}}(2018{\natexlab{a}})\citenamefont {{Delos}}, \citenamefont
  {{Erickcek}}, \citenamefont {{Bailey}},\ and\ \citenamefont
  {{Alvarez}}}]{Delos18_noucmh}%
  \BibitemOpen
  \bibfield  {author} {\bibinfo {author} {\bibfnamefont {M.~S.}\ \bibnamefont
  {{Delos}}}, \bibinfo {author} {\bibfnamefont {A.~L.}\ \bibnamefont
  {{Erickcek}}}, \bibinfo {author} {\bibfnamefont {A.~P.}\ \bibnamefont
  {{Bailey}}}, and\ \bibinfo {author} {\bibfnamefont {M.~A.}\ \bibnamefont
  {{Alvarez}}},\ }\bibfield  {title} {\enquote {\bibinfo {title} {{Are
  ultracompact minihalos really ultracompact?}}}\ }\href {\doibase
  10.1103/PhysRevD.97.041303} {\bibfield  {journal} {\bibinfo  {journal}
  {\prd}\ }\textbf {\bibinfo {volume} {97}},\ \bibinfo {eid} {041303} (\bibinfo
  {year} {2018}{\natexlab{a}})},\ \Eprint
  {http://arxiv.org/abs/1712.05421}{arXiv:1712.05421}\BibitemShut {NoStop}%
\bibitem [{\citenamefont {{Delos}}\ \emph {{\it
  et~al.}}(2018{\natexlab{b}})\citenamefont {{Delos}}, \citenamefont
  {{Erickcek}}, \citenamefont {{Bailey}},\ and\ \citenamefont
  {{Alvarez}}}]{Delos18_cheng}%
  \BibitemOpen
  \bibfield  {author} {\bibinfo {author} {\bibfnamefont {M.~S.}\ \bibnamefont
  {{Delos}}}, \bibinfo {author} {\bibfnamefont {A.~L.}\ \bibnamefont
  {{Erickcek}}}, \bibinfo {author} {\bibfnamefont {A.~P.}\ \bibnamefont
  {{Bailey}}}, and\ \bibinfo {author} {\bibfnamefont {M.~A.}\ \bibnamefont
  {{Alvarez}}},\ }\bibfield  {title} {\enquote {\bibinfo {title} {{Density
  profiles of ultracompact minihalos: Implications for constraining the
  primordial power spectrum}},}\ }\href {\doibase 10.1103/PhysRevD.98.063527}
  {\bibfield  {journal} {\bibinfo  {journal} {\prd}\ }\textbf {\bibinfo
  {volume} {98}},\ \bibinfo {eid} {063527} (\bibinfo {year}
  {2018}{\natexlab{b}})},\ \Eprint
  {http://arxiv.org/abs/1806.07389}{arXiv:1806.07389}\BibitemShut {NoStop}%
\bibitem [{\citenamefont {{Adamek}}\ \emph {{\it et~al.}}(2019)\citenamefont
  {{Adamek}}, \citenamefont {{Byrnes}}, \citenamefont {{Gosenca}},\ and\
  \citenamefont {{Hotchkiss}}}]{adamek19_pbhucmh}%
  \BibitemOpen
  \bibfield  {author} {\bibinfo {author} {\bibfnamefont {J.}~\bibnamefont
  {{Adamek}}}, \bibinfo {author} {\bibfnamefont {C.~T.}\ \bibnamefont
  {{Byrnes}}}, \bibinfo {author} {\bibfnamefont {M.}~\bibnamefont {{Gosenca}}},
  and\ \bibinfo {author} {\bibfnamefont {S.}~\bibnamefont {{Hotchkiss}}},\
  }\bibfield  {title} {\enquote {\bibinfo {title} {{WIMPs and stellar-mass
  primordial black holes are incompatible}},}\ }\href {\doibase
  10.1103/PhysRevD.100.023506} {\bibfield  {journal} {\bibinfo  {journal}
  {\prd}\ }\textbf {\bibinfo {volume} {100}},\ \bibinfo {eid} {023506}
  (\bibinfo {year} {2019})},\ \Eprint
  {http://arxiv.org/abs/1901.08528}{arXiv:1901.08528}\BibitemShut {NoStop}%
\bibitem [{\citenamefont {{Belikov}}\ and\ \citenamefont
  {{Silk}}(2014)}]{Belikov14spikeIGRB}%
  \BibitemOpen
  \bibfield  {author} {\bibinfo {author} {\bibfnamefont {A.}~\bibnamefont
  {{Belikov}}} and\ \bibinfo {author} {\bibfnamefont {J.}~\bibnamefont
  {{Silk}}},\ }\bibfield  {title} {\enquote {\bibinfo {title} {{Diffuse gamma
  ray background from annihilating dark matter in density spikes around
  supermassive black holes}},}\ }\href {\doibase 10.1103/PhysRevD.89.043520}
  {\bibfield  {journal} {\bibinfo  {journal} {\prd}\ }\textbf {\bibinfo
  {volume} {89}},\ \bibinfo {eid} {043520} (\bibinfo {year} {2014})},\ \Eprint
  {http://arxiv.org/abs/1312.0007}{arXiv:1312.0007}\BibitemShut {NoStop}%
\bibitem [{\citenamefont {{Lacroix}}\ and\ \citenamefont
  {{Silk}}(2018)}]{Lacroix18bh}%
  \BibitemOpen
  \bibfield  {author} {\bibinfo {author} {\bibfnamefont {T.}~\bibnamefont
  {{Lacroix}}} and\ \bibinfo {author} {\bibfnamefont {J.}~\bibnamefont
  {{Silk}}},\ }\bibfield  {title} {\enquote {\bibinfo {title}
  {{Intermediate-mass Black Holes and Dark Matter at the Galactic Center}},}\
  }\href {\doibase 10.3847/2041-8213/aaa775} {\bibfield  {journal} {\bibinfo
  {journal} {\apjl}\ }\textbf {\bibinfo {volume} {853}},\ \bibinfo {eid} {L16}
  (\bibinfo {year} {2018})},\ \Eprint
  {http://arxiv.org/abs/1712.00452}{arXiv:1712.00452}\BibitemShut {NoStop}%
\bibitem [{\citenamefont {{Cheng}}\ \emph {{\it et~al.}}(2020)\citenamefont
  {{Cheng}}, \citenamefont {{Li}}, \citenamefont {{Gan}}, \citenamefont
  {{Liang}}, \citenamefont {{Lu}},\ and\ \citenamefont {{Liang}}}]{cheng20}%
  \BibitemOpen
  \bibfield  {author} {\bibinfo {author} {\bibfnamefont {J.-G.}\ \bibnamefont
  {{Cheng}}}, \bibinfo {author} {\bibfnamefont {S.}~\bibnamefont {{Li}}},
  \bibinfo {author} {\bibfnamefont {Y.-Y.}\ \bibnamefont {{Gan}}}, \bibinfo
  {author} {\bibfnamefont {Y.-F.}\ \bibnamefont {{Liang}}}, \bibinfo {author}
  {\bibfnamefont {R.-J.}\ \bibnamefont {{Lu}}}, and\ \bibinfo {author}
  {\bibfnamefont {E.-W.}\ \bibnamefont {{Liang}}},\ }\bibfield  {title}
  {\enquote {\bibinfo {title} {{On the gamma-ray signals from UCMH/mini-spike
  accompanying the DAMPE 1.4 TeV e$^{+}$e$^{-}$ excess}},}\ }\href {\doibase
  10.1093/mnras/staa2092} {\bibfield  {journal} {\bibinfo  {journal} {Monthly
  Notices of the Royal Astronomical Society}\ }\textbf {\bibinfo {volume}
  {497}},\ \bibinfo {pages} {2486} (\bibinfo {year} {2020})}\BibitemShut
  {NoStop}%
\bibitem [{\citenamefont {{Xia}}\ \emph {{\it et~al.}}(2021)\citenamefont
  {{Xia}}, \citenamefont {{Shen}}, \citenamefont {{Pan}}, \citenamefont
  {{Feng}},\ and\ \citenamefont {{Fan}}}]{xzq2021}%
  \BibitemOpen
  \bibfield  {author} {\bibinfo {author} {\bibfnamefont {Z.-Q.}\ \bibnamefont
  {{Xia}}}, \bibinfo {author} {\bibfnamefont {Z.-Q.}\ \bibnamefont {{Shen}}},
  \bibinfo {author} {\bibfnamefont {X.}~\bibnamefont {{Pan}}}, \bibinfo
  {author} {\bibfnamefont {L.}~\bibnamefont {{Feng}}}, and\ \bibinfo {author}
  {\bibfnamefont {Y.-Z.}\ \bibnamefont {{Fan}}},\ }\bibfield  {title} {\enquote
  {\bibinfo {title} {{Investigating the dark matter minispikes with the
  gamma-ray signal from the halo of M31}},}\ }\href@noop {} {\bibfield
  {journal} {\bibinfo  {journal} {arXiv e-prints}\ ,\ \bibinfo {eid}
  {arXiv:2108.09204}} (\bibinfo {year} {2021})},\ \Eprint
  {http://arxiv.org/abs/2108.09204}{arXiv:2108.09204}\BibitemShut {NoStop}%
\bibitem [{\citenamefont {{DAMPE Collaboration}}\ \emph {{\it
  et~al.}}(2017)\citenamefont {{DAMPE Collaboration}}, \citenamefont
  {{Ambrosi}}, \citenamefont {{An}}, \citenamefont {{Asfandiyarov}},
  \citenamefont {{Azzarello}}, \citenamefont {{Bernardini}}, \citenamefont
  {{Bertucci}}, \citenamefont {{Cai}}, \citenamefont {{Chang}}, \citenamefont
  {{Chen}}, \citenamefont {{Chen}}, \citenamefont {{Chen}}, \citenamefont
  {{Chen}}, \citenamefont {{Cui}}, \citenamefont {{Cui}}, \citenamefont
  {{D'Amone}}, \citenamefont {{de Benedittis}}, \citenamefont {{De Mitri}},
  \citenamefont {{di Santo}}, \citenamefont {{Dong}}, \citenamefont {{Dong}},
  \citenamefont {{Dong}}, \citenamefont {{Dong}}, \citenamefont {{Donvito}},
  \citenamefont {{Droz}}, \citenamefont {{Duan}}, \citenamefont {{Duan}},
  \citenamefont {{Duranti}}, \citenamefont {{D'Urso}}, \citenamefont {{Fan}},
  \citenamefont {{Fan}}, \citenamefont {{Fang}}, \citenamefont {{Feng}},
  \citenamefont {{Feng}}, \citenamefont {{Fusco}}, \citenamefont {{Gallo}},
  \citenamefont {{Gan}}, \citenamefont {{Gao}}, \citenamefont {{Gao}},
  \citenamefont {{Gargano}}, \citenamefont {{Garrappa}}, \citenamefont
  {{Gong}}, \citenamefont {{Gong}}, \citenamefont {{Guo}}, \citenamefont
  {{Guo}}, \citenamefont {{Hu}}, \citenamefont {{Huang}}, \citenamefont
  {{Huang}}, \citenamefont {{Ionica}}, \citenamefont {{Jiang}}, \citenamefont
  {{Jiang}}, \citenamefont {{Jin}}, \citenamefont {{Kong}}, \citenamefont
  {{Lei}}, \citenamefont {{Li}}, \citenamefont {{Li}}, \citenamefont {{Li}},
  \citenamefont {{Li}}, \citenamefont {{Liang}}, \citenamefont {{Liang}},
  \citenamefont {{Liao}}, \citenamefont {{Liu}}, \citenamefont {{Liu}},
  \citenamefont {{Liu}}, \citenamefont {{Liu}}, \citenamefont {{Liu}},
  \citenamefont {{Loparco}}, \citenamefont {{Ma}}, \citenamefont {{Ma}},
  \citenamefont {{Ma}}, \citenamefont {{Ma}}, \citenamefont {{Ma}},
  \citenamefont {{Ma}}, \citenamefont {{Marsella}}, \citenamefont
  {{Mazziotta}}, \citenamefont {{Mo}}, \citenamefont {{Niu}}, \citenamefont
  {{Peng}}, \citenamefont {{Peng}}, \citenamefont {{Qiao}}, \citenamefont
  {{Rao}}, \citenamefont {{Salinas}}, \citenamefont {{Shang}}, \citenamefont
  {{H. Shen}}, \citenamefont {{Shen}}, \citenamefont {{Shen}}, \citenamefont
  {{Song}}, \citenamefont {{Su}}, \citenamefont {{Su}}, \citenamefont {{Sun}},
  \citenamefont {{Surdo}}, \citenamefont {{Teng}}, \citenamefont {{Tian}},
  \citenamefont {{Tykhonov}}, \citenamefont {{Vagelli}}, \citenamefont
  {{Vitillo}}, \citenamefont {{Wang}}, \citenamefont {{Wang}}, \citenamefont
  {{Wang}}, \citenamefont {{Wang}}, \citenamefont {{Wang}}, \citenamefont
  {{Wang}}, \citenamefont {{Wang}}, \citenamefont {{Wang}}, \citenamefont
  {{Wang}}, \citenamefont {{Wang}}, \citenamefont {{Wang}}, \citenamefont
  {{Wang}}, \citenamefont {{Wen}}, \citenamefont {{Wang}}, \citenamefont
  {{Wei}}, \citenamefont {{Wei}}, \citenamefont {{Wei}}, \citenamefont {{Wu}},
  \citenamefont {{Wu}}, \citenamefont {{Wu}}, \citenamefont {{Wu}},
  \citenamefont {{Wu}}, \citenamefont {{Xi}}, \citenamefont {{Xia}},
  \citenamefont {{Xin}}, \citenamefont {{Xu}}, \citenamefont {{Xu}},
  \citenamefont {{Xu}}, \citenamefont {{Xue}}, \citenamefont {{Yang}},
  \citenamefont {{Yang}}, \citenamefont {{Yang}}, \citenamefont {{Yang}},
  \citenamefont {{Yao}}, \citenamefont {{Yu}}, \citenamefont {{Yuan}},
  \citenamefont {{Yue}}, \citenamefont {{Zang}}, \citenamefont {{Zhang}},
  \citenamefont {{Zhang}}, \citenamefont {{Zhang}}, \citenamefont {{Zhang}},
  \citenamefont {{Zhang}}, \citenamefont {{Zhang}}, \citenamefont {{Zhang}},
  \citenamefont {{Zhang}}, \citenamefont {{Zhang}}, \citenamefont {{Zhang}},
  \citenamefont {{Zhang}}, \citenamefont {{Zhang}}, \citenamefont {{Zhang}},
  \citenamefont {{Zhang}}, \citenamefont {{Zhang}}, \citenamefont {{Zhang}},
  \citenamefont {{Zhang}}, \citenamefont {{Zhao}}, \citenamefont {{Zhao}},
  \citenamefont {{Zhao}}, \citenamefont {{Zhou}}, \citenamefont {{Zhou}},
  \citenamefont {{Zhu}}, \citenamefont {{Zhu}},\ and\ \citenamefont
  {{Zimmer}}}]{dampe17_nature}%
  \BibitemOpen
  \bibfield  {author} {\bibinfo {author} {\bibnamefont {{DAMPE Collaboration}}}
  {\it et~al.},\ }\bibfield  {title} {\enquote {\bibinfo {title} {{Direct
  detection of a break in the teraelectronvolt cosmic-ray spectrum of electrons
  and positrons}},}\ }\href {\doibase 10.1038/nature24475} {\bibfield
  {journal} {\bibinfo  {journal} {Nature}\ }\textbf {\bibinfo {volume} {552}},\
  \bibinfo {pages} {63} (\bibinfo {year} {2017})},\ \Eprint
  {http://arxiv.org/abs/1711.10981}{arXiv:1711.10981}\BibitemShut {NoStop}%
\bibitem [{\citenamefont {{Huang}}\ \emph {{\it et~al.}}(2018)\citenamefont
  {{Huang}}, \citenamefont {{Wu}}, \citenamefont {{Zhang}},\ and\ \citenamefont
  {{Zhou}}}]{huang18_dampe}%
  \BibitemOpen
  \bibfield  {author} {\bibinfo {author} {\bibfnamefont {X.-J.}\ \bibnamefont
  {{Huang}}}, \bibinfo {author} {\bibfnamefont {Y.-L.}\ \bibnamefont {{Wu}}},
  \bibinfo {author} {\bibfnamefont {W.-H.}\ \bibnamefont {{Zhang}}}, and\
  \bibinfo {author} {\bibfnamefont {Y.-F.}\ \bibnamefont {{Zhou}}},\ }\bibfield
   {title} {\enquote {\bibinfo {title} {{Origins of sharp cosmic-ray electron
  structures and the DAMPE excess}},}\ }\href {\doibase
  10.1103/PhysRevD.97.091701} {\bibfield  {journal} {\bibinfo  {journal}
  {Physical Review D}\ }\textbf {\bibinfo {volume} {97}},\ \bibinfo {eid}
  {091701} (\bibinfo {year} {2018})},\ \Eprint
  {http://arxiv.org/abs/1712.00005}{arXiv:1712.00005}\BibitemShut {NoStop}%
\bibitem [{\citenamefont {{Zhao}}\ \emph {{\it et~al.}}(2019)\citenamefont
  {{Zhao}}, \citenamefont {{Bi}}, \citenamefont {{Lin}},\ and\ \citenamefont
  {{Yin}}}]{zhaoyi2019}%
  \BibitemOpen
  \bibfield  {author} {\bibinfo {author} {\bibfnamefont {Y.}~\bibnamefont
  {{Zhao}}}, \bibinfo {author} {\bibfnamefont {X.-J.}\ \bibnamefont {{Bi}}},
  \bibinfo {author} {\bibfnamefont {S.-J.}\ \bibnamefont {{Lin}}}, and\
  \bibinfo {author} {\bibfnamefont {P.-F.}\ \bibnamefont {{Yin}}},\ }\bibfield
  {title} {\enquote {\bibinfo {title} {{Nearby dark matter subhalo that
  accounts for the DAMPE excess}},}\ }\href {\doibase
  10.1088/1674-1137/43/8/085101} {\bibfield  {journal} {\bibinfo  {journal}
  {Chinese Physics C}\ }\textbf {\bibinfo {volume} {43}},\ \bibinfo {eid}
  {085101} (\bibinfo {year} {2019})}\BibitemShut {NoStop}%
\bibitem [{\citenamefont {{Fillmore}}\ and\ \citenamefont
  {{Goldreich}}(1984)}]{1984ApJ...281....1F}%
  \BibitemOpen
  \bibfield  {author} {\bibinfo {author} {\bibfnamefont {J.~A.}\ \bibnamefont
  {{Fillmore}}} and\ \bibinfo {author} {\bibfnamefont {P.}~\bibnamefont
  {{Goldreich}}},\ }\bibfield  {title} {\enquote {\bibinfo {title}
  {{Self-similar gravitational collapse in an expanding universe}},}\ }\href
  {\doibase 10.1086/162070} {\bibfield  {journal} {\bibinfo  {journal}
  {Astrophys. J.}\ }\textbf {\bibinfo {volume} {281}},\ \bibinfo {pages} {1}
  (\bibinfo {year} {1984})}\BibitemShut {NoStop}%
\bibitem [{\citenamefont {{Bertschinger}}(1985)}]{1985ApJS...58...39B}%
  \BibitemOpen
  \bibfield  {author} {\bibinfo {author} {\bibfnamefont {E.}~\bibnamefont
  {{Bertschinger}}},\ }\bibfield  {title} {\enquote {\bibinfo {title}
  {{Self-similar secondary infall and accretion in an Einstein-de Sitter
  universe}},}\ }\href {\doibase 10.1086/191028} {\bibfield  {journal}
  {\bibinfo  {journal} {The Astrophysical Journal Supplement Series}\ }\textbf
  {\bibinfo {volume} {58}},\ \bibinfo {pages} {39} (\bibinfo {year}
  {1985})}\BibitemShut {NoStop}%
\bibitem [{\citenamefont {{Vogelsberger}}\ \emph {{\it
  et~al.}}(2009)\citenamefont {{Vogelsberger}}, \citenamefont {{White}},
  \citenamefont {{Mohayaee}},\ and\ \citenamefont
  {{Springel}}}]{vogelsberger09ucmhsim}%
  \BibitemOpen
  \bibfield  {author} {\bibinfo {author} {\bibfnamefont {M.}~\bibnamefont
  {{Vogelsberger}}}, \bibinfo {author} {\bibfnamefont {S.~D.~M.}\ \bibnamefont
  {{White}}}, \bibinfo {author} {\bibfnamefont {R.}~\bibnamefont {{Mohayaee}}},
  and\ \bibinfo {author} {\bibfnamefont {V.}~\bibnamefont {{Springel}}},\
  }\bibfield  {title} {\enquote {\bibinfo {title} {{Caustics in growing cold
  dark matter haloes}},}\ }\href {\doibase 10.1111/j.1365-2966.2009.15615.x}
  {\bibfield  {journal} {\bibinfo  {journal} {\mnras}\ }\textbf {\bibinfo
  {volume} {400}},\ \bibinfo {pages} {2174} (\bibinfo {year} {2009})},\ \Eprint
  {http://arxiv.org/abs/0906.4341}{arXiv:0906.4341}\BibitemShut {NoStop}%
\bibitem [{\citenamefont {{Ludlow}}\ \emph {{\it et~al.}}(2010)\citenamefont
  {{Ludlow}}, \citenamefont {{Navarro}}, \citenamefont {{Springel}},
  \citenamefont {{Vogelsberger}}, \citenamefont {{Wang}}, \citenamefont
  {{White}}, \citenamefont {{Jenkins}},\ and\ \citenamefont
  {{Frenk}}}]{ludlow10ucmhsim}%
  \BibitemOpen
  \bibfield  {author} {\bibinfo {author} {\bibfnamefont {A.~D.}\ \bibnamefont
  {{Ludlow}}}, \bibinfo {author} {\bibfnamefont {J.~F.}\ \bibnamefont
  {{Navarro}}}, \bibinfo {author} {\bibfnamefont {V.}~\bibnamefont
  {{Springel}}}, \bibinfo {author} {\bibfnamefont {M.}~\bibnamefont
  {{Vogelsberger}}}, \bibinfo {author} {\bibfnamefont {J.}~\bibnamefont
  {{Wang}}}, \bibinfo {author} {\bibfnamefont {S.~D.~M.}\ \bibnamefont
  {{White}}}, \bibinfo {author} {\bibfnamefont {A.}~\bibnamefont {{Jenkins}}},
  and\ \bibinfo {author} {\bibfnamefont {C.~S.}\ \bibnamefont {{Frenk}}},\
  }\bibfield  {title} {\enquote {\bibinfo {title} {{Secondary infall and the
  pseudo-phase-space density profiles of cold dark matter haloes}},}\ }\href
  {\doibase 10.1111/j.1365-2966.2010.16678.x} {\bibfield  {journal} {\bibinfo
  {journal} {\mnras}\ }\textbf {\bibinfo {volume} {406}},\ \bibinfo {pages}
  {137} (\bibinfo {year} {2010})},\ \Eprint
  {http://arxiv.org/abs/1001.2310}{arXiv:1001.2310}\BibitemShut {NoStop}%
\bibitem [{\citenamefont {{Ullio}}\ \emph {{\it et~al.}}(2002)\citenamefont
  {{Ullio}}, \citenamefont {{Bergstr{\"o}m}}, \citenamefont {{Edsj{\"o}}},\
  and\ \citenamefont {{Lacey}}}]{Ullio02dm}%
  \BibitemOpen
  \bibfield  {author} {\bibinfo {author} {\bibfnamefont {P.}~\bibnamefont
  {{Ullio}}}, \bibinfo {author} {\bibfnamefont {L.}~\bibnamefont
  {{Bergstr{\"o}m}}}, \bibinfo {author} {\bibfnamefont {J.}~\bibnamefont
  {{Edsj{\"o}}}}, and\ \bibinfo {author} {\bibfnamefont {C.}~\bibnamefont
  {{Lacey}}},\ }\bibfield  {title} {\enquote {\bibinfo {title} {{Cosmological
  dark matter annihilations into {\ensuremath{\gamma}} rays: A closer look}},}\
  }\href {\doibase 10.1103/PhysRevD.66.123502} {\bibfield  {journal} {\bibinfo
  {journal} {\prd}\ }\textbf {\bibinfo {volume} {66}},\ \bibinfo {eid} {123502}
  (\bibinfo {year} {2002})},\ \Eprint
  {http://arxiv.org/abs/astro-ph/0207125}{arXiv:astro-ph/0207125}\BibitemShut
  {NoStop}%
\bibitem [{\citenamefont {{Cirelli}}\ \emph {{\it et~al.}}(2012)\citenamefont
  {{Cirelli}}, \citenamefont {{Corcella}}, \citenamefont {{Hektor}},
  \citenamefont {{H{\"u}tsi}}, \citenamefont {{Kadastik}}, \citenamefont
  {{Panci}}, \citenamefont {{Raidal}}, \citenamefont {{Sala}},\ and\
  \citenamefont {{Strumia}}}]{pppc4}%
  \BibitemOpen
  \bibfield  {author} {\bibinfo {author} {\bibfnamefont {M.}~\bibnamefont
  {{Cirelli}}}, \bibinfo {author} {\bibfnamefont {G.}~\bibnamefont
  {{Corcella}}}, \bibinfo {author} {\bibfnamefont {A.}~\bibnamefont
  {{Hektor}}}, \bibinfo {author} {\bibfnamefont {G.}~\bibnamefont
  {{H{\"u}tsi}}}, \bibinfo {author} {\bibfnamefont {M.}~\bibnamefont
  {{Kadastik}}}, \bibinfo {author} {\bibfnamefont {P.}~\bibnamefont {{Panci}}},
  \bibinfo {author} {\bibfnamefont {M.}~\bibnamefont {{Raidal}}}, \bibinfo
  {author} {\bibfnamefont {F.}~\bibnamefont {{Sala}}}, and\ \bibinfo {author}
  {\bibfnamefont {A.}~\bibnamefont {{Strumia}}},\ }\bibfield  {title} {\enquote
  {\bibinfo {title} {{Erratum: PPPC 4 DM ID: a poor particle physicist cookbook
  for dark matter indirect detection Erratum: PPPC 4 DM ID: a poor particle
  physicist cookbook for dark matter indirect detection}},}\ }\href {\doibase
  10.1088/1475-7516/2012/10/E01} {\bibfield  {journal} {\bibinfo  {journal}
  {Journal of Cosmology and Astroparticle Physics}\ }\textbf {\bibinfo {volume}
  {2012}},\ \bibinfo {eid} {E01} (\bibinfo {year} {2012})}\BibitemShut
  {NoStop}%
\bibitem [{\citenamefont {{Bergstr{\"o}m}}\ \emph {{\it
  et~al.}}(2001)\citenamefont {{Bergstr{\"o}m}}, \citenamefont {{Edsj{\"o}}},\
  and\ \citenamefont {{Ullio}}}]{bergstrom01_od}%
  \BibitemOpen
  \bibfield  {author} {\bibinfo {author} {\bibfnamefont {L.}~\bibnamefont
  {{Bergstr{\"o}m}}}, \bibinfo {author} {\bibfnamefont {J.}~\bibnamefont
  {{Edsj{\"o}}}}, and\ \bibinfo {author} {\bibfnamefont {P.}~\bibnamefont
  {{Ullio}}},\ }\bibfield  {title} {\enquote {\bibinfo {title} {{Spectral
  Gamma-Ray Signatures of Cosmological Dark Matter Annihilations}},}\ }\href
  {\doibase 10.1103/PhysRevLett.87.251301} {\bibfield  {journal} {\bibinfo
  {journal} {\prl}\ }\textbf {\bibinfo {volume} {87}},\ \bibinfo {eid} {251301}
  (\bibinfo {year} {2001})},\ \Eprint
  {http://arxiv.org/abs/astro-ph/0105048}{arXiv:astro-ph/0105048}\BibitemShut
  {NoStop}%
\bibitem [{\citenamefont {{Steigman}}\ \emph {{\it et~al.}}(2012)\citenamefont
  {{Steigman}}, \citenamefont {{Dasgupta}},\ and\ \citenamefont
  {{Beacom}}}]{Steigman12thermalcs}%
  \BibitemOpen
  \bibfield  {author} {\bibinfo {author} {\bibfnamefont {G.}~\bibnamefont
  {{Steigman}}}, \bibinfo {author} {\bibfnamefont {B.}~\bibnamefont
  {{Dasgupta}}}, and\ \bibinfo {author} {\bibfnamefont {J.~F.}\ \bibnamefont
  {{Beacom}}},\ }\bibfield  {title} {\enquote {\bibinfo {title} {{Precise relic
  WIMP abundance and its impact on searches for dark matter annihilation}},}\
  }\href {\doibase 10.1103/PhysRevD.86.023506} {\bibfield  {journal} {\bibinfo
  {journal} {\prd}\ }\textbf {\bibinfo {volume} {86}},\ \bibinfo {eid} {023506}
  (\bibinfo {year} {2012})},\ \Eprint
  {http://arxiv.org/abs/1204.3622}{arXiv:1204.3622}\BibitemShut {NoStop}%
\bibitem [{\citenamefont {{Dom{\'\i}nguez}}\ \emph {{\it
  et~al.}}(2011)\citenamefont {{Dom{\'\i}nguez}}, \citenamefont {{Primack}},
  \citenamefont {{Rosario}}, \citenamefont {{Prada}}, \citenamefont
  {{Gilmore}}, \citenamefont {{Faber}}, \citenamefont {{Koo}}, \citenamefont
  {{Somerville}}, \citenamefont {{P{\'e}rez-Torres}}, \citenamefont
  {{P{\'e}rez-Gonz{\'a}lez}}, \citenamefont {{Huang}}, \citenamefont {{Davis}},
  \citenamefont {{Guhathakurta}}, \citenamefont {{Barmby}}, \citenamefont
  {{Conselice}}, \citenamefont {{Lozano}}, \citenamefont {{Newman}},\ and\
  \citenamefont {{Cooper}}}]{DominguezA2011}%
  \BibitemOpen
  \bibfield  {author} {\bibinfo {author} {\bibfnamefont {A.}~\bibnamefont
  {{Dom{\'\i}nguez}}} {\it et~al.},\ }\bibfield  {title} {\enquote {\bibinfo
  {title} {{Extragalactic background light inferred from AEGIS galaxy-SED-type
  fractions}},}\ }\href {\doibase 10.1111/j.1365-2966.2010.17631.x} {\bibfield
  {journal} {\bibinfo  {journal} {\mnras}\ }\textbf {\bibinfo {volume} {410}},\
  \bibinfo {pages} {2556} (\bibinfo {year} {2011})},\ \Eprint
  {http://arxiv.org/abs/1007.1459}{arXiv:1007.1459}\BibitemShut {NoStop}%
\bibitem [{\citenamefont {{Inoue}}\ \emph {{\it et~al.}}(2013)\citenamefont
  {{Inoue}}, \citenamefont {{Inoue}}, \citenamefont {{Kobayashi}},
  \citenamefont {{Makiya}}, \citenamefont {{Niino}},\ and\ \citenamefont
  {{Totani}}}]{Inoue13_ebl}%
  \BibitemOpen
  \bibfield  {author} {\bibinfo {author} {\bibfnamefont {Y.}~\bibnamefont
  {{Inoue}}}, \bibinfo {author} {\bibfnamefont {S.}~\bibnamefont {{Inoue}}},
  \bibinfo {author} {\bibfnamefont {M.~A.~R.}\ \bibnamefont {{Kobayashi}}},
  \bibinfo {author} {\bibfnamefont {R.}~\bibnamefont {{Makiya}}}, \bibinfo
  {author} {\bibfnamefont {Y.}~\bibnamefont {{Niino}}}, and\ \bibinfo {author}
  {\bibfnamefont {T.}~\bibnamefont {{Totani}}},\ }\bibfield  {title} {\enquote
  {\bibinfo {title} {{Extragalactic Background Light from Hierarchical Galaxy
  Formation: Gamma-Ray Attenuation up to the Epoch of Cosmic Reionization and
  the First Stars}},}\ }\href {\doibase 10.1088/0004-637X/768/2/197} {\bibfield
   {journal} {\bibinfo  {journal} {Astrophys. J.}\ }\textbf {\bibinfo {volume}
  {768}},\ \bibinfo {eid} {197} (\bibinfo {year} {2013})},\ \Eprint
  {http://arxiv.org/abs/1212.1683}{arXiv:1212.1683}\BibitemShut {NoStop}%
\bibitem [{\citenamefont {Gruppioni}\ \emph {{\it et~al.}}(2013)\citenamefont
  {Gruppioni} {\it et~al.}}]{Gruppioni:2013jna}%
  \BibitemOpen
  \bibfield  {author} {\bibinfo {author} {\bibfnamefont {C.}~\bibnamefont
  {Gruppioni}}  {\it et~al.},\ }\bibfield  {title} {\enquote {\bibinfo {title}
  {{The Herschel PEP/HerMES Luminosity Function. I: Probing the Evolution of
  PACS selected Galaxies to z\textasciitilde{}4}},}\ }\href {\doibase
  10.1093/mnras/stt308} {\bibfield  {journal} {\bibinfo  {journal} {Mon. Not.
  Roy. Astron. Soc.}\ }\textbf {\bibinfo {volume} {432}},\ \bibinfo {pages}
  {23} (\bibinfo {year} {2013})},\ \Eprint
  {http://arxiv.org/abs/1302.5209}{arXiv:1302.5209}\BibitemShut {NoStop}%
\bibitem [{\citenamefont {Chernoff}(1954)}]{Chernoff1954}%
  \BibitemOpen
  \bibfield  {author} {\bibinfo {author} {\bibfnamefont {H.}~\bibnamefont
  {Chernoff}},\ }\bibfield  {title} {\enquote {\bibinfo {title} {{On the
  Distribution of the Likelihood Ratio}},}\ }\href {\doibase
  10.1214/aoms/1177728725} {\bibfield  {journal} {\bibinfo  {journal} {The
  Annals of Mathematical Statistics}\ }\textbf {\bibinfo {volume} {25}},\
  \bibinfo {pages} {573 } (\bibinfo {year} {1954})}\BibitemShut {NoStop}%
\bibitem [{\citenamefont {{Yuan}}\ \emph {{\it et~al.}}(2017)\citenamefont
  {{Yuan}}, \citenamefont {{Feng}}, \citenamefont {{Yin}}, \citenamefont
  {{Fan}}, \citenamefont {{Bi}}, \citenamefont {{Cui}}, \citenamefont {{Dong}},
  \citenamefont {{Guo}}, \citenamefont {{Fang}}, \citenamefont {{Hu}},
  \citenamefont {{Huang}}, \citenamefont {{Lei}}, \citenamefont {{Li}},
  \citenamefont {{Lin}}, \citenamefont {{Liu}}, \citenamefont {{Ma}},
  \citenamefont {{Peng}}, \citenamefont {{Qiao}}, \citenamefont {{Shen}},
  \citenamefont {{Su}}, \citenamefont {{Wei}}, \citenamefont {{Xu}},
  \citenamefont {{Yue}}, \citenamefont {{Zang}}, \citenamefont {{Zhang}},
  \citenamefont {{Zhang}}, \citenamefont {{Zhang}}, \citenamefont {{Zhang}},\
  and\ \citenamefont {{Zhang}}}]{yuan2017}%
  \BibitemOpen
  \bibfield  {author} {\bibinfo {author} {\bibfnamefont {Q.}~\bibnamefont
  {{Yuan}}} {\it et~al.},\ }\bibfield  {title} {\enquote {\bibinfo {title}
  {{Interpretations of the DAMPE electron data}},}\ }\href@noop {} {\bibfield
  {journal} {\bibinfo  {journal} {arXiv e-prints}\ ,\ \bibinfo {eid}
  {arXiv:1711.10989}} (\bibinfo {year} {2017})},\ \Eprint
  {http://arxiv.org/abs/1711.10989}{arXiv:1711.10989}\BibitemShut {NoStop}%
\bibitem [{\citenamefont {{Cao}}\ \emph {{\it et~al.}}(2018)\citenamefont
  {{Cao}}, \citenamefont {{Feng}}, \citenamefont {{Guo}}, \citenamefont
  {{Shang}}, \citenamefont {{Wang}},\ and\ \citenamefont
  {{Wu}}}]{caojj18_dampe}%
  \BibitemOpen
  \bibfield  {author} {\bibinfo {author} {\bibfnamefont {J.}~\bibnamefont
  {{Cao}}}, \bibinfo {author} {\bibfnamefont {L.}~\bibnamefont {{Feng}}},
  \bibinfo {author} {\bibfnamefont {X.}~\bibnamefont {{Guo}}}, \bibinfo
  {author} {\bibfnamefont {L.}~\bibnamefont {{Shang}}}, \bibinfo {author}
  {\bibfnamefont {F.}~\bibnamefont {{Wang}}}, and\ \bibinfo {author}
  {\bibfnamefont {P.}~\bibnamefont {{Wu}}},\ }\bibfield  {title} {\enquote
  {\bibinfo {title} {{Scalar dark matter interpretation of the DAMPE data with
  U(1) gauge interactions}},}\ }\href {\doibase 10.1103/PhysRevD.97.095011}
  {\bibfield  {journal} {\bibinfo  {journal} {Physical Review D}\ }\textbf
  {\bibinfo {volume} {97}},\ \bibinfo {eid} {095011} (\bibinfo {year}
  {2018})},\ \Eprint
  {http://arxiv.org/abs/1711.11452}{arXiv:1711.11452}\BibitemShut {NoStop}%
\bibitem [{\citenamefont {{Pan}}\ \emph {{\it et~al.}}(2018)\citenamefont
  {{Pan}}, \citenamefont {{Zhang}},\ and\ \citenamefont {{Feng}}}]{panxu18}%
  \BibitemOpen
  \bibfield  {author} {\bibinfo {author} {\bibfnamefont {X.}~\bibnamefont
  {{Pan}}}, \bibinfo {author} {\bibfnamefont {C.}~\bibnamefont {{Zhang}}}, and\
  \bibinfo {author} {\bibfnamefont {L.}~\bibnamefont {{Feng}}},\ }\bibfield
  {title} {\enquote {\bibinfo {title} {{Interpretation of the DAMPE 1.4 TeV
  peak according to the decaying dark matter model}},}\ }\href {\doibase
  10.1007/s11433-018-9257-3} {\bibfield  {journal} {\bibinfo  {journal}
  {Science China Physics, Mechanics, and Astronomy}\ }\textbf {\bibinfo
  {volume} {61}},\ \bibinfo {eid} {101006} (\bibinfo {year}
  {2018})}\BibitemShut {NoStop}%
\bibitem [{\citenamefont {{Ghosh}}\ \emph {{\it et~al.}}(2018)\citenamefont
  {{Ghosh}}, \citenamefont {{Kumar}}, \citenamefont {{Marfatia}},\ and\
  \citenamefont {{Sand ick}}}]{Ghosh18}%
  \BibitemOpen
  \bibfield  {author} {\bibinfo {author} {\bibfnamefont {T.}~\bibnamefont
  {{Ghosh}}}, \bibinfo {author} {\bibfnamefont {J.}~\bibnamefont {{Kumar}}},
  \bibinfo {author} {\bibfnamefont {D.}~\bibnamefont {{Marfatia}}}, and\
  \bibinfo {author} {\bibfnamefont {P.}~\bibnamefont {{Sand ick}}},\ }\bibfield
   {title} {\enquote {\bibinfo {title} {{Searching for light from a dark matter
  clump}},}\ }\href {\doibase 10.1088/1475-7516/2018/08/023} {\bibfield
  {journal} {\bibinfo  {journal} {\jcap}\ }\textbf {\bibinfo {volume} {2018}},\
  \bibinfo {eid} {023} (\bibinfo {year} {2018})},\ \Eprint
  {http://arxiv.org/abs/1804.05792}{arXiv:1804.05792}\BibitemShut {NoStop}%
\bibitem [{\citenamefont {{Belotsky}}\ \emph {{\it et~al.}}(2019)\citenamefont
  {{Belotsky}}, \citenamefont {{Kamaletdinov}}, \citenamefont {{Laletin}},\
  and\ \citenamefont {{Solovyov}}}]{Belotsky19pdu}%
  \BibitemOpen
  \bibfield  {author} {\bibinfo {author} {\bibfnamefont {K.}~\bibnamefont
  {{Belotsky}}}, \bibinfo {author} {\bibfnamefont {A.}~\bibnamefont
  {{Kamaletdinov}}}, \bibinfo {author} {\bibfnamefont {M.}~\bibnamefont
  {{Laletin}}}, and\ \bibinfo {author} {\bibfnamefont {M.}~\bibnamefont
  {{Solovyov}}},\ }\bibfield  {title} {\enquote {\bibinfo {title} {{The DAMPE
  excess and gamma-ray constraints}},}\ }\href {\doibase
  10.1016/j.dark.2019.100333} {\bibfield  {journal} {\bibinfo  {journal}
  {Physics of the Dark Universe}\ }\textbf {\bibinfo {volume} {26}},\ \bibinfo
  {eid} {100333} (\bibinfo {year} {2019})},\ \Eprint
  {http://arxiv.org/abs/1904.02456}{arXiv:1904.02456}\BibitemShut {NoStop}%
\bibitem [{\citenamefont {{Catena}}\ and\ \citenamefont
  {{Ullio}}(2010)}]{localdmdens}%
  \BibitemOpen
  \bibfield  {author} {\bibinfo {author} {\bibfnamefont {R.}~\bibnamefont
  {{Catena}}} and\ \bibinfo {author} {\bibfnamefont {P.}~\bibnamefont
  {{Ullio}}},\ }\bibfield  {title} {\enquote {\bibinfo {title} {{A novel
  determination of the local dark matter density}},}\ }\href {\doibase
  10.1088/1475-7516/2010/08/004} {\bibfield  {journal} {\bibinfo  {journal}
  {\jcap}\ }\textbf {\bibinfo {volume} {2010}},\ \bibinfo {eid} {004} (\bibinfo
  {year} {2010})},\ \Eprint
  {http://arxiv.org/abs/0907.0018}{arXiv:0907.0018}\BibitemShut {NoStop}%
\bibitem [{\citenamefont {{Atoyan}}\ \emph {{\it et~al.}}(1995)\citenamefont
  {{Atoyan}}, \citenamefont {{Aharonian}},\ and\ \citenamefont
  {{V{\"o}lk}}}]{Atoyan95}%
  \BibitemOpen
  \bibfield  {author} {\bibinfo {author} {\bibfnamefont {A.~M.}\ \bibnamefont
  {{Atoyan}}}, \bibinfo {author} {\bibfnamefont {F.~A.}\ \bibnamefont
  {{Aharonian}}}, and\ \bibinfo {author} {\bibfnamefont {H.~J.}\ \bibnamefont
  {{V{\"o}lk}}},\ }\bibfield  {title} {\enquote {\bibinfo {title} {{Electrons
  and positrons in the galactic cosmic rays}},}\ }\href {\doibase
  10.1103/PhysRevD.52.3265} {\bibfield  {journal} {\bibinfo  {journal} {\prd}\
  }\textbf {\bibinfo {volume} {52}},\ \bibinfo {pages} {3265} (\bibinfo {year}
  {1995})}\BibitemShut {NoStop}%
\end{thebibliography}%
\end{document}